\documentclass[journal]{IEEEtran}
\usepackage{pbox}
\usepackage{enumerate}
\usepackage{cite}
\usepackage{graphicx}
\usepackage{graphicx, subfigure}
\usepackage{alltt}
\usepackage{amsmath}
\usepackage{amsopn}
\usepackage{amssymb}

\newtheorem{lemma}{Lemma}
\newtheorem{proposition}{Proposition}
\newtheorem{corollary}{Corollary}
\newtheorem{remark}{Remark}
\newtheorem{definition}{Definition}
\newtheorem{claim}{Claim}
\newtheorem{assumption}{Assumption}
\DeclareMathOperator{\mincut}{mincut}

\DeclareMathOperator{\out}{out}
\DeclareMathOperator{\inp}{in}

\DeclareMathOperator{\DC}{DC}
\DeclareMathOperator{\MBR}{MBR}
\DeclareMathOperator{\MSR}{MSR}

\DeclareMathOperator{\GF}{GF}

\newcommand{\gf}{\mathcal{G}_F}
\newcommand{\gfp}{\mathcal{G}_{F^+}}

\title{When and By How Much Can Helper Node Selection Improve Regenerating Codes?}

\author{Imad~Ahmad,~\IEEEmembership{Student Member,~IEEE,}
        and~Chih-Chun~Wang,~\IEEEmembership{Member,~IEEE}
\thanks{This work was supported in parts by NSF grants CCF-0845968, CNS-0905331, and CCF-1422997. Part of the results was presented in the 2014 Allerton Conference on Communication, Control, and Computing.}

\thanks{I. Ahmad and C.-C. Wang are with the School of Electrical and Computer Engineering, Purdue University, West Lafayette,
IN, 47906 USA e-mail: \{ahmadi,chihw\}@purdue.edu.}

}

\begin{document}
\maketitle
\thispagestyle{plain}
\pagestyle{plain}

\begin{abstract}
Regenerating codes (RCs) can significantly reduce the repair-bandwidth of distributed storage networks. Initially, the analysis of RCs was based on the assumption that during the repair process, the newcomer does not distinguish (among all surviving nodes) which nodes to access, i.e., the newcomer is oblivious to the set of helpers being used. Such a scheme is termed the {\em blind repair (BR)} scheme.  Nonetheless, it is intuitive in practice that the newcomer should choose to access only those ``good'' helpers. In this paper, a new characterization of the effect of choosing the helper nodes in terms of the storage-bandwidth tradeoff is given. Specifically, answers to the following fundamental questions are given: Under what conditions does proactively choosing the helper nodes improve the storage-bandwidth tradeoff? Can this improvement be analytically quantified?

\par
This paper answers the former question by providing a necessary and sufficient condition under which optimally choosing good helpers strictly improves the storage-bandwidth tradeoff. To answer the latter question, a low-complexity helper selection solution, termed the \emph{family repair (FR)} scheme, is proposed and the corresponding storage/repair-bandwidth curve is characterized. For example, consider a distributed storage network with $\mathbf{60}$ total number of nodes and the network is resilient against $\mathbf{50}$ node failures. If the number of helper nodes is $\mathbf{10}$, then the FR scheme and its variant demonstrate $\mathbf{27\%}$ reduction in the repair-bandwidth when compared to the BR solution. This paper also proves that under some design parameters, the FR scheme is indeed optimal among all helper selection schemes. An explicit construction of an exact-repair code is also proposed that can achieve the minimum-bandwidth-regenerating point of the FR scheme. The new exact-repair code can be viewed as a generalization of the existing \emph{fractional repetition} code.
\end{abstract}


\section{Introduction} \label{sec:intro}
\IEEEPARstart{T}{he} need for storing very large amounts of data reliably is one of the major reasons that has pushed for distributed storage systems. Examples of distributed storage systems include data centers \cite{ghemawat2003google} and peer-to-peer systems \cite{rhea2001maintenance,bhagwan2004total}.  One way to protect against data loss is by replication coding, i.e, if a disk in the network fails, it can be replaced and its data can be recovered from a replica disk. Another way is to use maximum distance separable (MDS) codes. Recently, regenerating codes (RCs) and its variants \cite{dimakis2010network,
wu2009reducing,rashmi2011optimal,shah2012distributed} have been used to further reduce the repair-bandwidth of MDS codes.

One possible mode of operation is to let the \emph{newcomer}, the node that replaces the failed node, {\em always} access/connect to all the remaining nodes. On the other hand, under some practical constraints we may be interested in letting the newcomer communicate with only a subset of the remaining nodes \cite{papailiopoulos2012locally}, termed the {\em helpers}. For example, reducing the number of helpers decreases I/O overhead during repair and thus mitigates one of the performance bottlenecks in cloud storage systems. In the original storage versus repair-bandwidth analysis of RCs \cite{dimakis2010network}, it is assumed that the newcomer does not distinguish/choose its helpers. We term such a solution the {\em blind repair (BR) scheme.}  Nonetheless, it is intuitive that the newcomer should choose to access only those ``good'' helpers of the remaining nodes. In fact, this idea of selecting good helpers exists even in replication codes, the simplest redundancy technique in the earliest literature of distributed storage systems.

To illustrate this, we consider a storage network with $4$ nodes numbered from $1$ to $4$. Suppose that we would like to protect against one node failure by replication. To that end, we first divide the file into two fragments, fragments $A$ and $B$, and we store fragment $A$ in node $1$ and fragment $B$ in node $2$. Each fragment is replicated once by storing a copy of fragment $A$ in node $3$ and a copy of fragment $B$ in node $4$.  If any one of the four nodes fails, then we can retrieve the entire file by accessing the intact fragments $A$ and $B$ in the remaining three nodes. The repair process of this replication scheme is also straightforward. Say node $4$ fails, the newcomer simply accesses node $2$ and restores fragment $B$. We observe that the newcomer only accesses the good helper (the one that stores the lost fragment) in this replication scheme. In this scheme, each node stores half of the file, and during the repair process, the newcomer accesses $1$ helper node and communicates half of the file. For comparison, if we apply the analysis of \cite{dimakis2010network} (also see our discussion in the next paragraph), we will see that if we use RCs to protect against one node failure, each node has to store the whole file and during the repair process, the newcomer accesses $1$ helper and communicates the entire file.  {\em The simplest replication code is twice more efficient than RCs in this example.}\footnote{One may think that this performance improvement over the blind repair (BR) scheme \cite{dimakis2010network} is due to that the parameter values $(n,k,d)=(4,3,1)$ are beyond what is originally considered for the regenerating codes (which requires $k\leq d$). In Appendix~\ref{app:example} and Section~\ref{sec:preview}, we provide other examples with $(n,k,d)=(6,3,3)$ and $(6,4,4)$, respectively, which show that a good helper selection can strictly outperform the BR solution in \cite{dimakis2010network} for $k\leq d$ as well.}

The reason why the replication code is the superior choice in the above example is that it only chooses the good helpers during the repair process, while the analysis in \cite{dimakis2010network} assumes a blind helper selection.\footnote{\label{footnote:extremes}Since our setting considers choosing the good helpers, it brings the two extremes: replication codes with helper selection and regenerating codes with blind helper selection, under the same analytical framework.} To illustrate this, suppose the newcomer does not choose good helper nodes but chooses the helpers blindly. One possibility is as follows. Suppose node $2$ fails first, and we let the new node $2$ choose node $1$ as the helper. Then suppose node $3$ fails and we let node $1$ again be the helper. Finally, suppose node $4$ fails and we let node $1$ be the helper. Since the content of all four nodes are now originating from the same node (node $1$), each node needs to store a complete copy of the file otherwise the network cannot tolerate the case when node $1$ fails.  As can be seen, blind repair is the main cause of the performance loss, i.e., every newcomer blindly requests help from the same node, node 1, which lacks the ``diversity'' necessary for implementing an efficient distributed storage system. Another insightful example with parameter values $(n,k,d)=(6,3,3)$ is provided in Appendix~\ref{app:example}.

The idea of choosing good helpers in RC has already been used in constructing exact-repair codes as in \cite{el2010fractional,papailiopoulos2012simple}. Under the subject of {\em locally repairable codes} some progress in analyzing this problem has been done on the minimum-storage point in \cite{gopalan2012locality,prakash2012optimal,papailiopoulos2012locally} when helper selection is fixed over time (See Section~\ref{subsec:comparison} for an in-depth comparison with these references). Reference \cite{el2010fractional} also observes that choosing good helpers
can strictly outperform BR at the minimum-bandwidth point. However, a complete characterization of the effect of choosing the helper nodes in RC, including {\em stationary} and {\em dynamic} helper selection, on the storage-bandwidth tradeoff is still lacking. This motivates the following open questions: Under what condition is it beneficial to proactively choose the helper nodes? Is it possible to analytically quantify the benefits of choosing the good helpers? Specifically, the answers to the aforementioned fundamental questions were still not known.

\par In this work, we answer the first question by providing a necessary and sufficient condition under which optimally choosing the helpers strictly improves the storage-bandwidth tradeoff. This new necessary and sufficient characterization of ``under what circumstances helper selection improves the performance'' is by far the most important contribution of this work since it provides a rigorous benchmark/guideline when designing the next-generation smart helper selection solutions. 

\par It is worth reemphasizing that which helpers are ``optimal'' at the current time slot $t$ depends on the history of the failure patterns and the helper choices for all the previous time slots $1$ to $(t-1)$, which makes it very difficult to quantify the corresponding performance. Therefore, even though our main result fully answers the question {\em whether} an optimal design can outperform the blind helper selection, the question {\em how} to design the optimal helper selection scheme remains largely open. As part of the continuing quest of designing high-performance helper selection methods, this work also proposes a low-complexity solution, termed the {\em family repair (FR) scheme}, that can harvest the benefits of (careful) helper selection without incurring any additional complexity when compared to a BR solution. We then characterize analytically the storage-bandwidth tradeoff of the FR scheme and its extension, the family-plus repair scheme, and prove that they are optimal (as good as any helper selection one can envision) in some cases and {\em weakly optimal} in general, see the discussion in Sections~\ref{sec:results} and \ref{sec:family-plus}. 

\par Finally, we provide in Section~\ref{sec:gfr} an explicit construction of an exact-repair code that can achieve the minimum-bandwidth-regenerating (MBR) points of the FR and family-plus repair schemes. The new MBR-point scheme is termed the \emph{generalized fractional repetition} code, which can be viewed as a generalization of the existing fractional repetition codes \cite{el2010fractional}.

Numerical computation shows that for many cases (different $(n,k,d)$ parameter values), the family-based schemes can reduce 40\% to 90\% of the repair-bandwidth of RCs when the same amount of storage space is used.

\section{Problem Statement} \label{sec:prob_stat}
\subsection{Functional-Repair Regenerating Codes with Dynamic Helper Selection}
Following the notation of the seminal paper \cite{dimakis2010network}, we denote the total number of nodes in a storage network by $n$ and the minimum number of nodes that are required to reconstruct the file by $k$. We denote by $d$ the number of helper nodes that a newcomer can access. From the above definitions, the $n$, $k$, and $d$ values must satisfy
\begin{align}
2\leq n,\quad 1\leq k\leq n,\quad\text{and}\quad 1\leq d\leq n-1.\label{eq:ccw1}
\end{align}
In all the results in this work, we assume {\em implicitly} that the $n$, $k$, and $d$ values satisfy\footnote{The following fact is proved in \cite{dimakis2010network}. Suppose $k>d$. If the storage $\alpha$ and the repair-bandwidth $\beta$ of each node allow the storage network to tolerate $(n-k)$ failed nodes using {\em blind-repair} (BR) regenerating codes, then the same storage network with BR codes can actually tolerate $(n-d)$ failed nodes. Therefore, any BR regenerating code that can support the values $(n,k,d)$ for some $k>d$ can also support the values $(n,d,d)$. By definition, any regenerating code that can support the values $(n,d,d)$ can also support the values $(n,k,d)$ for any $k>d$. This shows that for BR, the storage-bandwidth tradeoff of the values $(n,k,d)$ is identical to that of the values $(n,d,d)$ when $k>d$. This fact prompts the authors in \cite{dimakis2010network} to study only the case in which $k\leq d$ and use the results of $(n,d,d)$ as a substitute whenever we are considering the case of $k>d$. As will be seen later, the above equivalence between the $(n,k,d)$ and the $(n,d,d)$ cases when $k>d$ does not hold when considering non-blind helper selection. Therefore, throughout this paper, we do not assume $k\leq d$.

Also, in practice the parameter $k$ specifies the resilience of the system and the parameter $d$ specifies the repair cost. The choices of $k$ and $d$ values are generally orthogonal from a high-level design perspective. Any coupling between $k$ and $d$ is usually imposed by the kind of storage codes used, e.g., replication versus Reed-Solomon versus regenerating codes versus locally repairable codes. Since we are studying the most general form of helper-selection, we discard the assumption of $k\leq d$, which was originally used for the BR solution.} \eqref{eq:ccw1}. The overall file size is denoted by $\mathcal{M}$. The storage size for each node is $\alpha$, and during the repair process, the newcomer requests $\beta$ amount of traffic from each of the helpers. The total repair-bandwidth is thus $\gamma\stackrel{\Delta}{=}d\beta$.  We use the notation $(\cdot)^+$ to mean $(x)^+=\max(x,0)$. We also define the indicator function as follows
\begin{align}
1_{\{B\}}=
\begin{cases}1, \mbox{ if condition $B$ is true} \\
0, \mbox{ otherwise}.
\end{cases}
\end{align}

\par
In this work, we consider the helper selection/repair scheme in its most general form. Among all helper selection schemes, a special class, termed stationary repair schemes, is also studied. To distinguish the special class from the most general form, we use the term {\em dynamic repair} schemes whenever we are focusing on the most general type of helper selection schemes. In addition to studying the performance of any dynamic or stationary repair scheme, this work also proposes a new low-complexity solution, termed the family repair schemes. Detailed discussion of dynamic repair and stationary repair is provided in the following.

\subsection{Dynamic Versus Stationary Repair Schemes}

In general, the helper selection at current time $t$ can depend on the history of the failure patterns and the helper choices for all the previous time slots 1 to $(t-1)$. We call such a general helper selection scheme {\em the dynamic helper selection}. In contrast, a much simpler way of choosing the helpers, termed \emph{stationary helper selection} (or stationary repair scheme), is described as follows.

\par {\em Stationary Repair:} Each node index $i$ is associated with a set of indices $D_i$ where the size of $D_i$ is $d$. Whenever node $i$ fails, the newcomer (for node $i$) simply accesses those helpers $j$ in $D_i$ and requests $\beta$ amount of data from each helper. It is called stationary since the helper choices $\{D_1,D_2,\dots,D_n\}$ are fixed and do not evolve over time. As can be easily seen, the stationary repair scheme is a special case of (dynamic) helper selection, which incurs zero additional complexity when compared to the BR solution.

For any helper selection scheme $A$ and given system parameters $(n,k,d,\alpha,\beta)$, we say that the corresponding RC with helper selection scheme $A$ ``satisfies the reliability requirement'' if it is able to protect against any failure pattern/history while being able to reconstruct the original file from arbitrary $k$ surviving nodes. We consider exclusively single failure at any given time. The setting of multiple simultaneous failed nodes \cite{el2010fractional,shum2013cooperative,kamath2013codes} is beyond the scope of this work.

\subsection{Information Flow Graphs and the Existing Results}\label{subsec:ifg_existing}
As in \cite{dimakis2010network}, the performance of a distributed storage system can be characterized by the concept of information flow graphs (IFGs). This IFG depicts the storage in the network and the communication that takes place during repair as will be described in the following.

\begin{figure}[h!]
\centering
\includegraphics[width=0.45\textwidth]{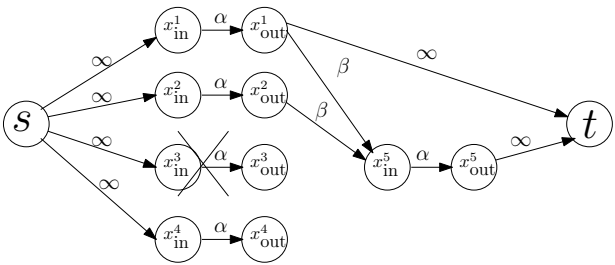} 
\caption{An example of the information flow graph with $(n,k,d)=(4,2,2)$.}
\label{fig:ifg}
\end{figure}

As shown in Fig~\ref{fig:ifg}, an IFG has three different kinds of nodes. It has a single \emph{source} node $s$ that represents the source of the data object. It also has nodes $x_{\inp}^i$ and $x_{\out}^i$ that represent storage node $i$ of the IFG. A storage node is split into two nodes so that the IFG can represent the storage capacity of the nodes. We often refer to the pair of nodes $x_{\inp}^i$ and $x_{\out}^i$ simply by storage node $i$. In addition to those nodes, the IFG has \emph{data collector} (DC) nodes. Each data collector node is connected to a set of $k$ active storage nodes, which represents the party that is interested in extracting the original data object initially produced by the source $s$. Fig.~\ref{fig:ifg} illustrates one such data collector, denoted by $t$, which connects to $k=2$ storage nodes. A more detailed description of the IFG is provided as follows.

The IFG evolves with time. In the first stage of an information flow graph, the source node $s$ communicates the data object to all the initial nodes of the storage network. We represent this communication by edges of infinite capacity as this stage of the IFG is virtual. See Fig.~\ref{fig:ifg} for illustration. This stage models the encoding of the data object over the storage network. To represent storage capacity, an edge of capacity $\alpha$ connects the input node of storage nodes to the corresponding output node. When a node fails in the storage network, we represent that by a new stage in the IFG where, as shown in Fig.~\ref{fig:ifg}, the newcomer connects to its helpers by edges of capacity $\beta$ resembling the amount of data communicated from each helper. We note that although the failed node still exists in the IFG, it cannot participate in helping future newcomers. Accordingly, we refer to failed nodes by \emph{inactive} nodes and existing nodes by \emph{active} nodes. By the nature of the repair problem, the IFG is always acyclic.

Intuitively, each IFG reflects one unique history of the failure patterns and the helper selection choices from time $1$ to $(t-1)$ \cite{dimakis2010network}. Consider any given helper selection scheme $A$ which can be either dynamic or stationary. Since there are infinitely many different failure patterns (since we consider $t=1$ to $\infty$), there are infinitely many IFGs corresponding to the same given helper selection scheme $A$. We denote the collection of all such IFGs by $\mathcal{G}_A(n,k,d,\alpha,\beta)$. We define $\mathcal{G}(n,k,d,\alpha,\beta)=\bigcup_{\forall A}\mathcal{G}_A(n,k,d,\alpha,\beta)$ as the union over all possible helper selection schemes $A$. We sometimes drop the input argument and use $\mathcal{G}_A$ and $\mathcal{G}$ as shorthands.

Given an IFG $G\in \mathcal{G}$, we use $\DC(G)$ to denote the collection of all ${n\choose k}$ {\em data collector nodes} in $G$ \cite{dimakis2010network}. Each data collector $t\in\DC(G)$ represents one unique way of choosing $k$ out of $n$ active nodes when reconstructing the file. Given an IFG $G\in\mathcal{G}$ and a data collector $t\in \DC(G)$, we use $\mincut_G(s,t)$ to denote the {\em minimum cut value} \cite{west2001introduction} separating $s$, the root node (source node) of $G$, and $t$.

The key reason behind representing the repair problem by an IFG is that it casts the problem as a multicast scenario \cite{dimakis2010network}. This allows for invoking the results of network coding in \cite{ahlswede2000network}, \cite{ho2006random}. More specifically, for any helper scheme $A$ and given system parameters $(n,k,d,\alpha,\beta)$, the results in \cite{ahlswede2000network} prove that the following condition is \emph{necessary} for the RC with helper selection scheme $A$ to satisfy the reliability requirement.
\begin{align}
\min_{G\in \mathcal{G}_A}\min_{t\in \DC(G)}\mincut_G(s,t)\geq \mathcal{M}. \label{eq:condition}
\end{align}
If we limit our focus to the blind repair scheme, then the above necessary condition becomes
\begin{align}
\min_{G\in \mathcal{G}}\min_{t\in \DC(G)}\mincut_G(s,t)\geq \mathcal{M}.\label{eq:condition-BR}
\end{align}
Reference \cite{dimakis2010network} found a closed-form expression of the LHS of \eqref{eq:condition-BR}
\begin{align}
\min_{G\in \mathcal{G}}\min_{t\in \DC(G)}\mincut_G(s,t)=\sum_{i=0}^{k-1} \min((d-i)^+\beta,\alpha),\label{eq:ex_low_b}
\end{align}
which allows us to numerically check whether \eqref{eq:condition-BR} is true (or equivalently whether ``\eqref{eq:ex_low_b} $\geq \mathcal{M}$'') for any $(n,k,d,\alpha,\beta)$ values. Being a necessary condition for the blind repair scheme implies that whenever ``\eqref{eq:ex_low_b} $<\mathcal{M}$'' there exists a bad helper selection scheme $A$ for which the reliability requirement cannot be met.

\par Reference \cite{wu2010existence} further proves that \eqref{eq:condition-BR} is not only necessary but also sufficient for the existence of a blind RC with some finite field $\GF(q)$ that satisfies the reliability requirement. Namely, as long as ``\eqref{eq:ex_low_b} $\geq \mathcal{M}$" is true, then there exists a RC that meets the reliability requirement even for the worst possible helper selection scheme (since we take the minimum over $\mathcal{G}$).

\subsection{The Minimum-Bandwidth and Minimum-Storage Points}
\label{sec:br_extreme_points} 

Fix the values of $(n,k,d)$, ``\eqref{eq:ex_low_b} $\geq \mathcal{M}$'' describes the storage-bandwidth tradeoff ($\alpha$ versus $\beta$) of the BR scheme. Two points on a storage-bandwidth tradeoff curve are of special interest: the minimum-bandwidth regenerating code (MBR) point and the minimum-storage regenerating code (MSR) point where the former has the smallest possible repair-bandwidth (the $\beta$ value) and the latter has the smallest possible storage per node (the $\alpha$ value). The expressions of the MBR and MSR points ($\alpha_{\MBR}$,$\gamma_{\MBR}$) and ($\alpha_{\MSR}$,$\gamma_{\MSR}$) of the BR scheme are derived in \cite{dimakis2010network}:

\begin{align}\label{eq:mbr_br}
\alpha_{\MBR}&=\gamma_{\MBR}= \nonumber\\
&\frac{2d\mathcal{M}}{\min(d,k)(2d-\min(d,k)+1)} 
\end{align}
and
\begin{align}
\alpha_{\MSR}&=\frac{\mathcal{M}}{\min(d,k)}, \label{eq:msr_br_alpha} \\
\gamma_{\MSR}&=\frac{d\mathcal{M}}{\min(d,k)(d-\min(d,k)+1)} \label{eq:msr_br_gamma}.
\end{align}

\subsection{Characterizing the RC with Helper Selection Scheme $A$}\label{subsec:characterizing}
In contrast with the existing results on the BR scheme that hold for the {\em worst} possible helper selection scheme, this work focuses on any given helper selection scheme $A$ and studies the impact of the given helper selection scheme on the storage-bandwidth tradeoff of the corresponding regenerating codes.  To facilitate the discussion, we assume the following statement holds for the given helper selection $A$.
\begin{assumption} \label{ass:finite_field}
\eqref{eq:condition} is not only necessary but also {\em sufficient} for the existence of an RC with helper selection scheme $A$ that satisfies the reliability requirement.
\end{assumption}

This assumption allows us to use \eqref{eq:condition} as the complete characterization for the RC with a given helper selection scheme $A$. We then note that it is possible mathematically that when focusing on $\mathcal{G}_A$ ($\mathcal{G}_A$ is by definition a strict subset of $\mathcal{G}$) we may have
\begin{align}
\min_{G\in \mathcal{G}_A}\min_{t\in \DC(G)}\mincut_G(s,t)> \min_{G\in \mathcal{G}}\min_{t\in \DC(G)}\mincut_G(s,t). \label{eq:outperform}
\end{align}
If \eqref{eq:outperform} is true, then the given helper selection scheme $A$ strictly outperforms the BR solution. Whether (or under what condition) \eqref{eq:outperform} is true and how much the gap can be are the two main focuses of this work.

\begin{remark}
As discussed in Section~\ref{subsec:ifg_existing}, the necessary direction of Assumption~\ref{ass:finite_field} is always true \cite{ahlswede2000network}. The sufficient direction of Assumption~\ref{ass:finite_field} is equivalent to the following statement: For any helper selection scheme $A$ and any $(n,k,d,\alpha, \beta)$ values satisfying \eqref{eq:condition}, there exists a finite field $\GF(q)$ such that the corresponding RC satisfies the reliability requirement. Many similar statements have been proved in the existing works\footnote{In fact, there is not yet any example in which the min-cut-based characterization is provably not achievable by any finite field. }  (e.g.,~\cite{wu2010existence}). However, rigorous proofs are still  needed for the sufficiency direction of Assumption~\ref{ass:finite_field} and we leave them as future directions of this work. On the other hand, we have proved the following partial statement in Section~\ref{sec:gfr}. 
\begin{quote} {\em Sufficiency for the MBR points:} For the two helper selection schemes proposed in this work, termed the family repair and the family repair plus schemes, if the $(\alpha,\beta)$ values correspond to the minimum-bandwidth regenerating (MBR) point of the corresponding storage-bandwidth tradeoff, then Assumption~\ref{ass:finite_field} is provably true.
\end{quote}
As will be discussed in Section~\ref{subsec:mbr}, the MBR point is the point when good helper selection results in the largest improvement over the blind repair scheme. Since our focus is on quantifying the benefits of helper selection, the above partial statement proved in Section~\ref{sec:gfr} is sufficient for our discussion.
\end{remark}

\subsection{Comparison to Locally Repairable Codes}\label{subsec:comparison}
\begin{table*}[t]
\caption{The comparison table among blind-repair regenerating codes, locally repairable codes, and the smart-repair regenerating codes.}
\begin{center}
\begin{tabular}{| p{2cm} || p{4.5cm} | p{4.5cm} | p{4.5cm} |}
\hline
    &Original RC \cite{dimakis2010network,wu2009reducing,rashmi2009explicit,shah2012interference,rashmi2011optimal}& Locally Repairable Codes \cite{gopalan2012locality,prakash2012optimal,papailiopoulos2012locally,kamath2013codes,rawat2012optimal,kamath2013explicit}&
Dynamic Helper Selection\\
\hline\hline

Repair Mode& Functional/Exact-Repair & Exact-Repair & Functional\footnotemark{} Repair\\ \hline

Helper Selection&Blind&Stationary (Fixed over time)& Dynamic (helper choices may depend on failure history)\\ \hline

$(n,k,d)$ range&
\begin{enumerate}
\item[(1)] Designed for $k\leq d$.
\item[(2)] Can still be applied to the case of $k>d$ with reduced efficiency.
\end{enumerate}
&\begin{enumerate}
\item[(1)] Designed for $k>d$. 
\item[(2)] Can still be applied to the case of $k\leq d$ with reduced efficiency.
\end{enumerate}
&Allow for arbitrary $(n,k,d)$ values\\\hline

Contribution&Storage/repair-bandwidth tradeoff for the worst possible helper selection&Storage/repair-bandwidth characterization for the specific stationary helper selection of the proposed exact-repair local code, which may/may not be optimal&First exploration of the storage/repair-bandwidth tradeoff for the optimal dynamic helper selection\\

\hline
\end{tabular}
\label{tab:comparison}
\end{center}
\end{table*}

Recall that RCs are distributed storage codes that minimize the repair-bandwidth (given a storage constraint). In comparison, {\em locally repairable codes (LRC)}, recently introduced in \cite{gopalan2012locality}, are codes that minimize the number of helpers participating in the repair of a failed node. LRCs were proposed to address the disk I/O overhead problem that the repair process can entail on a storage network since the number of helpers participating in the repair of a failed node is proportional to the amount of disk I/O needed during repair. Subsequent development has been done on LRCs in \cite{prakash2012optimal,papailiopoulos2012locally,kamath2013codes,rawat2012optimal,kamath2013explicit}.

\par In Table~\ref{tab:comparison}, we compare the setting of the original RCs, LRCs, and the dynamic helper selection considered in this work. As first introduced in \cite{dimakis2010network}, original RCs were proposed under the functional-repair scenario, i.e., nodes of the storage network are allowed to store any combination of the original packets as long as the reliability requirement is statisfied. In subsequent works \cite{wu2009reducing,rashmi2009explicit,shah2012interference,rashmi2011optimal,
cadambe2013asymptotic,shah2012distributed}, RCs were considered under the exact-repair scenario in which nodes have to store the same original packets at any given time. In contrast, LRCs are almost always considered under the exact-repair scenario. However, in this work, for RCs with dynamic helper selection, we consider functional-repair as the mode of repair as we aim at understanding the absolute benefits/limits of helper selection in RCs. Albeit our setting is under functional-repair, in Section~\ref{sec:gfr}, we are able to present an explicit construction of exact-repair codes that achieve the optimal or weakly optimal minimum-bandwidth point of the functional-repair. For comparison, existing works \cite{rashmi2011optimal,el2010fractional} design an exact-repair scheme that achieves the minimum-bandwidth regenerating (MBR) point of the ``blind-functional-repair''. The main difference is that our exact-repair construction achieves the MBR point of the ``smart-functional-repair''. 

\par Table~\ref{tab:comparison} also summarizes the differences between RCs, LRCs, and smart helper RCs in terms of the helper selection mechanisms. The original RCs are codes that do not perform helper selection at all, i.e., BR, while LRCs are codes that can perform stationary helper selection only. In this work, we consider the most general setting in which codes are allowed to have dynamic helper selection. Surprisingly, we are able to find a stationary helper selection scheme that is weakly optimal among all dynamic schemes and strictly optimal for a range of $(n,k,d)$ values. 

\par Another dimension in this comparison table is the $(n,k,d)$ values that each of the three codes addresses. The original RCs were designed for storage networks with large $d$ values as they perform rather poorly when applied to small $d$ values. LRCs, on the other hand, are designed for small $d$ values, and for that reason, they perform poorly when $d$ is large. In contrast, the codes we present in this work are designed for arbitrary $(n,k,d)$ values. 
\par The comparison above illustrates the main differences in the goals/contributions of each scenario. Namely, the original RCs are concerned with the storage/repair-bandwidth tradeoff for the worst possible helper selection. LRCs, however, are concerned with only data storage (ignoring repair-bandwidth) of the codes when restricting to stationary helper selection and exact-repair. Some recent developments \cite{kamath2013codes,kamath2013explicit} in LRCs consider using RCs in the construction of the codes therein (as local codes) in an attempt to examine the repair-bandwidth performance of LRCs. This approach, however, is not guaranteed to be optimal in terms of storage/repair-bandwidth tradeoff. 

\par In this work, we present the first exploration of the optimal storage-bandwidth tradeoff for RCs that allow {\em dynamic helper selection} for arbitrary $(n,k,d)$ values, including both the cases of $k\gg d$ and $k\ll d$. The closest setting in the existing literature is in a very recent work in \cite{hollmann2014minimum}. That work finds upper bounds on the file size $\mathcal{M}$ when $\alpha=d\beta$ and $\alpha=\beta$ for functional-repair with dynamic helper selection. However, \cite{hollmann2014minimum} considers the case of $k=n-1$ only. Also, it is not clear whether the provided upper bounds for $k=n-1$ are tight or not. A byproduct of the results of this work shows that the upper bounds in \cite{hollmann2014minimum} are tight in some cases and loose in others, see Corollary~\ref{cor:existing_loose} and Propositions~\ref{prop:optimal_2} and~\ref{prop:family-plus_optimal}. 

\section{Preview Of The Results} \label{sec:preview}
\footnotetext{{A (weakly) optimal exact-repair code construction is also provided in Section\ref{sec:gfr}}}

In the following, we give a brief preview of our results through concrete examples to illustrate the main contributions of this work. Although we only present here specific examples as a preview, the main results in Section~\ref{sec:results} are for general $(n,k,d)$ values.

\par\emph{Result~1:} For $(n,k,d)=(6,3,4)$, RCs with BR are absolutely optimal, i.e., there exists no RCs with dynamic helper selection that can outperform BR. Since LRCs with symmetric repair can be viewed as a specially-designed stationary helper selection with exact-repair, this also implies that for $(n,k,d)=(6,3,4)$ there exists no LRCs with symmetric repair-bandwidth per node that can outperform BR.

\par
\begin{figure}[h!]
\centering
\includegraphics[width=0.475\textwidth]{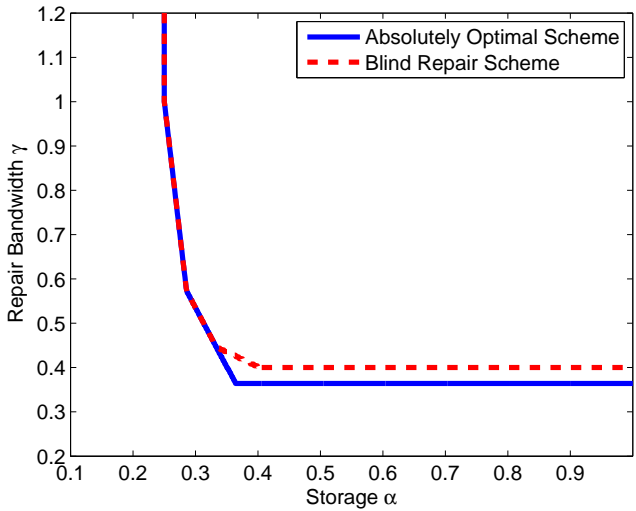}
\caption{Storage-bandwidth tradeoff curves of RCs with BR versus RCs with the absolutely optimal scheme (FR) for $(n,k,d)=(6,4,4)$ and file size $\mathcal{M}=1$.}
\label{fig:storage_vs_bandwidth_(6-4-4)}
\end{figure}

\emph{Result~2:} For $(n,k,d)=(6,4,4)$, the RCs with family repair (FR) proposed in this paper are absolutely optimal in terms of the storage-bandwidth tradeoff among all RCs with dynamic helper selection. In Fig.~\ref{fig:storage_vs_bandwidth_(6-4-4)}, the storage-bandwidth tradeoff curve of the FR scheme, the optimal helper selection scheme, is plotted against the BR scheme with file size $\mathcal{M}=1$. In Section~\ref{sec:gfr}, we provide an explicit construction of an exact-repair code that can achieve $(\alpha,\gamma)=(\frac{4}{11},\frac{4}{11})$, the MBR point of the storage-bandwidth tradeoff curve of the FR scheme in Fig.~\ref{fig:storage_vs_bandwidth_(6-4-4)}. If we take a closer look at Fig.~\ref{fig:storage_vs_bandwidth_(6-4-4)}, there are 3 corner points on the FR scheme curve and they are $(\alpha,\gamma)=(0.25,1)$, $(\frac{2}{7},\frac{4}{7})$, and $(\frac{4}{11},\frac{4}{11})$. Since the two corners  $(\alpha,\gamma)=(0.25,1)$ and $(\frac{2}{7},\frac{4}{7})$ can be achieved by the scheme in \cite{wu2010existence} and the new corner point $(\alpha,\gamma)=(\frac{4}{11},\frac{4}{11})$ is proved to be achievable in Proposition~\ref{prop:gfr_rec}, we can thus achieve the entire optimal tradeoff curve in Fig.~\ref{fig:storage_vs_bandwidth_(6-4-4)} by space-sharing while no other scheme can do better, as proved in Proposition~\ref{prop:optimal}. In fact, for $(n,k,d)=(6,4,4)$, the random LRCs in \cite{papailiopoulos2012locally} designed for $\gamma=\infty$ have to satisfy $\mathcal{M}\leq k\alpha=4\alpha$, i.e., can at most perform as good as the MSR point $(\alpha,\gamma)=(0.25,1)$ of the BR scheme. Moreover, the LRCs utilizing MBR codes in \cite{kamath2013explicit} perform equally to the MBR point $(\alpha,\gamma)=(0.4,0.4)$ of the BR scheme. Both LRC constructions in \cite{kamath2013explicit} and \cite{papailiopoulos2012locally} are strictly suboptimal and perform worse than the proposed family repair scheme, which is provably optimal for $(n,k,d)=(6,4,4)$.

\par
\begin{figure}[h!]
\centering
\includegraphics[width=0.475\textwidth]{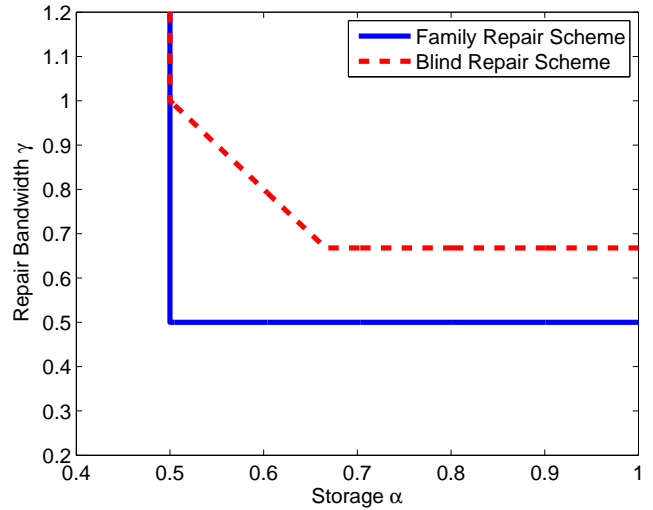}
\caption{Storage-bandwidth tradeoff curves of RCs with BR versus RCs with FR for $(n,k,d)=(5,3,2)$ and file size $\mathcal{M}=1$.}
\label{fig:storage_vs_bandwidth_(5-3-2)}
\end{figure}

\emph{Result~3:} For $(n,k,d)=(5,3,2)$, we do not know what is the absolutely optimal dynamic helper selection scheme. On the other hand, the proposed FR scheme again outperforms the BR scheme. Fig.~\ref{fig:storage_vs_bandwidth_(5-3-2)} shows a tradeoff curve comparison between the FR scheme and the BR scheme. An interesting phenomenon is that the tradeoff curve of the FR scheme has only one corner point $(\alpha,\gamma)=(0.5,0.5)$ and we can achieve this point by an exact-repair scheme, see Proposition~\ref{prop:gfr_rec}. Note that this exact-repair scheme for $(\alpha,\gamma)=(0.5,0.5)$ has the same storage consumption as the MSR point of the original RC ($(\alpha,\gamma)=(0.5,1)$) while using strictly less than the bandwidth of the MBR point of the original RC ($(\alpha,\gamma)=(\frac{2}{3},\frac{2}{3})$). Since the tradeoff curve of the FR scheme has only 1 corner point, it also suggests that with smart helper selection, it is possible to achieve minimum-storage (MSR) and minimum-bandwidth (MBR) simultaneously.

\par
\begin{figure}[h!]
\centering
\includegraphics[width=0.475\textwidth]{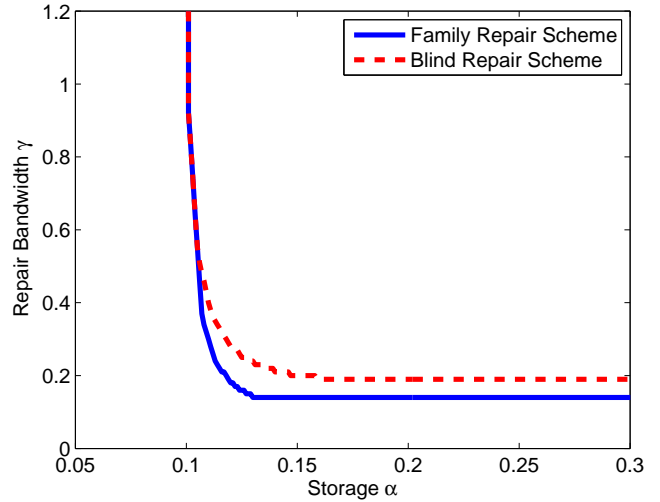}
\caption{Storage-bandwidth tradeoff curves of RCs with BR versus RCs with FR for $(n,k,d)=(20,10,10)$ and file size $\mathcal{M}=1$.}
\label{fig:storage_vs_bandwidth_(20-10-10)}
\end{figure}

\emph{Resul~4:} For $(n,k,d)=(20,10,10)$, we do not know what is the absolutely optimal dynamic helper selection scheme. We, however, have that the FR scheme again outperforms the BR scheme.  Fig.~\ref{fig:storage_vs_bandwidth_(20-10-10)} shows a tradeoff curve comparison between the FR scheme and the BR scheme.

\par
\begin{figure}[h!]
\centering
\includegraphics[width=0.475\textwidth]{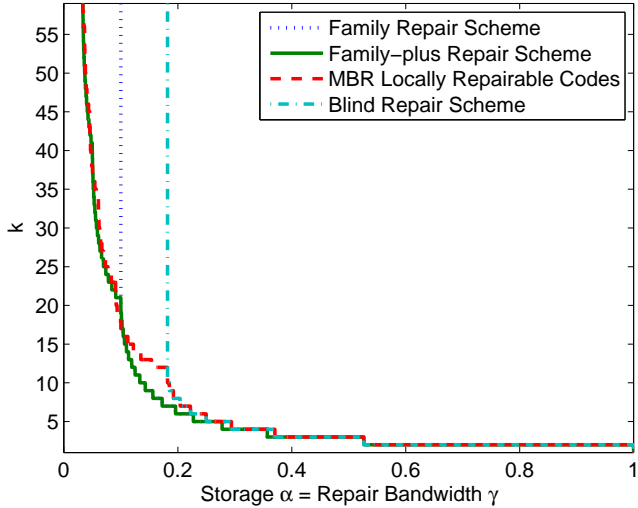}
\caption{The $k$ value versus repair-bandwidth $\gamma$ curve comparison at the MBR point for $(n,d)=(60,10)$ and file size $\mathcal{M}=1$.}
\label{fig:k_vs_gamma_(60-10)}
\end{figure}

\emph{Result~5:} For $(n,d)=(60,10)$, we do not know what is the absolutely optimal dynamic helper selection. However, in Fig.~\ref{fig:k_vs_gamma_(60-10)}, we plot a $k$ versus repair-bandwidth curve to compare the blind repair scheme to the FR scheme when restricting to the minimum-bandwidth (MBR) points. The curve of the MBR LRCs in \cite{kamath2013explicit} is also provided in the same figure. Note that the family-plus repair scheme in the figure, described in Section~\ref{sec:family-plus}, is an extension of the FR scheme to cover the case when $n\gg d$. Examining Fig.~\ref{fig:k_vs_gamma_(60-10)}, we can see that the BR scheme performs very poorly compared to the other codes when $k$ is large. Comparing the plots of the family-plus repair scheme to the plot of the MBR LRCs, we can see that the MBR LRCs perform equally when $k$ is very large but performs poorly otherwise (say when $k=10$). From this, we see that RCs with the family-plus repair scheme perform well for arbitrary $(n,k,d)$ values as discussed in Table~\ref{tab:comparison}.

\par\emph{Result~6:} Although the main focus of this work is on investigating the benefits of helper selection, a byproduct of our results is a new explicit construction of locally repairable codes (LRCs) for arbitrary $(n,k,d,\alpha,\beta)$ values satisfying $\alpha=d\beta$. Numerically, the proposed LRCs demonstrate good performance in all $(n,k,d)$ cases. Analytically, it achieves  the absolutely optimal MBR points (using the smallest possible bandwidth among all dynamic helper selection schemes) for all $(n,k,d,\alpha,\beta)$ values satisfying (i) $n\neq 5$, $k=n-1$, and $d=2$; (ii) $n$ is even, $k=n-1$, and $d=3$; (iii) $n\notin \{7,9\}$, $k=n-1$, and $d=4$; (iv) $n$ is even, $n\notin \{8,14\}$, $k=n-1$, and $d=5$; and (v) $n\notin \{10,11,13\}$, $k=n-1$, and $d=6$. This result is the combination of Proposition~\ref{prop:family-plus_optimal} and the explicit code construction in Section~\ref{sec:gfr}.

\section{The Main Results} 
\label{sec:results}
Our main results include two parts. In Section~\ref{subsec:when}, we answer the question ``When is it beneficial to choose the good helpers?'' In Section~\ref{subsec:how_much}, we quantify the potential benefits of good helper selection by characterizing the storage-bandwidth tradeoff of the family repair (FR) scheme proposed in Section~\ref{subsec:desc_fr}. Since the FR scheme is a special example of the general dynamic helper selection, the improvement of the FR scheme over the blind repair (BR) scheme serves as a lower bound for the improvement of the optimal dynamic repair scheme over the BR scheme.

\par It is worth noting that the first part, answering when it is beneficial to choose good helpers, is of more importance since it completely solves an open fundamental problem. At the same time, the second part can be viewed as an attempt towards finding the optimal helper selection schemes for general $(n,k,d)$ values. For comparison, the existing LRC constructions \cite{papailiopoulos2012locally,kamath2013explicit} are other ways of designing smart helper repair solutions for a subset of $(n,k,d)$ values.

\subsection{When Is It Beneficial to Choose the Good Helpers?}\label{subsec:when}

Recall that we only consider $(n,k,d)$ values that satisfy \eqref{eq:ccw1}.

\begin{proposition}  \label{prop:comparison}
If at least one of the following two conditions is true:  (i) $d=1$, $k=3$, and $n$ is odd; and (ii) $k\leq \left\lceil \frac{n}{n-d}\right\rceil$, then for any arbitrary dynamic helper selection scheme $A$ and any arbitrary $(\alpha,\beta)$ values, we have
\begin{align} \label{eq:neg}
\min_{G\in\mathcal{G}_A}\min_{t\in \DC(G) } \mincut_G(s,t)=\sum_{i=0}^{k-1}\min ((d-i)^+\beta,\alpha).
\end{align}
That is, even the best dynamic repair scheme cannot do better than the BR solution. Conversely, for any $(n,k,d)$ values that satisfy neither (i) nor (ii), there exists a helper selection scheme $A$ and a pair of $(\alpha,\beta)$ values such that

\begin{align} \label{eq:pos}
\min_{G\in\mathcal{G}_A}\min_{t\in \DC(G) } \mincut_G(s,t)>\sum_{i=0}^{k-1}\min ((d-i)^+\beta,\alpha).
\end{align}
Moreover, for the same $(\alpha,\beta)$ values and the same helper selection scheme $A$ that satisfy \eqref{eq:pos}, if the file size $\mathcal{M}$ also satisfies \eqref{eq:condition}, then there exists a finite field $\text{GF}(q)$ such that we can explicitly construct an RC that meets the reliability requirement.
\end{proposition}

\par The proof of Proposition~\ref{prop:comparison} is presented in Section~\ref{subsec:comparison_proof}.

\par By noticing that the right-hand sides of \eqref{eq:neg} and \eqref{eq:pos} are identical to \eqref{eq:ex_low_b}, Proposition~\ref{prop:comparison} thus answers the central question: Under what conditions is it beneficial to choose the good helpers?

\subsection{The Family Repair Schemes and Their Notation} \label{subsec:desc_fr}
To quantify the benefits of smart helper selection, we propose a new helper selection scheme, which is termed the \emph{family repair (FR) scheme} and is a sub-class of stationary repair schemes. To describe the FR scheme, we first arbitrarily sort all storage nodes and denote them by $1$ to $n$. We then define a {\em complete family} as a group of $(n-d)$ physical nodes. The first $(n-d)$ nodes are grouped as the first complete family and the second $(n-d)$ nodes are grouped as the second complete family and so on and so forth. In total, there are $\left\lfloor \frac{n}{n-d}\right\rfloor$ complete families. The remaining $n\bmod(n-d)$ nodes are grouped as an {\em incomplete family}. The helper set $D_i$ of any node $i$ in a complete family contains all the nodes {\em not} in the same family of node $i$. That is, a newcomer only seeks help from {\em outside} its family. The intuition is that we would like each family to preserve as much information (or equivalently as diverse information) as possible. To that end, we design the helper selection sets such that each newcomer refrains from requesting help from its own family. For any node in the incomplete family,\footnote{\label{footnote:incomplete}All the concepts and intuitions are based on complete families. The incomplete family is used to make the scheme consistent and applicable to the case when $n\bmod(n-d)\neq 0$. } we set the corresponding $D_i=\{1,\cdots, d\}$.

\par For example, suppose that $(n,d)=(8,5)$. There are $2$ complete families, $\{1,2,3\}$ and $\{4,5,6\}$, and $1$ incomplete family, $\{7,8\}$. Then if node $4$ fails, the corresponding newcomer will access nodes $\{1,2,3,7,8\}$ for repair since nodes 1, 2, 3, 7, and 8 are outside the family of node 4. If node $7$ (a member of the incomplete family) fails, the newcomer will access nodes $1$ to $5$ for repair.

By the above definitions, we have in total $\left\lceil\frac{n}{n-d}\right\rceil$ number of families, which are indexed from $1$ to $\left\lceil\frac{n}{n-d}\right\rceil$. However, since the incomplete family has different properties from the complete families, we replace the index of the incomplete family with $0$. Therefore, the family indices become from $1$ to $c\stackrel{\Delta}{=}\left\lfloor\frac{n}{n-d}\right\rfloor$ and then $0$, where $c$ is the index of the last Complete family. If there is no incomplete family, we simply omit the index $0$. Moreover, by our construction, any member of the incomplete family has $D_i=\{1,\cdots, d\}$. That is, it will request help from {\em all} the members of the first $(c-1)$ complete families, {\em but only from} the first $d-(n-d) (c-1)=n\bmod(n-d)$ members of the last complete family. Among the $(n-d)$ members in the last complete family, we thus need to distinguish those members who will be helpers for incomplete family members, and those who will not. Therefore, {\em we add a negative sign to the family indices of those who will ``not'' be helpers for the incomplete family.}

\par From the above discussion, we can now list the family indices of the $n$ nodes as an $n$-dimensional {\em family index vector}. Consider the same example as listed above where $(n,d)=(8,5)$. There are two complete families, nodes 1 to 3 and nodes 4 to 6. Nodes 7 and 8 belong to the incomplete family and thus have family index 0. The third member of the second complete family, node $6$, is not a helper for the incomplete family members, nodes $7$ and $8$, since both $D_7=D_8=\{1,\cdots, d\}=\{1,2,\cdots, 5\}$. Therefore, we replace the family index of node 6 by $-2$. In sum, the {\em family index vector} of this $(n,d)=(8,5)$ example becomes $(1,1,1,2,2,-2, 0,0)$. Mathematically, we can write the family index vector as
\begin{align}
\left(\overbrace{1,\cdots, 1}^{n-d}, \right. \overbrace{2,\cdots, 2}^{n-d}&, \cdots, \overbrace{ c, \cdots, c}^{ n\bmod(n-d)},\nonumber\\
&\left.\overbrace{ -c, \cdots, -c}^{n-d-(n\bmod(n-d))} , \overbrace{0,\cdots, 0}^{n\bmod (n-d)}\right).\label{eq:ccw5}
\end{align}

\begin{figure}[h!]
\centering
\includegraphics[width=0.475\textwidth]{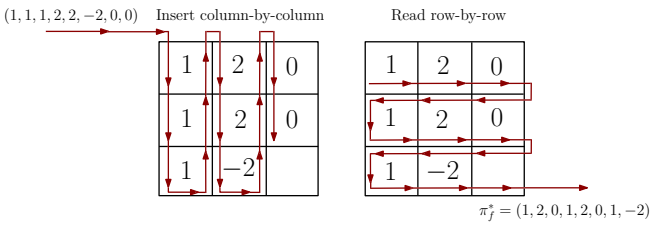}
\caption{The construction of the RFIP for $(n,d)=(8,5)$.}
\label{fig:rfip}
\end{figure}

A {\em family index permutation} is a permutation of the family index vector defined in \eqref{eq:ccw5}, which we denote by $\pi_f$. Continuing from the previous example, one instance of family index permutations is $\pi_f=(1,1,0,2,0,-2,1,2)$. A rotating family index permutation (RFIP) $\pi_f^*$ is a special family index permutation that puts the family indices of \eqref{eq:ccw5} in an $(n-d)\times \left\lceil \frac{n}{n-d}\right\rceil$ table column-by-column and then reads it row-by-row. Fig.~\ref{fig:rfip} illustrates the construction of the RFIP for the case of $(n,d)=(8,5)$. The input is the family index vector $(1,1,1,2,2,-2,0,0)$ and the output RFIP  $\pi_f^*$ is $(1,2,0, 1,2,0,1,-2)$.

\subsection{Quantifying the benefits of the Family Repair scheme}\label{subsec:how_much}

To quantify the gap in \eqref{eq:pos} (or equivalently the gap in \eqref{eq:outperform}) for the best dynamic helper selection scheme, we analyze the performance of the stationary/FR schemes and use it as a lower bound for the gap of \eqref{eq:pos}.
\begin{proposition} \label{prop:low_b_gen}
Consider any stationary repair scheme $A$ and denote its collection of helper sets by $\{D_1,D_2,\dots,D_n\}$. We then have  
\begin{align}
\min_{G\in \mathcal{G}_A}\min_{t\in \DC(G)}\mincut(s,t)\geq \min_{\mathbf{r}\in R}\sum_{i=1}^{k}\min ((d-z_i(\mathbf{r}))\beta,\alpha) \label{eq:low_b_gen},
\end{align}
where $\mathbf{r}$ is a $k$-dimensional integer-valued vector, $R=\{(r_1,r_2,\cdots,r_k):\forall i \in\{1,\cdots,k\}, 1\leq r_i\leq n\}$ and $z_i(\mathbf{r})=|\{r_j:j<i, r_j\in D_{r_i}\}|$. For example, suppose $n=6$,  $k=4$, $D_3=\{1,4\}$, and $\mathbf{r}=(1,2,1,3)$, then we have $r_4=3$ and $z_4(\mathbf{r})=|\{r_j: j<4, r_j\in D_3\}|= 1$. (The double appearances of $r_1=r_3=1$ are only counted as one.)
\end{proposition}

\par The proof of Proposition~\ref{prop:low_b_gen} is relegated to Appendix~\ref{app:low_b_gen}.

\par Proposition~\ref{prop:low_b_gen} above establishes a lower bound on the cut capacity of any stationary repair scheme. Therefore, when designing any stationary scheme, one simply needs to choose $(n,k,d,\alpha,\beta)$ values and the helper sets $D_i$ so that the right-hand side of \eqref{eq:low_b_gen} is no less than the file size $\mathcal{M}$. However, since we do not have equality in \eqref{eq:low_b_gen}, the above construction is sufficient but not necessary. That is, we may be able to use smaller $\alpha$ and $\beta$ values while still guaranteeing that the resulting stationary regenerating code meets the reliability requirement.

\par When we focus on the family repair scheme introduced in Section~\ref{subsec:desc_fr}, a special example of stationary repair, the inequality \eqref{eq:low_b_gen} can be further sharpened to the following equality.
\begin{proposition} \label{prop:low_b}
Consider any given FR scheme $F$ with the corresponding IFGs denoted by $\mathcal{G}_F(n,k,d,\alpha,\beta)$. We have that
\begin{align}
\min_{G\in\mathcal{G}_F} \min_{t\in\DC(G)}&\mincut_G(s,t) = \nonumber\\
&\min_{\forall \pi_f} \sum_{i=1}^{k}\min \left(\left(d-y_i(\pi_f)\right)\beta,\alpha\right),\label{eq:low_b}
\end{align}
where $\pi_f$ can be any family index permutation and $y_i(\pi_f)$ is computed as follows. If the $i$-th coordinate of $\pi_f$ is $0$, then $y_i(\pi_f)$ returns the number of $j$ satisfying both (i) $j<i$ and (ii) the $j$-th coordinate $>0$. If the $i$-th coordinate of $\pi_f$ is not $0$, then $y_i(\pi_f)$ returns the number of $j$ satisfying both (i) $j<i$ and (ii) the absolute value of the $j$-th coordinate of $\pi_f$ and the absolute value of the $i$-th coordinate of $\pi_f$ are different.  For example, if $\pi_f=(1,2,-2,1,0,0,1,2)$, then $y_6(\pi_f)= 3$ and $y_8(\pi_f)= 5$.
\end{proposition}

\par The proof of Proposition~\ref{prop:low_b} is presented in Section~\ref{subsec:low_b_proof}.

\begin{remark} In general, the minimum cut of an IFG may exist in the interior of the graph. When computing the min-cut value in the left-hand side of \eqref{eq:low_b_gen}, we generally need to exhaustively consider all possible cuts for any $G\in {\mathcal{G}}_A$, which is why we have to choose $\mathbf{r}\in R$ in \eqref{eq:low_b_gen} that allows for repeated values in the coordinates of $\mathbf{r}$ and we can only prove the inequality (lower bound) in \eqref{eq:low_b_gen}.
\end{remark}

\par Recall that the family index permutation $\pi_f$ is based on the family index vector of all ``currently active nodes.'' Proposition~\ref{prop:low_b} thus implies that when focusing on the family repair scheme $F$, we can reduce the search scope and consider only those cuts that directly separate $k$ currently active nodes from the rest of the IFG (see \eqref{eq:low_b}). This allows us to explicitly compute the corresponding min-cut value with equality.

\par Combining Proposition~\ref{prop:low_b} and \eqref{eq:condition}, we can derive the new storage-bandwidth tradeoff ($\alpha$ vs.\ $\beta$) for the FR scheme. For example, Fig.~\ref{fig:storage_vs_bandwidth_(20-10-10)} plots $\alpha$ versus $\gamma\stackrel{\Delta}{=}d\beta$ for the $(n,k,d)$ values $(20,10,10)$ with file size $\mathcal{M}=1$. As can be seen in Fig.~\ref{fig:storage_vs_bandwidth_(20-10-10)}, the MBR point (the smallest $\gamma$ value) of the FR scheme uses only 72\% of the repair-bandwidth of the MBR point of the BR scheme ($\gamma_{\MBR}=0.13$ vs.\ $0.18$). It turns out that for any $(n,k,d)$ values, the biggest improvement always happens at the MBR point.\footnote{If we compare the min-cut value of FR in \eqref{eq:low_b} with the min-cut value of BR in \eqref{eq:ex_low_b}, we can see that the greatest improvement happens when the new term $(d-y_i(\pi_f))\beta\leq \alpha$ for all $i$. These are the mathematical reasons why the MBR point sees the largest improvement.} The intuition is that choosing the good helpers is most beneficial when the per-node storage $\alpha$ is no longer a bottleneck (thus the MBR point).

\subsection{The MBR and MSR Points of the FR Scheme} \label{subsec:mbr}
\par The right-hand side of \eqref{eq:low_b} involves taking the minimum over a set of $\mathcal{O}\left(\left(\frac{n}{n-d}\right)^k\right)$ entries. As a result, computing the entire storage-bandwidth tradeoff is of complexity $\mathcal{O}\left(\left(\frac{n}{n-d}\right)^k\right)$. The following proposition shows that if we are interested in the most beneficial point, the MBR point, then we can compute the corresponding $\alpha$ and $\beta$ values in polynomial time.

\begin{proposition} \label{prop:mbr}
For the MBR point of \eqref{eq:low_b}, i.e., when $\alpha$ is sufficiently large, the minimizing family index permutation is the RFIP $\pi_f^*$ defined in Section~\ref{subsec:desc_fr}. That is, the $\alpha$, $\beta$, and
$\gamma$ values of the MBR point can be computed by
\begin{align} \label{eq:gamma}
\alpha_{\MBR}=\gamma_{\MBR}=d\beta_{\MBR}=\frac{d\mathcal{M}}{\sum_{i=1}^{k} (d-y_i(\pi_f^*)) }.
\end{align}
\end{proposition}

\par The proof of Proposition~\ref{prop:mbr} is relegated to Appendix~\ref{app:mbr_proof}.

\par We use Proposition~\ref{prop:mbr} to plot the reliability requirement $k$ versus the repair-bandwidth $\gamma$ for the MBR point when $(n,d)=(60,10)$  in Fig.~\ref{fig:k_vs_gamma_(60-10)}. Since the network is protected against $(n-k)$ simultaneous node failures, the larger the $k$, the less resilient is the network, and the smaller the necessary repair-bandwidth $\gamma=d\beta$ to maintain the network. As can be seen in Fig.~\ref{fig:k_vs_gamma_(60-10)}, for $k\geq 19$, the FR scheme needs only $58\%$ of the repair-bandwidth of the BR solution. Even for the case of $k=10$, i.e.,  $(n,k,d)=(60,10,10)$ which is still within the range of the parameter values ($k\leq d$) considered by the BR scheme,  the FR scheme needs only $73\%$ of the repair-bandwidth of the BR solution. 

Unfortunately, we do not have a general formula for the least beneficial point, the MSR point, of the FR scheme. Our best knowledge for computing the MSR point is the following

\begin{proposition}\label{prop:msr}
For arbitrary $(n,k,d)$ values, the minimum-storage of \eqref{eq:low_b} is $\alpha_{\MSR}= \frac{\mathcal{M}}{\min(d,k)}$. If the $(n,k,d)$ values also satisfy $d\geq k$, then the corresponding $\beta_{\MSR}=\frac{\mathcal{M}}{k(d-k+1)}$. If $d<k$, then the corresponding $\beta_{\MSR}\leq \frac{\mathcal{M}}{d}$.
\end{proposition}
\par The proof of Proposition~\ref{prop:msr} is relegated to Appendix~\ref{app:msr_proof}.
\par By Proposition~\ref{prop:msr}, we can quickly compute $\alpha_{\MSR}$ and $\beta_{\MSR}$ when $d\geq k$. If $d<k$, then we still have $\alpha_{\MSR}=\frac{\mathcal{M}}{\min(d,k)}$ but we do not know how to compute the exact value of $\beta_{\MSR}$ other than directly applying the formula in Proposition~\ref{prop:low_b}. 

\begin{remark}
If we compare the expressions of Proposition~\ref{prop:msr} and the MSR point of the BR scheme provided in \eqref{eq:msr_br_alpha} and \eqref{eq:msr_br_gamma} of Section~\ref{sec:br_extreme_points}, Proposition~\ref{prop:msr} implies that the FR scheme does not do better than the BR scheme at the MSR point when $d\geq k$. However, it is still possible that the FR scheme can do better than the BR scheme at the MSR point when $d<k$. One such example is the example we considered in Section~\ref{sec:preview} when $(n,k,d)=(5,3,2)$. For this example, we have $\alpha_{\MSR}=\frac{\mathcal{M}}{2}$, $\beta_{\MSR}=\frac{\mathcal{M}}{4}$, and  $\gamma_{\MSR}=\frac{\mathcal{M}}{2}$ for the FR scheme where $\beta_{\MSR}=\frac{\mathcal{M}}{4}$ is derived by searching over all family index permutations $\pi_f$ in \eqref{eq:low_b}. For comparison, the BR scheme has $\alpha_{\MSR}=\frac{\mathcal{M}}{2}$, $\beta_{\MSR}=\frac{\mathcal{M}}{2}$, and  $\gamma_{\MSR}=\mathcal{M}$. This shows that the FR scheme can indeed do better at the MSR point when $d<k$ in terms of the repair-bandwidth although we do not have a closed-form expression for this case.
\end{remark}

\subsection{Is the family repair scheme optimal?}
The results presented above quantify the performance benefits of one particular helper selection scheme, the FR scheme. When compared to the BR scheme, the improvement of the FR scheme can be substantial for some $(n,k,d)$ value combinations. At the same time, it is still important to see how close to optimal is the FR scheme among all, stationary or dynamic, helper selection schemes. In the following, we prove that the FR scheme is indeed optimal for some $(n,k,d)$ values. 

\begin{proposition} \label{prop:optimal} For the $(n,k,d)$ values satisfying simultaneously the following three conditions (i) $d$ is even, (ii) $n=d+2$, and (iii) $k=\frac{n}{2}+1$; we have
\begin{align} \label{eq:tight}
\min_{G\in \mathcal{G}_F}\min_{t\in \DC(G)}\mincut_G(s,t) \geq \min_{G\in \mathcal{G}_A}\min_{t\in \DC(G)} \mincut_G(s,t)
\end{align}
for any arbitrary dynamic helper selection scheme $A$ and any arbitrary $(\alpha,\beta)$ values.
\end{proposition}

\par The proof of Proposition~\ref{prop:optimal} is presented in Section~\ref{subsec:optimal_proof}.
\par Note that for any $(n,k,d)$ values satisfying conditions (i) to (iii) in Proposition~\ref{prop:optimal}, they must also satisfy neither (i) nor (ii) in Proposition~\ref{prop:comparison}. As a result, by Proposition~\ref{prop:comparison}, there exists some helper selection scheme that strictly outperforms the BR scheme. Proposition~\ref{prop:optimal} further establishes that among all those schemes strictly better than the BR scheme, the FR scheme is indeed optimal.

\par We also note that \cite[Theorem~5.4]{hollmann2014minimum} proves that when $k=n-1$ and $\alpha=\beta$, no dynamic helper selection scheme can protect a file of size $>\frac{nd\alpha}{d+1}$. Combining Propositions~\ref{prop:low_b} and~\ref{prop:optimal}, we can strictly sharpen this result for the case of $(n,k,d)=(4,3,2)$ and $\alpha=\beta$.

\begin{corollary}\label{cor:existing_loose}
When $(n,k,d)=(4,3,2)$ and $\alpha=\beta$, no dynamic helper scheme can protect a file of size $\mathcal{M}>2\alpha$, for which \cite[Theorem~5.4]{hollmann2014minimum} only proves that no scheme can protect a file of size $\mathcal{M}>\frac{8\alpha}{3}$.
\end{corollary}

\begin{IEEEproof}
By Proposition~\ref{prop:low_b}, when $(n,k,d)=(4,3,2)$ and $\alpha=\beta$, the FR scheme can protect a file of size $2\alpha$. We then notice that $(n,k,d)=(4,3,2)$ satisfies Proposition~\ref{prop:optimal} and therefore the FR scheme is optimal. As a result, no scheme can protect a file of size $\mathcal{M}>2\alpha$.
\end{IEEEproof}

\par Proposition~\ref{prop:optimal} shows that for certain $(n,k,d)$ value combinations, the FR scheme is optimal for the entire storage-bandwidth tradeoff curve. If we only focus on the MBR point, we can also have the following optimality result.

\begin{proposition} \label{prop:optimal_2} Consider $k=n-1$ and $\alpha=d\beta$. For the $(n,k,d)$ values satisfying $n\bmod(n-d)=0$, we have
\begin{align} \label{eq:tight_2}
\min_{G\in \mathcal{G}_F}\min_{t\in \DC(G)}\mincut_G(s,t)&=\frac{n\alpha}{2} \nonumber\\
&\geq \min_{G\in \mathcal{G}_A}\min_{t\in \DC(G)} \mincut_G(s,t)
\end{align}
for any arbitrary dynamic helper selection scheme $A$.
\end{proposition}

\begin{IEEEproof}
\cite[Theorem~5.2]{hollmann2014minimum} proved that for $k=n-1$ and $\alpha=d\beta$,
\begin{align}\label{eq:optimal_2_proof_4}
\min_{G\in \mathcal{G}_A}\min_{t\in \DC(G)} \mincut_G(s,t)\leq \frac{nd\beta}{2}
\end{align}
for any arbitrary dynamic helper selection scheme $A$. As a result, we only need to prove that when $n\bmod(n-d)=0$, the min-cut of the FR scheme equals $\frac{nd\beta}{2}$.

Since $\alpha=d\beta$, we know by Proposition~\ref{prop:mbr} that
\begin{align} \label{eq:optimal_2_proof_1}
\min_{G\in \mathcal{G}_F}\min_{t\in \DC(G)}\mincut_G(s,t)= \sum_{i=1}^{n-1} (d-y_i(\pi_f^*)) \beta.
\end{align}
Now, when $n\bmod(n-d)=0$, we have no incomplete family in the FR scheme and the RFIP has the following form
\begin{align}
\pi_f^*=(1,2,\cdots,c,1,2,\cdots,c,\cdots,1,2,\cdots,c)\label{eq:rfip_optimal_2},
\end{align}
where recall that $c=\left\lfloor\frac{n}{n-d}\right\rfloor=\frac{n}{n-d}$. Using \eqref{eq:rfip_optimal_2}, we get that
\begin{align}\label{eq:yi_optimal_2}
y_i(\pi_f^*)=i-1-\left\lfloor \frac{i-1}{c}\right\rfloor.
\end{align}
The reason behind \eqref{eq:yi_optimal_2} is the following. Examining the definition of $y_i(\cdot)$, we can see that $y_i(\cdot)$ counts all the coordinates $j<i$ of $\pi_f^*$ that have a family index different than the family index at the $i$-th coordinate. For each coordinate $i$, with the aid of \eqref{eq:rfip_optimal_2}, there are $\left\lfloor \frac{i-1}{c}\right\rfloor$ coordinates in $\pi_f^*$ preceding it with the same family index. Therefore, in total there are $i-1-\left\lfloor \frac{i-1}{c}\right\rfloor$ coordinates in $\pi_f^*$ preceding the $i$-th coordinate with a different family index, thus, we get \eqref{eq:yi_optimal_2}.

\par By \eqref{eq:optimal_2_proof_1} and \eqref{eq:yi_optimal_2}, we get 
\begin{align}
\min_{G\in \mathcal{G}_F}\min_{t\in \DC(G)}&\mincut_G(s,t)=\sum_{i=0}^{n-2}\left(d-i+\left\lfloor\frac{i}{\frac{n}{n-d}}\right\rfloor\right)\beta \nonumber\\
&=\sum_{i=0}^{n-1}\left(d-i+\left\lfloor\frac{i}{\frac{n}{n-d}}\right\rfloor\right)\beta\label{eq:n-1_n-2}\\
&=\left(nd-\frac{(n-1)n}{2}+\sum_{i=0}^{n-1}\left\lfloor\frac{i}{\frac{n}{n-d}}\right\rfloor\right)\beta\nonumber\\
&=\left(nd-\frac{(n-1)n}{2}+\frac{n}{n-d}\sum_{i=0}^{n-d-1}i\right) \beta\nonumber\\
&=\left(nd-\frac{(n-1)n}{2}+\frac{n(n-d-1)}{2}\right)\beta\nonumber\\
&=\frac{nd\beta}{2}, \nonumber
\end{align}
where we get \eqref{eq:n-1_n-2} by the fact that $d-(n-1)+\left\lceil \frac{n-1}{c}\right\rceil =d-(n-1)+(n-d-1)=0$. The proof is thus complete
\end{IEEEproof}

Proposition~\ref{prop:optimal_2} establishes again that the FR scheme is optimal, among all dynamic helper schemes, for $k=n-1$ and $\alpha=d\beta$ whenever $n\bmod(n-d)=0$. We will show in Section~\ref{sec:family-plus} that  the FR scheme and its extension, the family-plus repair scheme, are actually also {\em weakly optimal} for general $(n,k,d)$ values. The definition of weak optimality will be provided in Proposition~\ref{prop:weak}.

\section{Family-plus Repair Scheme} \label{sec:family-plus}
In the FR scheme, there are $\left\lfloor\frac{n}{n-d}\right\rfloor$ complete families and $1$ incomplete family 
(if $n\bmod (n-d)\neq 0$). For the scenario in which the $n$ and $d$ values are comparable, we have many complete families and the FR solution harvests almost all of the benefits of choosing good helpers, see the discussion of Proposition~\ref{prop:optimal} for which $n=d+2$. However, when $n$ is large but $d$ is small, we have only one complete family and one incomplete family. Therefore, even though the FR scheme still substantially outperforms the BR scheme, see Fig.~\ref{fig:k_vs_gamma_(60-10)} for the case of $(n,d)=(60,10)$, the performance of the FR scheme is far from optimal due to having only $1$ complete family. In this section, we propose the {\em family-plus repair} scheme that further improves the storage-bandwidth tradeoff when $n$ is large but $d$ is small.

The main idea is as follows. We first partition the $n$ nodes into several disjoint groups of $2d$ nodes and one disjoint group of $n_{\text{remain}}$ nodes. The first type of groups is termed the regular group while the second group is termed the remaining group. If we have to have one remaining group (when $n\bmod (2d)\neq 0$), then we enforce the size of the remaining group to be as small as possible but still satisfying $n_\text{remain}\geq 2d+1$. For example, if $d=2$ and $n=8$, then we will have 2 regular groups and no remaining group since $n\bmod (2d)=0$. If $d=2$ and $n=9$, then we choose $1$ regular group $\{1,2,3,4\}$ and $1$ remaining group $\{5,6,7,8,9\}$ since we need to enforce $n_\text{remain}\geq 2d+1$.

After the partitioning, we apply the FR scheme to the individual groups. For example, if $d=2$ and $n=8$, then we have two regular groups $\{1,2,3,4\}$ and $\{5,6,7,8\}$. Applying the FR scheme to the first group means that nodes $1$ and $2$ form a family and nodes $3$ and $4$ form another family. Whenever node $1$ fails, it will access helpers from outside its family, which means that it will access nodes $3$ and $4$. Node $1$ will never request help from any of nodes $5$ to $8$ as these nodes are not in the same group as node $1$. Similarly, we apply the FR scheme to the second group $\{5,6,7,8\}$. All the FR operations are always performed within the same group.

\par Another example is when $d=2$ and $n=9$. In this case, we have 1 regular group $\{1,2,3,4\}$ and 1 remaining group $\{5,6,7,8,9\}$. In the remaining group, $\{5,6,7\}$ will form a complete family and $\{8,9\}$ will form an incomplete family. If node 6 fails, it will request help from both nodes 8 and 9. If node 9 fails, it will request help from nodes $\{5,6\}$, the first $d=2$ nodes of this group. Again, all the repair operations for nodes 5 to 9 are completely separated from the operations of nodes 1 to 4. The above scheme is termed the \emph{family-plus repair scheme}.

One can easily see that when $n\leq 2d$, there is only one group and the family-plus repair scheme collapses to the FR scheme. When $n>2d$, there are approximately $\frac{n}{2d}$ regular groups, each of which contains two complete families. Therefore, the construction of the family-plus repair scheme ensures that there are many complete families even for the scenario of $n\gg d$. In the following proposition, we characterize the performance of the family-plus repair scheme.

\begin{proposition} \label{prop:low_b_plus}
Consider any given $(n,k,d)$ values and the family-plus repair scheme $F^+$. Suppose we have $B$ groups in total (including both regular and remaining groups) and each group has $n_b$ number of nodes for $b=1$ to $B$. Specifically, if the $b$-th group is a regular group, then $n_b=2d$. If the $b$-th group is a remaining group (when $n\bmod (2d)\neq 0$), then $n_b=n-2d(B-1)$. We use ${\mathcal{G}}_{F^+}(n,k,d,\alpha,\beta)$ to denote the collection of IFGs generated by the family-plus repair scheme. We have that
\begin{align}\label{eq:low_b_plus}
\min_{G\in\gfp} &\min_{t\in\DC(G)} \mincut(s,t) = \nonumber \\
& \min_{\mathbf{k}\in K} \sum_{b=1}^{B} \min_{H\in \gf(n_b, k_b,d,\alpha,\beta)} \min_{t_b\in \DC(H)} \mincut_H(s,t_b),
\end{align}
where $\mathbf{k}$ is a $B$-dimensional integer-valued vector, $K=\{(k_1,k_2,\cdots, k_B): \forall b\in\{1,\cdots, B\}, 0\leq k_b\leq n_b, \sum_{b=1}^{B}k_b=k\}$. Note that for any given $\bf{k}$, the right-hand side of \eqref{eq:low_b_plus} can be evaluated by Proposition~\ref{prop:low_b}. 
\end{proposition}

\begin{IEEEproof} Observe that any IFG $G\in\gfp$ is a union of $B$ parallel IFGs that are in $\gf(n_b,\cdot,d,\alpha,\beta)$ where ``$\cdot$'' means that we temporarily ignore the placement of the data collectors. For any data collector $t$ in $G_{F^+}$, we use $k_b$ to denote the number of active nodes that $t$ accesses in group $b$. Therefore, the $\mincut_G(s,t)$ is simply the summation of the $\mincut_H(s,t_b)$ for all $b\in \{1,\cdots, B\}$ where $t_b$ corresponds to the ``sub-data-collector'' of group $b$. By further minimizing over all possible data collectors $t$ (thus minimizing over $\{k_b\}$), we get \eqref{eq:low_b_plus}.
\end{IEEEproof}

To evaluate the right-hand side of \eqref{eq:low_b_plus}, we have to try all possible choices of the $\mathbf{k}$ vectors and for each given $\mathbf{k}$, we evaluate each of the $B$ summands by Proposition~\ref{prop:low_b}, which requires checking all $n_b!$ different family index permutations. On the other hand, for the MBR point of the family-plus repair scheme, we can further simplify the computation complexity following similar arguments as used in Proposition~\ref{prop:mbr}.
\begin{corollary}\label{cor:mbr_plus}
The MBR point of the family-plus repair scheme is
\begin{align} 
\alpha_{\MBR}=\gamma_{\MBR}=d\beta_{\MBR}\nonumber
\end{align}
and $\beta_{\MBR}$ can be computed by solving the following equation 
\begin{align}\label{eq:gamma_plus}
\Bigg(&1_{\{n\bmod(2d)\neq 0\}}\cdot \sum_{i=0}^{\min(k,2d-1)-1}\left(d-i+\left\lfloor\frac{i}{2}\right\rfloor\right)+\nonumber\\
&d^2\left\lfloor \frac{(k-n_l)^+}{2d}\right\rfloor+ \sum_{i=0}^{q}\left(d-i+\left\lfloor\frac{i}{2}\right\rfloor\right)\Bigg) \beta_{\MBR}=\mathcal{M},
\end{align}
where $\mathcal{M}$ is the file size,
\begin{align}
&q=((k-n_l)^+\bmod(2d))-1, \text{ and}\nonumber\\ 
&n_l=
\begin{cases}
n_{\text{remain}},& \text{ if } n\bmod(2d)\neq 0\\
0,& \text{ otherwise}.
\end{cases}\nonumber
\end{align}
\end{corollary}
\par The proof of Corollary~\ref{cor:mbr_plus} is relegated to Appendix~\ref{app:mbr_plus_proof}.

In Fig.~\ref{fig:k_vs_gamma_(60-10)}, we plot the $k$ vs. $\gamma$ curves for the BR, the FR, and the family-plus repair schemes for the case of $(n,d)=(60,10)$ using \eqref{eq:mbr_br}, Proposition~\ref{prop:mbr}, and Corollary~\ref{cor:mbr_plus}, respectively. As can be seen, when $k=40$, the repair-bandwidth of the family-plus repair scheme is only $28\%$ of the repair-bandwidth of the BR scheme (cf.\ the repair-bandwidth of the FR scheme is $58\%$ of the repair-bandwidth of the BR scheme). This demonstrates the benefits of the family-plus repair scheme, which creates as many complete families as possible by further partitioning the nodes into several disjoint groups.

\par We are now ready to state the weak optimality of the family-plus repair scheme for all $(n,k,d)$ values.

\begin{proposition}\label{prop:weak} Consider a family-plus repair scheme denoted by $F^+$, and its corresponding collection of IFGs ${\mathcal{G}}_{F^+}(n,k,d,\alpha,\beta)$. For any $(n,k,d)$ values satisfying neither of the (i) and (ii) conditions in Proposition~\ref{prop:comparison}, there exists a pair $(\alpha,\beta)$ such that

\begin{align}\label{eq:weak}
\min_{G\in\mathcal{G}_{F^+}}\min_{t\in \DC (G) } \mincut_G(s,t)>\sum_{i=0}^{k-1}\min((d-i)^+\beta,\alpha).
\end{align}
\end{proposition}

\par The proof of Proposition~\ref{prop:weak} is relegated to Appendix~\ref{app:weak_proof}.

\par Propositions~\ref{prop:weak} and~\ref{prop:comparison} jointly show that whenever helper selection can improve the performance, so can the family-plus repair scheme. We term this property the ``weak optimality.''
\par Note that although the FR scheme in Section~\ref{subsec:desc_fr} is optimal for some $(n,k,d,\alpha,\beta)$ value combinations, the FR scheme is {\em not} weakly optimal, i.e., Proposition~\ref{prop:weak} does not hold for the FR scheme. By introducing the additional partitioning step, the family-plus scheme is monotonically better than the FR scheme when\footnote{The proof is provided in Appendix~\ref{app:weak_proof}.} $\alpha=d\beta$, and is guaranteed to be weakly optimal. Moreover, in addition to the cases of $(n,k,d,\alpha,\beta)$ values for which FR is optimal (so is the family-plus scheme since the family-plus scheme is monotonically better in those cases), the family-plus scheme is optimal for some additional $(n,k,d,\alpha,\beta)$ values.

\begin{proposition}\label{prop:family-plus_optimal}
Consider $k=n-1$ and $\alpha=d\beta$ and a family-plus repair scheme that divides $n$ nodes into $B$ groups with $n_1$ to $n_B$ nodes. If $n_b\bmod(n_b-d)=0$ for all $b=1$ to $B$, then we have
\begin{align} \label{eq:tight_2_plus}
\min_{G\in \gfp}\min_{t\in \DC(G)}\mincut_G(s,t)&=\frac{n\alpha}{2} \nonumber\\
&\geq \min_{G\in \mathcal{G}_A}\min_{t\in \DC(G)} \mincut_G(s,t)
\end{align}
for any arbitrary dynamic helper selection scheme $A$.
\end{proposition}

\begin{remark}
Thus far, our family-plus scheme assumes all but one group have $n_b=2d$ nodes and the remaining group has $n_b=n_{\text{remain}}\geq 2d+1$ nodes. One possibility for further generalization is to allow arbitrary $n_b$ choices. It turns out that Proposition~\ref{prop:family-plus_optimal} holds even for any arbitrary choices of $n_b$ values. For example, for the case of $(n,k,d)=(19,18,4)$ and $\alpha=d\beta$, the generalized family-plus scheme is absolutely optimal if we divide the 19 nodes into 3 groups of $(n_1,n_2,n_3)=(8,6,5)$. Also, one can prove that for any $(n,k,d,\alpha,\beta)$ values satisfying $n\neq 5$, $k=n-1$, $d=2$, and $\alpha=d\beta$, we can always find some $(n_1,\cdots,n_B)$ such that the generalized family-plus repair scheme is absolutely optimal. See Result~6 in Section~\ref{sec:preview} for some other $(n,k,d)$ value combinations for which the generalized family-plus scheme is optimal. 
\end{remark}

\begin{proof}
By Proposition~\ref{prop:low_b_plus} and the fact that $k=n-1$, we must have all but one $k_b=n_b$ and the remaining one $k_b=n_b-1$. Without loss of generality, we assume $k_1=n_1-1$ and all other $k_b=n_b$ for $b=2$ to $B$ for the minimizing $\mathbf{k}$ vector in \eqref{eq:low_b_plus}. Since $n_1\bmod(n_1-d)=0$, by Proposition~\ref{prop:optimal_2}, the first summand of \eqref{eq:low_b_plus} must be equal to $\frac{n_1\alpha}{2}$.

\par For the case of $b=2$ to $B$, we have $k_b=n_b$ instead of $k_1=n_1-1$. However, if we examine the proof of Proposition~\ref{prop:optimal_2}, we can see that Proposition~\ref{prop:optimal_2} holds even for the case of $k=n$ since (i) when compared to the case of $k=n-1$, the case of $k=n$ involves one additional summand $(d-y_n(\pi_f^*))\beta$ in \eqref{eq:optimal_2_proof_1} and (ii) $(d-y_n(\pi_f^*))=0$. By applying Proposition~\ref{prop:optimal_2} again, the $b$-th summand of \eqref{eq:low_b_plus}, $b=2$ to $B$, must be $\frac{n_b\alpha}{2}$ as well.

\par Finally, by Proposition~\ref{prop:low_b_plus}, we have the equality in \eqref{eq:tight_2_plus}

\begin{align}
\min_{G\in \gfp}\min_{t\in \DC(G)}\mincut_G(s,t)=\sum_{b=1}^B\frac{n_b\alpha}{2}=\frac{n\alpha}{2}\label{eq:family-plus_optimal2}.
\end{align}
The inequality in \eqref{eq:tight_2_plus} is by \cite[Theorem~5.4]{hollmann2014minimum}. The proof is thus complete.
\end{proof}

Before closing this section, we should mention that a similar scheme to the family-plus repair scheme was devised in \cite{papailiopoulos2012locally} for the MSR point when $n$ is a multiple of $(d+1)$. In that scheme the nodes are divided into groups of $(d+1)$ nodes. Whenever a node fails, its set of helpers is the set of $d$ remaining nodes in the same group. This can be viewed as a special example of the generalized family-plus repair scheme by choosing $n_b=d+1$ for all $b=1$ to $B$. Each group thus has $\frac{n_b}{n_b-d}=n_b=d+1$ complete families and each family contains only $n_b-d=1$ node. As we saw for the family-plus repair scheme above, the scheme in \cite{papailiopoulos2012locally} can be easily analyzed by noticing that the IFGs representing this scheme consist of $\frac{n}{d+1}$ parallel graphs with parameters $(n,d)=(d+1,d)$. By similar analysis as in Corollary~\ref{cor:mbr_plus}, it is not hard to find the MBR point of this scheme which is
\begin{align} \label{eq:gamma_other}
\gamma_{\MBR}=d\mathcal{M}\left(\left\lfloor \frac{k}{d+1} \right\rfloor \frac{(d+1)d}{2}+\frac{2dr-r^2+r}{2}\right)^{-1},
\end{align}
where $r=k-\left\lfloor \frac{k}{d+1}\right\rfloor(d+1)$. 

\par Note that unlike the construction in \cite{papailiopoulos2012locally} that requires each group to have $(d+1)$ nodes and thus requires $n\bmod(d+1)=0$, our construction and analysis hold for arbitrary ways\footnote{Our construction and analysis work for arbitrary $n_b$ partitions. On the other hand, the optimality guarantee in Proposition~\ref{prop:family-plus_optimal} only holds when $n_b\bmod(n_b-d)=0$ for all $b$.} of partitioning $n$ nodes into separate groups of $n_b$ nodes, $b=1$ to $B$. Also, our analysis in this work has characterized the entire storage-bandwidth tradeoff. For comparison, \cite{papailiopoulos2012locally} analyzed it only for for the MSR point. In summary, the result in this work is a much more general code construction and analysis for arbitrary $(n,k,d)$ values.

\par Also note that in addition to deriving the entire storage-bandwidth
tradeoff of the proposed family-based schemes, one
main contribution of this work is to successfully position
the family-based schemes in the context of characterizing the
benefits of optimal helper selection of regenerating codes, e.g., Propositions~\ref{prop:optimal}, \ref{prop:optimal_2}, \ref{prop:weak}, and \ref{prop:family-plus_optimal}.

\section{Some Major Proofs}
\subsection{Proof of Proposition~\ref{prop:comparison}}\label{subsec:comparison_proof}
Before presenting the proof of Proposition~\ref{prop:comparison}, we introduce the following definition and lemma.
\begin{definition}A set of $m$ active storage nodes (input-output pairs) of an IFG is called an $m$-set if the following conditions are satisfied simultaneously.  (i) Each of the $m$ active nodes has been repaired at least once; and (ii) Jointly the $m$ nodes satisfy the following property: Consider any two distinct active nodes $x$ and $y$ in the $m$-set and without loss of generality assume that $x$ was repaired before $y$. Then there exists an edge in the IFG that connects $x_\text{out}$ and $y_\text{in}$.
\end{definition}

\begin{lemma} \label{lem:set} Fix the helper selection scheme $A$. Consider an arbitrary $G\in G_A(n,k,d,\alpha,\beta)$ such that each active node in $G$ has been repaired at least once. Then there exists a $\left\lceil \frac{n}{n-d}\right\rceil$-set in $G$.
\end{lemma}

{\em Proof of Lemma~\ref{lem:set}:}
We prove this lemma by proving the following stronger claim: Consider any integer value $m\geq 1$. There exists an $m$-set in every group of $(m-1)(n-d)+1$ active nodes of which each active node has been repaired at least once. Since the $G$ we consider has $n$ active nodes, the above claim implies that $G$ must contain a $\left\lceil \frac{n}{n-d}\right\rceil$-set.

\par We prove this claim by induction on the value of $m$. When $m=1$, by the definition of the $m$-set, any group of 1 active node in $G$ forms a 1-set. The claim thus holds naturally.

\par Suppose the claim is true for all $m<m_0$, we now claim that in every group of $(m_0-1)(n-d)+1$ active nodes of $G$ there exists an $m_0$-set. The reason is as follows. Given an arbitrary, but fixed group of $(m_0-1)(n-d)+1$ active nodes, we use $y$ to denote the youngest active node in this group  (the one which was repaired last). Obviously, there are $(m_0-1)(n-d)$ active nodes in this group other than $y$. On the other hand, since any newcomer accesses $d$ helpers out of $n-1$ surviving nodes, during its repair, node $y$ was able to avoid connecting to at most $(n-1)-d$ surviving nodes (the remaining active nodes). Therefore, out of the remaining $(m_0-1)(n-d)$ active nodes in this group, node $y$ must be connected to at least $((m_0-1)(n-d))-(n-1-d)=(m_0-2)(n-d)+1$ of them. By induction, among those $\geq (m_0-2)(n-d)+1$ nodes, there exists an $(m_0-1)$-set. Since, by our construction, $y$ is connected to {\em all} nodes in this $(m_0-1)$-set, node $y$ and this $(m_0-1)$-set jointly form an $m_0$-set. The proof of this claim is complete.

{\em Proof of Proposition~\ref{prop:comparison}:}
\par We first prove the forward direction. Assume condition (ii) holds and consider an IFG $G\in \mathcal{G}_A$ in which every active node has been repaired at least once. By Lemma~\ref{lem:set}, there exists a $\left\lceil \frac{n}{n-d} \right\rceil$-set in $G$. Since condition (ii) holds, we can consider a data collector of $G$ that connects to $k$ nodes out of this $\left\lceil \frac{n}{n-d} \right\rceil$-set. Call this data collector $t$. If we focus on the edge cut that separates source $s$ and the $k$ node pairs connected to $t$, one can use the same analysis as in \cite[Lemma 2]{dimakis2010network} and derive ``$\mincut(s,t)\leq \sum_{i=0}^{k-1}\min ((d-i)^+\beta,\alpha)$''  for the given $G\in G_A$ and the specific choice of $t$. Therefore, we have
\begin{align}\label{eq:worse_br}
\min_{G\in\mathcal{G}_A}\min_{t\in \DC(G) } \mincut_G(s,t)\leq \sum_{i=0}^{k-1}\min((d-i)^+\beta,\alpha).
\end{align}
On the other hand, by definition we have
\begin{align}
\min_{G\in\mathcal{G}_A}\min_{t\in \DC(G) } \mincut_G(s,t)\geq \min_{G\in\mathcal{G}}\min_{t\in \DC(G) } \mincut_G(s,t). \label{eq:new2_better}
\end{align}
Then by \eqref{eq:worse_br}, \eqref{eq:new2_better}, and \eqref{eq:ex_low_b}, we have proved that whenever condition (ii) holds, the equality \eqref{eq:neg} is true.

Now, assume condition (i) holds. We first state the following claim and use it to prove \eqref{eq:neg}. 

\begin{claim} \label{clm:vertex-cut} 
For any given dynamic helper selection scheme $A$ and the corresponding collection of IFGs $\mathcal{G}_A$, we can always find a $G^*\in \mathcal{G}_A$ such that there exists a set of 3 active nodes in $G^*$, denoted by $x$, $y$, and $z$ such that the following three properties hold simultaneously. (a) $x$ is repaired before $y$, and $y$ is repaired before $z$; (b) $(x_\text{out},y_\text{in})$ is an edge in $G^*$; and (c) either $(x_\text{out}, z_\text{in})$ is an edge in $G^*$ or $(y_\text{out},z_\text{in})$ is an edge in $G^*$. 
\end{claim}

\par Suppose the above claim is true. We let $t^*$ denote the data collector that is connected to $\{x,y,z\}$. By properties (a) to (c) we can see that node $x$ is a vertex-cut separating source $s$ and the data collector $t^*$. The min-cut value separating $s$ and $t^*$ thus satisfies $\mincut_{G^*}(s,t^*)\leq\min(d\beta,\alpha)=\sum_{i=0}^{k-1}\min((d-i)^+\beta,\alpha)$ for $G^*\in G_A$ and the specific choice of $t$, where the inequality follows from $x$ being a vertex-cut separating $s$ and $t^*$ and the equality follows from that condition (i) being true implies $d=1$ and $k=3$. By the same arguments as used in proving the case of condition (ii), we thus have \eqref{eq:neg} when condition (i) holds.

\par We prove Claim~\ref{clm:vertex-cut} by explicit construction. Start from any $G\in \mathcal{G}_A$ with all $n$ nodes have been repaired at least once. We choose one arbitrary active node in $G$ and denote it by $w^{(1)}$. We let $w^{(1)}$ fail and denote the newcomer that replaces $w^{(1)}$ by $y^{(1)}$. The helper selection scheme $A$ will choose a helper node (since $d=1$) and we denote that helper node as $x^{(1)}$. The new IFG after this failure and repair process is denoted by $G^{(1)}$. By our construction $x^{(1)}$, as an existing active node, is repaired before the newcomer $y^{(1)}$ and there is an edge $(x^{(1)}_\text{out},y^{(1)}_\text{in})$ in $G^{(1)}$.

Now starting from $G^{(1)}$, we choose another $w^{(2)}$, which is not one of $x^{(1)}$ and $y^{(1)}$ and let this node fail. Such $w^{(2)}$ always exists since $n$ is odd by condition (i). We use $y^{(2)}$ to denote the newcomer that replaces $w^{(2)}$. The helper selection scheme $A$ will again choose a helper node based on the history of the failure pattern. We denote the new IFG (after the helper selection chosen by scheme $A$) as $G^{(2)}$. If the helper node of $y^{(2)}$ is $x^{(1)}$, then the three nodes $(x^{(1)},y^{(1)}, y^{(2)})$ are the $(x,y,z)$ nodes satisfying properties (a), (b) and the first half of (c). If the helper node of $y^{(2)}$ is $y^{(1)}$, then the three nodes $(x^{(1)},y^{(1)}, y^{(2)})$ are the $(x,y,z)$ nodes satisfying properties (a), (b) and the second half of (c).  In both cases, we can stop our construction and let $G^*=G^{(2)}$ and we say that the construction is complete in the second round. Suppose neither of the above two is true, i.e., the helper of $y^{(2)}$ is neither $x^{(1)}$ nor $y^{(1)}$. Then, we denote the helper of $y^{(2)}$ by $x^{(2)}$. Note that after this step, $G^{(2)}$ contains two disjoint pairs of active nodes such that there is an edge $(x^{(m)}_\text{out}, y^{(m)}_\text{in})$ in $G^{(2)}$ for $m=1,2$.

\par We can repeat this process for the third time by failing a node $w^{(3)}$ that is none of $\{x^{(m)},y^{(m)}:\forall m=1,2\}$. We can always find such a node $w^{(3)}$ since $n$ is odd when condition (i) holds. Again, let $y^{(3)}$ denote the newcomer that replaces $w^{(3)}$ and the scheme $A$ will choose a helper for $y^{(3)}$. The new IFG after this failure and repair process is denoted by $G^{(3)}$.  If the helper of $y^{(3)}$ is $x^{(m)}$ for some $m=1,2$, then the three nodes $(x^{(m)},y^{(m)}, y^{(3)})$ are the $(x,y,z)$ nodes satisfying properties (a), (b) and the first half of (c).  If the helper node of $y^{(3)}$ is $y^{(m)}$ for some $m=1,2$, then the three nodes $(x^{(m)},y^{(m)}, y^{(3)})$ are the $(x,y,z)$ nodes satisfying properties (a), (b) and the second half of (c).  In both cases, we can stop our construction and let $G^*=G^{(3)}$ and we say that the construction is complete in the third round. If neither of the above two is true, then we denote the helper of $y^{(3)}$ by $x^{(3)}$, and repeat this process for the fourth time and so on so forth.

\par We now observe that since $n$ is odd, if the construction is not complete in the $m_0$-th round, we can always start the $(m_0+1)$-th round since we can always find a node $w^{(m_0+1)}$ that is none of $\{x^{(m)},y^{(m)}:\forall m=1,2,\cdots, m_0\}$.   On the other hand, we cannot repeat this process indefinitely since we only have a finite number of $n$ active nodes in the network. Therefore, the construction must be complete in the $\tilde{m}$-th round for some finite $\tilde{m}$. If the helper of $y^{(\tilde{m})}$ is $x^{(m)}$ for some $m=1,2,\cdots \tilde{m}-1$,  then the three nodes $(x^{(m)},y^{(m)}, y^{(\tilde{m})})$ are the $(x,y,z)$ nodes satisfying properties (a), (b) and the first half of (c).  If the helper node of $y^{(\tilde{m})}$ is $y^{(m)}$ for some $m=1,2,\cdots, \tilde{m}-1$, then the three nodes $(x^{(m)},y^{(m)}, y^{(\tilde{m})})$ are the $(x,y,z)$ nodes satisfying properties (a), (b) and the second half of (c). Let $G^*=G^{(\tilde{m})}$  denote the final IFG. The explicit construction of $G^*$ and the corresponding $(x,y,z)$ nodes is thus complete.

The backward direction \eqref{eq:pos} is a direct result of Proposition~\ref{prop:weak}. The proof of Proposition~\ref{prop:weak} is relegated to Appendix~\ref{app:weak_proof}.

\subsection{Proof of Proposition~\ref{prop:low_b}} \label{subsec:low_b_proof}
The outline of the proof is as follows.

\par Part I: We will first show that
\begin{align}\label{eq:new2}
\min_{G\in\mathcal{G}_F} \min_{t\in\DC(G)}&\mincut_G(s,t) \leq  \nonumber\\
&\min_{\forall \pi_f} \sum_{i=1}^{k}\min \left(\left(d-y_i(\pi_f)\right)\beta,\alpha\right).
\end{align}
The proof of Part I is provided in Appendix~\ref{app:eq:new2}.

\par Part II: By definition, the family repair scheme is a stationary repair scheme. Thus, \eqref{eq:low_b_gen} is also a lower bound on all IFGs in $\gf$ and we quickly have
\begin{align}\label{eq:upper_lower}
\min_{\mathbf{r}\in R}\sum_{i=1}^{k}\min ((d-z_i(\mathbf{r}))\beta,\alpha)\leq& \nonumber\\
\min_{G\in\mathcal{G}_F} \min_{t\in\DC(G)}\mincut_G(s,t)& \leq  \nonumber\\
\min_{\forall \pi_f} \sum_{i=1}^{k} \min &\left(\left(d-y_i(\pi_f)\right)\beta,\alpha\right).
\end{align}
The remaining step is to prove that
\begin{align}\label{new3}
\min_{\mathbf{r}\in R}\sum_{i=1}^{k}\min ((d-z_i(\mathbf{r}))\beta,\alpha)=\nonumber&\\
\min_{\forall \pi_f} \sum_{i=1}^{k}\min &\left(\left(d-y_i(\pi_f)\right)\beta,\alpha\right).
\end{align}
Once we prove \eqref{new3}, we have \eqref{eq:low_b} since \eqref{eq:upper_lower} is true. The proof is then complete.

\par The proof of Part II (i.e., \eqref{new3}) is as follows. To that end, we first prove that with the helper sets $D_1$ to $D_n$ specified in a family repair scheme, we have
\begin{align}\label{new4}
\text{LHS of \eqref{eq:new2}}=\min_{\mathbf{r}\in R_2}\sum_{i=1}^{k}\min ((d-z_i(\mathbf{r}))\beta,\alpha)
\end{align}
where $R_2=\{(r_1,r_2,\cdots,r_k): \forall i,j\in\{1,\cdots,k\},1 \leq r_i\leq n, r_i\neq r_j \text{ if } i\neq j\}$. That is, when evaluating the LHS of \eqref{new4}, we can minimize over $R_2$ instead of over $R=\{1,\cdots, n\}^k$. We prove \eqref{new4} by proving that for any $\mathbf{r}\in R$ we can always find a vector $\mathbf{r'} \in R_2$ such that
\begin{align}
\sum_{i=1}^{k}\min ((d-z_i(\mathbf{r}))\beta,\alpha) \geq   \sum_{i=1}^{k}\min ((d-z_i(\mathbf{r'}))\beta,\alpha).\label{new5}
\end{align}
Equation~\eqref{new5} implies that at least one of the minimizing $\mathbf{r}^*\in R$ of the LHS of \eqref{new3} is also in $R_2$. We thus have \eqref{new4}. The proof of \eqref{new5} is provided in Appendix~\ref{app:procedure}.

\par We now notice that any $\mathbf{r} \in R_2$ corresponds to the first $k$ coordinates of a permutation of the node indices $(1,2,3,\cdots, n)$. For easier reference, we use $\overline{\mathbf{r}} $ to represent an $n$-dimensional permutation vector such that the first $k$ coordinates of $\overline{\mathbf{r}}$ match $\mathbf{r}$. One can view $\overline{\mathbf{r}}$ as the extended version of $\mathbf{r}$ from a partial $k$-dimensional permutation to a complete $n$-dimensional permutation vector.  Obviously, the choice of $\overline{\mathbf{r}}$ is not unique.  The following discussion holds for any $\overline{\mathbf{r}}$.

\par For any $\mathbf{r}\in R_2$, we first find its extended version $\overline{\mathbf{r}}$. We then construct $\pi_f$ from $\overline{\mathbf{r}}$ by transcribing the permutation of the node indices $\overline{\mathbf{r}}$ to the corresponding family indices. For example, consider the parameter values $(n,k,d)=(8,4,5)$. Then, one possible choice of $\mathbf{r}\in R_2$ is $\mathbf{r}=(3,5,2,4)$ and a corresponding $\overline{\mathbf{r}}$ is $(3,5,2,4,1,6,7,8)$. The transcribed family index vector is $\pi_f=(1,2,1,2,1,-2,0,0)$. We now argue that $z_i(\mathbf{r})=y_i(\pi_f)$ for all $i=1$ to $k$. The reason is that the definition of $y_i(\pi_f)$ is simply a transcribed version of the original definition of $z_i(\mathbf{r})$ under the node-index to family-index translation. In sum, the above argument proves that for any $\mathbf{r}\in R_2$, there exists a $\pi_f$ satisfying

\begin{align}
\sum_{i=1}^{k}\min((d-&z_i(\mathbf{r}))\beta,\alpha)=\sum_{i=1}^{k}\min \left(\left(d-y_i(\pi_f)\right)\beta,\alpha\right).\nonumber
\end{align}
As a result, we have 
\begin{align}
\min_{\mathbf{r}\in R_2}\sum_{i=1}^{k}\min((d-&z_i(\mathbf{r}))\beta,\alpha)\geq \nonumber\\
&\min_{\forall \pi_f} \sum_{i=1}^{k}\min \left(\left(d-y_i(\pi_f)\right)\beta,\alpha\right).\label{eq:new-CCW-001}
\end{align}
Jointly, \eqref{eq:new-CCW-001}, \eqref{new4}, and \eqref{eq:upper_lower} imply \eqref{new3}. The proof of Proposition~\ref{prop:low_b} is thus complete.

\subsection{Proof of Proposition~\ref{prop:optimal}}\label{subsec:optimal_proof}
We first introduce the following corollary that will be used shortly to prove Proposition~\ref{prop:optimal}.

\begin{corollary} \label{cor:low_b} For any $(n,k,d)$ values satisfying $d\geq 2$ and $k=\left\lceil \frac{n}{n-d}\right\rceil + 1$, we consider the corresponding IFGs $\gf (n,k,d,\alpha,\beta)$ generated by the family repair scheme $F$. We then have that
\begin{align}
\min_{G\in\gf}\min_{t\in\DC(G)}\mincut(s,t) = \min_{2\leq m\leq k} C_m, \label{eq:low_b_spec}
\end{align}
where $C_m=\sum_{i=0}^{k-1}\min ((d-i)\beta,\alpha)1_{\{i\neq m-1\}} +  \min((d-m+2)\beta,\alpha)$ for $2\leq m\leq k$.
\end{corollary}
\par The proof of Corollary~\ref{cor:low_b} is relegated to Appendix~\ref{app:cor_proof}.

\par We now prove Proposition~\ref{prop:optimal} by proving the following. Consider any fixed $(n,k,d)$ values that satisfy the three conditions of Proposition~\ref{prop:optimal} and any $G\in \mathcal{G}(n,k,d,\alpha,\beta)$ where all the active nodes of $G$ have been repaired at least once. We will prove the statement that such $G$ satisfies that there exists $\frac{n}{2}$ different data collectors, denoted by $t_2,\cdots,t_{\frac{n}{2}+1}\in \DC(G)$, such that
\begin{align} \label{eq:tight_proof}
\mincut_G(s,t_m)\leq C_{m}, \text{ for } 2\leq m \leq\frac{n}{2}+1,
\end{align}
where $C_m$ is defined as in Corollary~\ref{cor:low_b}. Note that the above statement plus Corollary~\ref{cor:low_b} immediately prove Proposition~\ref{prop:optimal} since it says that no matter how we design the helper selection scheme $A$, the resulting $G$ (still belongs to $\mathcal{G}(n,k,d,\alpha,\beta)$) will have $\min_{t\in \DC(G)}\mincut_G(s,t)\leq \min_{2\leq m\leq k} C_m$. 

\par We now prove the above statement. We start with the following definition.

\begin{definition}\label{def:mp_set}A set of $m$ active storage nodes (input-output pairs) of an IFG is called an $(m,p)$-set if the following conditions are satisfied simultaneously. (i) Each of the $m$ active nodes has been repaired at least once; (ii) The chronologically $p$-th node in the $m$ nodes, call it $z$, satisfies that $z_{\inp}$ is connected to at least $p-2$ older nodes of the $m$ nodes; and (iii) Jointly the $m$ nodes satisfy the following property: For any two distinct active nodes $x$ and $y$ in the set of $m$-active nodes such that $y$ is younger than $x$ and $y\neq z$, there exists an edge in the IFG that connects $x_{\out}$ and $y_{\inp}$.
\end{definition}

We now prove the following claim, which will later be used to prove the desired statement.
\begin{claim}
Consider any $G\in\mathcal{G}(n,k,d,\alpha,\beta)$ where $(n,k,d)$ satisfy the three conditions of Proposition~\ref{prop:optimal} and all the active nodes of $G$ have been repaired at least once. In any $l$ active nodes of $G$, where $l$ is an even integer value satisfying $4\leq l\leq n$, there exists a $(\frac{l}{2}+1,p)$-set for all $2\leq p \leq \frac{l}{2}+1$.
\end{claim}

\begin{IEEEproof}
We prove this claim by induction on $l$. We first prove that the claim holds for $l=4$. Consider any set $H_1$ of 4 active nodes of $G$. We will prove the existence of a $(3,2)$-set and a $(3,3)$-set, separately.
\begin{itemize}
\item Existence of a $(3,2)$-set: First, call the chronologically fourth active node of $G$, $u$. Since $d=n-2$, $u$ can avoid at most 1 active node during repair and $u$ is thus connected to at least $3-1=2$ older active nodes in $H_1$. Pick two nodes that $u$ is connected to and call this set of two nodes $V$. Then, we claim that $\{u\}\cup V$ forms a $(3,2)$-set. The reason is the following. Let $v_1$ and $v_2$ denote the two nodes in $V$ and without loss of generality, we assume $v_1$ is older than $v_2$. We have that $u$ is connected to $v_1$ and $v_2$. One can verify that $\{v_1,v_2,u\}$ satisfy the properties (i), (ii), and (iii) of Definition~\ref{def:mp_set} since the second oldest node $z=v_2$. Therefore, $\{v_1,v_2,u\}$ form a $(3,2)$-set. Note that $v_2$ may or may not be connected to $v_1$.
\item Existence of a $(3,3)$-set: Call the chronologically third and fourth active nodes of $H_1$, $v$ and $w$, respectively. Observe that $v$ is connected to at least $2-1=1$ older active node since $d=n-2$ and $v$ can avoid at most one active node during repair. There are only two cases in this scenario: Case~1, $v$ is connected to both the chronologically first and second active nodes; Case~2, $v$ is connected to only one of the chronologically first and second active nodes. Call the active node that $v$ is connected to by $u$ (in Case~1, $u$ can be either the first or the second active node). Then, we claim that $\{u,v,w\}$ is a $(3,3)$-set. This can be proved by verifying that $\{u,v,w\}$ satisfy the properties (i), (ii), and (iii) of Definition~\ref{def:mp_set} based on the following observations. The third oldest node is $z=w$ in this construction. Since $d=n-2$, $w$ can avoid connecting to at most one of its older active nodes. Therefore, $w$ must be connected to at least one of $u$ and $v$. Condition~(ii) in Definition~\ref{def:mp_set} thus holds. Lastly, $u_{\out}$ and $v_{\inp}$ are connected by our construction of $u$, which means that condition~(iii) in Definition~\ref{def:mp_set} holds. 
\end{itemize}

Now, assume that the claim holds for $l\leq l_0-2$. Consider any set of $l_0$ active nodes of $G$ and call it $H_2$. Since $d=n-2$, each node can avoid connecting to at most 1 active node. Therefore, the youngest node in $H_2$, call it $x$, is connected to $l_0-2$ older nodes in $H_2$. Call this set of $(l_0-2)$ nodes, $V_2$. We assumed that the claim holds for $l\leq l_0-2$, this tells us that in $V_2$ there exists an $(\frac{l_0}{2},p)$-set for all $2\leq p \leq \frac{l_0}{2}$. Moreover, for any $(\frac{l_0}{2},p)$-set in $V_2$ with $2\leq p \leq \frac{l_0}{2}$, denoted by $V_3$, we argue that the set $V_3\cup \{x\}$ is a $(\frac{l_0}{2}+1,p)$-set in $H_2$. The reason is that the $p$-th oldest node in $V_3\cup\{x\}$ must be in $V_3$ since $2\leq p \leq \frac{l_0}{2}$. Also, node $x$ is connected to all nodes in $V_2\supseteq V_3$. Therefore, $V_3\cup\{x\}$ satisfies properties (i) to (iii) in Definition~\ref{def:mp_set} and thus form a $(\frac{l_0}{2}+1,p)$-set. 

We are now left with proving that there exists a $(\frac{l_0}{2}+1,\frac{l_0}{2}+1)$-set in $H_2$. By the claim in the proof of Lemma~\ref{lem:set}, there exists an $m$-set in any $(l_0-1)$ active nodes provided that $m$ satisfies $2(m-1)+1\leq l_0-1$. Since $2(\frac{l_0}{2}-1)+1=l_0-1$, there exists a $\frac{l_0}{2}$-set in the oldest $(l_0-1)$ active nodes of $H_2$. Denote this $\frac{l_0}{2}$-set by $V_4$. We argue that $V_4\cup\{x\}$ form a $(\frac{l_0}{2}+1,\frac{l_0}{2}+1)$-set where $x$ is the youngest node in $H_2$. The reason is as follows. Condition~(ii) holds since $x$ can avoid connecting to at most one node that is older, and thus must connect to $(\frac{l_0}{2}-1)$ nodes in this set. Condition~(iii) in Definition~\ref{def:mp_set} holds obviously since $x$ is the youngest node (the $(\frac{l_0}{2}+1)$-th node  chronologically) and the first $\frac{l_0}{2}$ nodes are fully connected as they form an $\frac{l_0}{2}$-set. Hence, the proof of this claim is complete.
\end{IEEEproof}

By the above claim, we have that for any $G\in\mathcal{G}(n,k,d,\alpha,\beta)$ where all the active nodes of $G$ have been repaired at least once there exist all $(\frac{n}{2}+1,p)$-sets for all $2\leq p \leq \frac{n}{2}+1$. We then assign one data collector to each of these $(\frac{n}{2}+1,p)$-sets and denote it by $t_p$, for $p=2$ to $\frac{n}{2}+1$. In total, there are $\frac{n}{2}$ data collectors. 

\par We now apply a similar analysis as in the proof of \cite[Lemma 2]{dimakis2010network} to prove \eqref{eq:tight_proof}. Consider the case of $t_p$. We need to prove that 
\begin{align} \label{eq:tight_proof_p}
\mincut_G(s,t_p)\leq C_{p},
\end{align}
where $t_p$ is the data collector connecting to a $(\frac{n}{2}+1,p)$-set. Denote the storage nodes (input-output pair) of this $(\frac{n}{2}+1,p)$-set by $1,2,\dots,\frac{n}{2}+1$. Define cut $(U,\overline{U})$ between $t_p$ and $s$ as the following: for each $i\in \{0,1,\dots,\frac{n}{2}\}\backslash (p-1)$, if $\alpha \leq (d-i)\beta$ then we include $x_{\out}^{i+1}$ in $\overline{U}$; otherwise, we include both $x_{\out}^{i+1}$ and $x_{\inp}^{i+1}$ in $\overline{U}$. For $i=p-1$, if $\alpha \leq (d-p+2)\beta$, then we include $x_{\out}^p$ in $\overline{U}$; otherwise, we include both $x_{\out}^p$ and $x_{\inp}^p$ in $\overline{U}$. It is not hard to see that the cut-value of the cut $(U,\overline{U})$ is equal to $C_p$. Therefore, we get \eqref{eq:tight_proof_p}. Since \eqref{eq:tight_proof_p} is for general $p$, we get \eqref{eq:tight_proof} and the proof is hence complete.

\section{Generalized Fractional Repetition Codes} \label{sec:gfr}
All the previous analysis assumes that the cut-value condition alone is sufficient for deciding whether one can construct the regenerating code under a given helper selection scheme, i.e., Assumption~\ref{ass:finite_field} in Section~\ref{subsec:characterizing}. In this section, we describe an explicit construction of an exact-repair code, termed \emph{generalized fractional repetition code}, that achieves the MBR point of the FR scheme and can be easily modified to achieve the MBR point of the family-plus repair scheme as well. Since the benefits of helper selection are greatest at the MBR point, our construction completes our mission of understanding under what condition helper selection improves the performance of regenerating codes and how much improvement one can expect from helper selection.

\subsection{The Description of the Generalized Fractional Repetition Code}
\par Our construction idea is based on fractional repetition codes \cite{el2010fractional}. Before describing the generalized fractional repetition codes, we list some notational definitions. We denote the set of nodes of complete family $i$ by $N_i$. For the last complete family, i.e., $i=c$ where $c=\left\lfloor\frac{n}{n-d}\right\rfloor$, we split its nodes into two disjoint node sets, $N_{-c}$ is the set of nodes in family $c$ that is not in the helper set of the incomplete family nodes and $N_c$ is the set of the remaining nodes of this complete family. We denote the set of nodes in the incomplete family by $N_0$. The set of all nodes in the network is denoted by $N$. For example, if $(n,d)=(7,4)$, then we have $c=2$ complete families $\{1,2,3\}$ and $\{4,5,6\}$,  and 1 incomplete family $\{7\}$. Furthermore, we have $N_1=\{1,2,3\}$, $N_2=\{4\}$, $N_{-2}=\{5,6\}$; $N_0=\{7\}$. In short, $N_x$ contains the nodes that have family index $x$. Moreover, we assume throughout this section that $\beta=1$ and $\alpha=d\beta=d$, i.e., one packet is communicated per helper and $d$ packets are stored in each node since the generalized fractional repetition code we describe does not require sub-packetizing. 

\par The goal of generalized fractional repetition codes is to protect a file of size 
\begin{align} \label{eq:file_size_construction}
\mathcal{M}=\sum_{i=1}^{k} \left(d-y_i(\pi_f^*)\right) \text { packets}
\end{align}
against any $(n-k)$ simultaneous failures. From \eqref{eq:file_size_construction}, we can easily see that the larger the $k$ value, the more relaxed the reliability requirement is, and the larger the file size $\mathcal{M}$ the generalized fractional repetition code can protect. 

\par To handle all possible $(n,k,d)$ values, the construction of the generalized fractional repetition code is quite complicated. The core idea of these codes stems from a graph representation of the distributed storage system. Although the proposed generalized fractional repetition codes can still be constructed without the aid of this graph, the graph representation is inevitable for gaining intuition about their construction and facilitating their analysis. For that reason, we base our detailed discussion of the generalized fractional repetition codes on the graph. In the following, we start the description of these codes by introducing their graph representation.

\par {\bf The graph representation:}  Each physical node in the distributed storage system is represented by a vertex in the graph, which is denoted by $G=(V,E)$ where $V$ denotes the set of vertices of $G$ and $E$ denotes its set of edges. As will be described, the graph consists of two disjoint groups of edges. Graph $G$ has the following properties:

\begin{enumerate}
\item $V=\{1,2,\cdots, n\}$. Each vertex $i$ in $V$ corresponds to physical node $i$ in $N$. For convenience, throughout our discussion, we simply say vertex $i\in N_x$ if the physical node that vertex $i$ corresponds to is in $N_x$.
\item Two vertices $i\in N_x$ and $j\in N_y$ are connected by an edge in $E$ if $|x|\neq|y|$ and $(x,y)\notin \{(0,-c),(-c,0)\}$. The collection of all those edges is denoted by $\bar{E}$.
\item Two vertices $i\in N_0$ and $j\in N_{-c}$ are connected by an edge in $E$. The collection of all those edges is denoted by $\tilde{E}$.
\item From the above construction, we have $E=\bar{E}\cup\tilde{E}$. We further assume that all the edges are undirected and there are no parallel edges in $G$.
\end{enumerate}

\begin{figure}[h!] 
\centering
\includegraphics[width=0.475\textwidth]{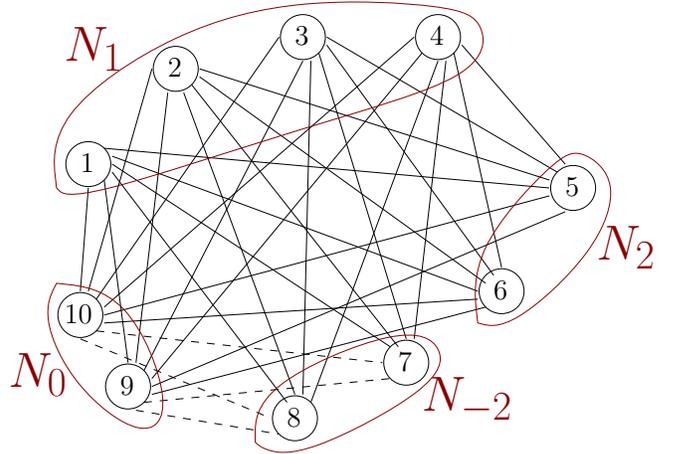} 
\caption{A graph representation of the generalized fractional repetition code for $(n,d)=(10,6)$.}
\label{fig:gfr_graph}
\end{figure}

Fig.~\ref{fig:gfr_graph} illustrates the graph representation for the generalized fractional repetition code with $(n,d)=(10,6)$. We graphically represent edges in $\bar{E}$ by solid lines and edges in $\tilde{E}$ by dashed lines.

\par For any physical node $i$, we use $FI(i)$ to denote the family index of $i$. We define the following three sets:
\begin{align}
{\sf IJ}^{[1]}&= \nonumber\\
&\{(i,j): 1\leq i< j\leq n, 1\leq|FI(i)|<|FI(j)| \leq c\}\nonumber\\
{\sf IJ}^{[2]}&= \{(i,j): 1\leq i<j \leq n, 1\leq FI(i)\leq c,FI(j)=0\}\nonumber\\
{\sf IJ}^{[3]}&= \{(i,j): 1\leq j<i \leq n, FI(i)=0, FI(j)=-c\}\nonumber.
\end{align}

One can easily verify that the union of the first two sets, ${\sf IJ}^{[1]}\cup{\sf IJ}^{[2]}$, can be mapped bijectively to the edge set $\bar{E}$, and the third set ${\sf IJ}^{[3]}$ can be mapped bijectively to the edge set $\tilde{E}$. The difference between sets ${\sf IJ}^{[1]}$ to ${\sf IJ}^{[3]}$ and $\bar{E}$, $\tilde{E}$, and $E$ is that the sets ${\sf IJ}^{[1]}$ to ${\sf IJ}^{[3]}$ focus on {\em ordered pairs} while the edges in $E$ correspond to unordered vertex pairs (undirected edges). Also, we can see that there are $\frac{(n-|N_0|)(d-|N_0|)}{2}$ pairs in ${\sf IJ}^{[1]}$, $d|N_0|$ pairs in ${\sf IJ}^{[2]}$, and $|N_{-c}|\cdot|N_0|$ pairs in ${\sf IJ}^{[3]}$. Thus, in total, there are
\begin{align}\label{eq:total_packets}
\frac{(n-|N_0|)(d-|N_0|)}{2}+d|N_0|+|N_{-c}|\cdot |N_0|
\end{align}
distinct pairs in the overall index set ${\sf IJ}^{[1]}\cup {\sf IJ}^{[2]}\cup {\sf IJ}^{[3]}$. This implies that the total number of edges of graph $G$ is $|E|=\frac{(n-|N_0|)(d-|N_0|)}{2}+d|N_0|+|N_{-c}|\cdot |N_0|$.

\par {\bf Coded packets generation:} Each edge of graph $G$ corresponds to one coded packet that is stored in the distributed storage system. More specifically, each edge $(i,j)\in \bar{E}$ represents a packet $P_{(i,j)}$ that is stored in the two physical nodes $i$ and $j$, i.e., both nodes $i$ and $j$ store an identical copy of the packet $P_{(i,j)}$. On the other hand, each edge $(i,j)\in\tilde{E}$ represents a packet $\tilde{P}_{(i,j)}$ that is only stored in one of its two vertices, the corresponding vertex in $N_{-c}$. One can verify by examining the ${\sf IJ}^{[1]}$ to ${\sf IJ}^{[3]}$ index sets defined previously that each physical node stores exactly $\alpha=d$ packets. 

\par We now describe how to generate the $|{\sf IJ}^{[1]}|+|{\sf IJ}^{[2]}|+|{\sf IJ}^{[3]}|$ coded packets (the $P_{(i,j)}$ and $\tilde{P}_{(i,j)}$ packets depending on whether $(i,j)\in\bar{E}$ or $(i,j)\in\tilde{E}$) from the to-be-protected file of $\mathcal{M}$ packets, where $\mathcal{M}$ is specified by \eqref{eq:file_size_construction}. To that end, we impose the following two properties on the coded packets of the edges.

\par {\bf Property~1:} Any coded packet $\tilde{P}_{(i_0,j_0)}$ corresponding to some $(i_0,j_0)\in {\sf IJ}^{[3]}$ is a linear combination of the $P_{(j_1,i_0)}$ for all $j_1$ satisfying $(j_1,i_0)\in {\sf IJ}^{[2]}$. In total, there are $d$ such $j_1$ indices. Specifically, the packet corresponding to $\tilde{P}_{(i_0,j_0)}$ is stored only in node $j_0$ since $(i_0,j_0)\in\tilde{E}$ and $\tilde{P}_{(i_0,j_0)}$ is a linear combination of the $d$ packets stored in node $i_0$.

\par We now describe the second required property. Recall that there are $|N_0|=n\bmod (n-d)$ nodes in the incomplete family and they are nodes $c(n-d)+1$ to $c(n-d)+|N_0|$ where $c$ is the family index of the last complete family. For any subset of the total $|E|$ packets, define $a_m$, $m=1$ to $|N_0|$, as the number of packets that correspond to all edges in $E=\bar{E}\cup\tilde{E}$ connected to the vertex $(c(n-d)+m)\in N_0$. Define $a_0$ as the number of packets in this subset that correspond to edges that are not connected to any of the vertices in $N_0$. Define $\mathsf{a.count}\stackrel{\Delta}{=} a_0+\sum_{m=1}^{|N_0|}\min(a_m,d)$. In sum, we can compute a value $\mathsf{a.count}$ from any subset of edges.

\par {\bf Property~2:} The $|E|$ coded packets satisfy that we must be able to reconstruct the original file from any subset of packets (edges) that satisfies $\mathsf{a.count} \geq \mathcal{M}$.

\par We now argue that we can always find a set of $|E|$ coded packets that satisfy the above two properties. Specifically, we can use a two-phase approach to generate the packets. We first independently and uniformly randomly generate $|\bar{E}|=\frac{(n-|N_0|)(d-|N_0|)}{2}+d|N_0|$ linearly encoded packets from the $\mathcal{M}$ packets of the original file. These packets are fixed and arbitrarily assigned to the edges in $\bar{E}$ (one for each edge). After this first step, all physical nodes store exactly $d$ packets except those nodes in $N_{-c}$, each of which now stores exactly $(d-|N_0|)$ packets. Now, from each node in $u\in N_0$, we generate independently and uniformly a random set of $|N_{-c}|$ linearly encoded packets from the $d$ packets stored in $u$. We fix these newly generated packets and assign them arbitrarily to each of the $|N_{-c}|$ edges in $\{(u,w)\in \tilde{E}:\forall w\in N_{-c}\}$. Specifically, these $|N_{-c}|$ packets will now be stored in node $w\in N_{-c}$, one for each $w$. Repeat this construction for all $u\in N_0$. After this second step, each edge in $\bar{E}\cup \tilde{E}$ has been assigned one distinct coded packet and each node in $N=N_1\cup\cdots N_{c}\cup N_{-c}\cup N_0$ now stores exactly $d$ packets. After the initial random-construction phase, we enter the second phase, the verification phase. In this phase, we fix the packets and deterministically check whether they satisfy Property~2 (by our construction the coded packets always satisfy Property~1). The following lemma states that with high probability, the randomly generated packets will satisfy Property~2.

\begin{lemma}\label{lem:gfr_existence}
When $\text{GF}(q)$ is large enough, with close-to-one probability, the above random construction will satisfy Property~2.
\end{lemma}

\par The proof of Lemma~\ref{lem:gfr_existence} is relegated to Appendix~\ref{app:gfr_existence}.

\par Lemma~\ref{lem:gfr_existence} implies that with high-probability, the random construction will lead to a deterministic set of coded packets that satisfies Properties 1 and 2. In the rare event that the random construction does not satisfy Property~2, we simply repeat the random construction until we find a set of coded packets that satisfies Properties 1 and 2. Note that this construction is performed off-line during the design stage. Once the coded packets are found by random construction, we will fix the coded packets for future use. Also, the construction is not unique. We may be able to use some other method of construction.\footnote{The computational complexity during the design stage is not the main focus in this work. Therefore, we opted to use the random code construction to demonstrate the existence of a desired code. For practical implementation, some finite-algebra-based construction could drastically reduce the complexity of the construction.} All our subsequent discussion holds as long as the final coded packets satisfy Properties~1 and 2. 

\par We now provide a detailed example on the construction of a generalized fractional repetition code. Suppose $(n,k,d)=(7,4,4)$. Then, there are two complete families $\{1,2,3\}$ and $\{4,5,6\}$ and $1$ incomplete family $\{7\}$. We will have that the RFIP is $\pi_f^*=(1,2,0,1,-2,1,-2)$ and the file size is $\mathcal{M}=11$ packets, see \eqref{eq:file_size_construction}. By \eqref{eq:total_packets}, we have $|E|=15$, $|\bar{E}|=13$, and $|\tilde{E}|=2$. Then, we choose $\text{GF}(128)$ and randomly generate the first $|\bar{E}|=13$ packets and their coding vectors are

\begin{align}
(i,j) \quad & \text{ Coding vector for } P_{(i,j)}\nonumber\\
(1,7) \quad & (1,0,0,0,0,0,0,0,0,0,0) \nonumber\\
(2,7)\quad & (0,1,0,0,0,0,0,0,0,0,0) \nonumber\\
(3,7)\quad & (0,0,1,0,0,0,0,0,0,0,0) \nonumber\\
(4,7)\quad & (0,0,0,1,0,0,0,0,0,0,0) \nonumber\\
(1,4)\quad & (0,0,0,0,1,0,0,0,0,0,0) \nonumber\\
(1,5)\quad & (0,0,0,0,0,1,0,0,0,0,0) \nonumber\\
(1,6)\quad & (0,0,0,0,0,0,1,0,0,0,0) \nonumber\\
(2,4)\quad & (0,0,0,0,0,0,0,1,0,0,0) \nonumber\\
(2,5)\quad & (0,0,0,0,0,0,0,0,1,0,0) \nonumber\\
(2,6)\quad & (0,0,0,0,0,0,0,0,0,1,0) \nonumber\\
(3,4)\quad & (0,0,0,0,0,0,0,0,0,0,1) \nonumber\\
(3,5)\quad & (21,56,81,119,67,80,87,118,19,51,39) \nonumber\\
(3,6)\quad & (88,114,62,103,41,70,49,114,86,106,14)\nonumber.
\end{align}

Then, we generate the additional $\tilde{E}$ packets by mixing the packets in any given $u\in N_0$. The newly generated coding vectors are 
 
\begin{align}
(i,j) \quad & \text{ Coding vector for }\tilde{P}_{(i,j)}\nonumber\\
(7,5)\quad & (35,98,27,4,0,0,0,0,0,0,0) \nonumber\\
(7,6)\quad & (55,119,33,72,0,0,0,0,0,0,0)\nonumber.
\end{align}
One can easily verify, with the aid of a computer, that both Properties 1 and 2 hold for the above choices of coded packets (coding vectors).  

\par The correctness of the proposed generalized fractional repetition codes for FR will be proved in Section~\ref{subsec:gfr_proof}. 

\par We note that the generalized fractional repetition codes described above can be modified and used to construct an explicit exact-repair code that can achieve the MBR point of the family-plus repair scheme. This is achieved by first applying the same graph construction of the above generalized fractional repetition codes to each group of the family-plus repair scheme, i.e., the edge representation of each group consists of the two edge sets $\bar{E}$ and $\tilde{E}$. Then, since the repair of the family-plus scheme occurs within each group separately, we enforce Property~1 for each individual group so that we can maintain the exact-repair property. Finally, we need to ensure that any subset of $k$ nodes (which could be across multiple groups) can be used to reconstruct the original file. Therefore, we have to ensure that the coded packets satisfy a modified version of Property~2. 

\par In the following we briefly describe how to do this modification with a slight abuse of notation. Recall that in the family-plus repair scheme, only the incomplete group has an incomplete family. Denote the set of incomplete family nodes in the incomplete group by $M_0$ and the graph of the incomplete group by $G_{\text{inc}}=(V_{\text{inc}},E_{\text{inc}})$. The new property imposed on the packets becomes

{\bf Modified Property~2:} Index the vertices in $M_0\subset V_{\text{inc}}$ by $\{u_1,u_2,\cdots,u_{|M_0|}\}$. For any given subset of the total packets (across all groups) and any given $m$ satisfying $1\leq m\leq |M_0|$, define $a_m$ as the number of packets in this subset that correspond to the edges in $E_{\text{inc}}=\bar{E}_{\text{inc}}\cup\tilde{E}_{\text{inc}}$ that are incident to vertex $u_m\in M_0$. Define $a_0$ as the number of the other packets in this subset, i.e., those packets not corresponding to any edges that are incident to $M_0$. Define $\mathsf{a.count}\stackrel{\Delta}{=} a_0+\sum_{m=1}^{|M_0|}\min(a_m,d)$. Then we must be able to reconstruct the original file of size $\mathcal{M}$ if $\mathsf{a.count}\geq \mathcal{M}$.

We can again use the concept of random linear network coding to prove the existence of a code satisfying Property~1 and the modified Property~2 in a similar way as in Lemma~\ref{lem:gfr_existence}. The correctness of the proposed generalized fractional repetition codes for family-plus repair schemes can be proved in a similar way as when proving the correctness for family repair schemes provided in Section~\ref{subsec:gfr_proof}. We omit the detailed proofs since the proofs are simple extensions of the proofs we provide for the FR scheme with only the added notational complexity of handling different groups of nodes in the family-plus repair schemes.

\par We also note that the proposed code construction is termed the generalized fractional repetition codes because it borrows the main ingredient of representing the code construction as a graph with each edge representing a packet. Such a representation leads to straightforward arguments that the proposed codes can be exactly repaired by communicating the missing copy from the other helper. On the other hand, the proposed solution has the new ingredient of the edges in $\tilde{E}$ which allows the code construction to handle arbitrary parameter values while still being an exact-repair code. One major contribution of the code construction in this work is to put the generalized fractional repetition codes in the context of quantifying the benefits of intelligent helper node selection and to show that the generalized  fractional repetition codes achieve the MBR point of the FR scheme predicted by the pure min-cut-value-based characterization.

\par The remaining part of Section~\ref{sec:gfr} is dedicated exclusively to proving that the generalized fractional repetition code is a legitimate exact-repair regenerating code that achieves the MBR point of the FR scheme described in Proposition~\ref{prop:mbr}. Practitioners may consider skipping the proofs and go directly to the conclusion section, Section~\ref{sec:conc}.

\begin{remark} The original fractional repetition code in \cite{el2010fractional} is an explicit exact-repair code for the case when the product $nd$ is even, but \cite{el2010fractional} does not provide any construction when $nd$ is odd. Moreover, the performance of the construction of \cite{el2010fractional} depends heavily on ``the underlying regular graph.'' Since \cite{el2010fractional} does not discuss how to choose the regular graphs, it is not clear how to optimize the performance of the fractional repetition codes in \cite{el2010fractional}. For comparison,  our construction is an exact-repair code applicable to {\em all possible $(n,k,d)$ combinations}; we provide a new way of optimally designing the regular and possibly irregular graphs,  and prove that our construction always achieves the MBR point of the FR scheme.  
\end{remark}

\subsection{Proofs for the GFR Code}\label{subsec:gfr_proof}

In this subsection, we first argue that the above generalized fractional repetition code can be exactly repaired using the FR scheme. First, consider the case that node $i$ fails for some $i\in N_1\cup N_2\cup \cdots\cup N_c\cup N_0$ (those in $N\backslash N_{-c}$). The $d$ packets stored in node $i$ thus need to be repaired. We then notice that the $d$ packets in node $i$ correspond to the $d$ edges in $\bar{E}$ that are incident to node $i$. Therefore, each of those $d$ packets to be repaired is stored in another node $j$. Also by our construction, the neighbors of node $i$ are indeed the helper set $D_i$ of the FR scheme. Therefore, the newcomer $i$ can use the FR scheme to decide which nodes to be the helpers and request the helpers to send the intact copies of the to-be-repaired $d$ packets (one intact copy from each of the helpers). 

\par For example, suppose we reconsider the example above where $(n,k,d)=(7,4,4)$. Node $4\in N_2$ stores the $d=4$ packets corresponding to edges $(4,1)$, $(4,2)$, $(4,3)$, and $(4,7)$. Suppose that node 4 fails. Since each of the nodes $\{1,2,3,7\}$ store one of the packets of node 4 and node 4 can receive one packet from each of the $d=4$ helper nodes during repair, node 4 can always restore the exact packets $P_{(4,1)}$, $P_{(4,2)}$, $P_{(4,3)}$, and $P_{(4,7)}$ that it initially stored. Observe that in the same way, all nodes in $N_1\cup N_2\cup \cdots\cup N_c\cup N_0$ can be repaired exactly. Therefore, we are left to show how nodes in the set $N_{-c}$ can be repaired exactly. 

\par Suppose node $i$ in $N_{-c}$ fails. We again notice that $(d-n\bmod(n-d))$ of its $d$ packets correspond to edges in $\bar{E}$ and their corresponding neighbors are also in the helper set $D_i$ of the FR scheme. Therefore, the newcomer $i$ can use the FR scheme to decide which nodes to be the helpers and request $(d-n\bmod(n-d))$ out of its $d$ helpers to send one of the to-be-repaired packets. If we dig deeper, those $(d-n\bmod(n-d))$ helpers are the nodes that have family indices belonging to $\{1,\cdots, c-1\}$. 

\par To restore the remaining $n\bmod(n-d)$ packets, we notice first that by our construction, these packets correspond to the edges in $\{(i,w)\in \tilde{E}:w\in N_0\}$. By our code construction, for any $w_0\in N_0$, $\tilde{P}_{(i,w_0)}$ is a linear combination of the $d$ packets $\{P_{(w_0,j)}:(w_0,j)\in \bar{E}, j=1,2,\cdots, d\}$ stored in node $w_0\in N_0$. Thus, during repair, newcomer $i$ can ask physical node $w_0$ to compute the packet $\tilde{P}_{(i,w_0)}$ and send the final result. In a similar fashion, newcomer $i\in N_{-c}$ can repair all other packets $\tilde{P}_{(i,w)}$ for all $w\in N_0$. Therefore, newcomer $i$ can exactly repair all the remaining $n\bmod(n-d)$ packets as well.  

\par Considering the same example above, node $6\in N_{-2}$ can restore packets corresponding to $\{(6,1),(6,2),(6,3)\}\subseteq \bar{E}$ by receiving copies of these packets from nodes $\{1,2,3\}$ and can request the packet of edge $(6,7)\in\tilde{E}$ from node $7\in N_0$. Node $7$ can generate that packet $\tilde{P}_{(6,7)}$ by computing the corresponding linear combination from the packets it stores, i.e., the packets $P_{(7,1)}$, $P_{(7,2)}$, $P_{(7,3)}$, and $P_{(7,4)}$. This shows that nodes in $N_{-c}$ can also be exactly repaired, hence, all the nodes in a generalized fractional repetition code can be exactly repaired when following the FR helper selection scheme.

\par The following proposition shows that the generalized fractional repetition code with FR helper selection can protect against any $(n-k)$ simultaneous failures.
\begin{proposition}\label{prop:gfr_rec}
Consider the generalized fractional repetition code with any given $(n,k,d)$ values. For any arbitrary selection of $k$ nodes, one can use all the $kd$ packets stored in these $k$ nodes (some of them are identical copies of the same coded packets) to reconstruct the original $\mathcal{M}$ file packets.
\end{proposition}

Since the $\alpha$, $\beta$, and $\mathcal{M}$ values in \eqref{eq:file_size_construction} match the MBR point of the FR scheme, Proposition~\ref{prop:gfr_rec} shows that the explicitly constructed generalized fractional repetition code indeed achieves the MBR point of the FR scheme predicted by the min-cut-based analysis. 

\par The rest of this section is dedicated to the proof of Proposition~\ref{prop:gfr_rec}.
\begin{IEEEproof}
Consider an arbitrarily given set of $k$ nodes in the distributed storage network, denoted by $S$. Denote nodes in $S$ that belong to $N_i$ by $S_i\stackrel{\Delta}{=}S\cap N_i$. We now consider the set of edges that are incident to the given node set $S$, i.e., those edges have at least one end being in $S$ and each of the edges corresponds to a distinct packet stored in nodes $S$. Recall that for any set of edges, we can compute the corresponding $\mathsf{a.count}$ value as defined in Property~2 of our code construction. The following is a procedure, termed {\sc Count}, that computes the value $\mathsf{a.count}$ of the edges incident to $S$: 

\begin{enumerate}
\item We first define $G_1=(V_1,E_1)=G=(V,E)$ as the original graph representation of the generalized fractional repetition code. Choose an arbitrary order for the vertices in $S$ such that all nodes in $S_{-c}$ come last. Call the $i$-th vertex in the order by $v_i$. That is, we have that $S_{-c}=\{v_i:k-|S_{-c}|+1\leq i\leq k\}$ and $S_1\cup\cdots\cup S_c\cup S_0=\{v_i:1\leq i\leq k-|S_{-c}| \}$. 
\item Set $e(S)=0$, where $e(S)$ will be used to compute $\mathsf{a.count}$.

Now, do the following step sequentially for $i=1$ to $|S|=k$: 
\item Consider vertex $v_i$. We first compute
\begin{align}\label{eq:inter9}
x_i=&|\{(v_i,j)\in E_i \cap \bar{E}:j\in N\}|+\nonumber\\
&1_{\{v_i\in S_{-c}\}}\cdot\sum_{u\in N_0} 1_{\{(u,v_i) \in E_i\cap \tilde{E}\}} \cdot \nonumber\\
&1_{\{|\{(u,j)\in E_i :j\in N\}|>|N_{-c}| \}}. 
\end{align}
Once $x_i$ is computed, update $e(S)=e(S)+x_i$. Remove all the edges incident to $v_i$ from $G_i$. Denote the new graph by $G_{i+1}=(V_{i+1},E_{i+1})$.
\end{enumerate}

\par Intuitively, we first ``count'' the number of edges in $G_i$ that belongs to $\bar{E}$ and is connected to the target vertex $v_i$, namely, the $|\{(v_i,j)\in E_i \cap \bar{E}:j\in N\}|$ term in \eqref{eq:inter9}. Then, if the target vertex $v_i\in S_{-c}$, we compute one more term in the following way. For each edge $(u,v_i)\in E_i\cap \tilde{E}$, if the following inequality holds, we also count this specific $(u,v_i)$ edge: 

\begin{align}
|\{(u,j)\in E_i:j\in N\}|>|N_{-c}|.
\end{align}
That is, we check how many edges (including those in $\bar{E}$ and in $\tilde{E}$) are connected to $u$. We count the single edge $(u,v_i)$ if there are still at least $(|N_{-c}|+1)$ edges in $E_i$ that are connected to $u$. Collectively, this additional counting mechanism for the case of $v_i\in S_{-c}$ gives the second term in \eqref{eq:inter9}. After counting the edges incident to $v_i$, we remove those edges from future counting rounds (rounds $>i$) so that we do not double count the edges in any way.

\begin{claim}\label{clm:proc_count}
After finishing the subroutine {\sc Count}, the final $e(S)$ value is exactly the value of $\mathsf{ a.count}$. 
\end{claim}
\emph{Proof of Claim~\ref{clm:proc_count}:}
\par The proof of the above claim is as follows. We first note that in the subroutine, we order the nodes in $S$ in the specific order such that all nodes in $S_{-c}$ are placed last. Therefore, in the beginning of the subroutine {\sc Count}, all the $v_i$ vertices do not belong to $S_{-c}$. Therefore, the second term in \eqref{eq:inter9} is zero. Since $v_i\notin S_{-c}$, all the edges connected to $v_i$ are in $\bar{E}$. The first term of \eqref{eq:inter9} thus ensures that we count all those edges in this subroutine. Since we remove those counted edges in each step (from $G_i$ to $G_{i+1}$), we do not double count any of the edges. Therefore, before we start to encounter a vertex $v_i\in S_{-c}$, the subroutine correctly counts the number of edges incident to the $v_{j}$ for all $1\leq j<i$. 

\par We now consider the second half of the subroutine, i.e., when $v_i\in S_{-c}$. We then notice that the subroutine still counts all those edges in $\bar{E}$ through the first term in \eqref{eq:inter9}. The only difference between {\sc Count} and a regular counting procedure is the second term in \eqref{eq:inter9}. That is, when counting any edge in $\tilde{E}$, we need to first check whether the total number of edges in $G_i$ incident to $u$ is greater than $|N_{-c}|$. To explain why we have this {\em conditional counting} mechanism, 
we notice that in the original graph $G$, each node $u\in N_0$ has $|\{(u,j)\in \bar{E}:j\in N\}|=d$ and $|\{(u,j)\in\tilde{E}:j\in N\}|=|N_{-c}|$. Therefore, the total number of edges connected to $u$ is $|\{(u,j)\in E:j\in N\}|=d+|N_{-c}|$. Note that during the counting process, those counted edges are removed from the graph during each step. Since $G_i$ is the remaining graph after removing all those counted edges in the previous $(i-1)$ steps, if we still have $|\{(u,j)\in E_i:j\in N\}|>|N_{-c}|$, then it means that we have only removed strictly less than $(d+|N_{-c}|)-|N_{-c}|=d$ number of edges in the previous $(i-1)$ counting rounds. The above argument thus implies that in the previous $(i-1)$ counting rounds, we have only counted $<d$ edges that are incident to node $u$.

\par Without loss of generality, we assume that $u$ is the $m$-th node of $N_0$. Then it means that the $a_m$ value (the number of edges connected to $u$) computed thus far (until the beginning of the $i$-th counting round) is still strictly less than $d$. Therefore, when computing the objective value $\mathsf{a.count}=a_0+\sum_m\min(a_m,d)$, the to-be-considered edge $(v_i,u) $ in the second term of \eqref{eq:inter9} will increment $a_m$ value by 1 and thus increment $\mathsf{a.count}$ by 1. Since our goal is to correctly compute the $\mathsf{a.count}$ value by this subroutine, the subroutine needs to include this edge into the computation, which leads to the second term in \eqref{eq:inter9}. 

\par On the other hand, if the total number of edges in $G_i$ that are adjacent to $u$ is $\leq |N_{-c}|$, it means that we have removed $\geq (d+|N_{-c}|)-|N_{-c}|=d$ number of edges in the previous counting rounds. That is, when counting those edges adjacent to $u$, we have already included/encountered $\geq d$ such edges in the previous $(i-1)$ rounds. As a result, the corresponding $a_m$ value is $\geq d$. Therefore, when computing the objective value $\mathsf{a.count}=a_0+\sum_m\min(a_m,d)$, the to-be-considered edge $(v_i,u) $ will increment the value of $a_m$ by 1 but {\em will not} increment the $\mathsf{a.count}$ value. In the subroutine {\sc Count}, we thus do not count the edges in $\tilde{E}_i$ anymore, which leads to the second term in \eqref{eq:inter9}. 

\par The new constraint put in Step~3 thus ensures that the final output $e(S)$ is the value of $\mathsf{a.count}$. We now need to prove that for any set $S$ of $k$ nodes, the corresponding $e(S)\geq\mathcal{M}$. Assuming this is true, we can then invoke Property~2, which guarantees that we can reconstruct the $\mathcal{M}$ packets of the original file from the coded packets stored in $S$. 

\par The proof of $e(S)\geq\mathcal{M}$ consists of two additional claims.
\begin{claim}\label{clm:edges_count}
Suppose there exists a node $a\in S_{-c}$ and a node $b\in N_c\backslash S_c$. Then
\begin{align} \label{eq:claim1}
e(S)=e(S\cup \{b\}\backslash a).
\end{align}
\end{claim}
Claim~\ref{clm:edges_count} will be used to prove the following claim. 
\begin{claim} \label{clm:gfr}
For any arbitrarily given set $S$, there exists an $\tilde{\mathbf{r}} \in R=\{(r_1,r_2,\cdots,r_k):\forall i \in\{1,\cdots,k\}, 1\leq r_i\leq n\}$ such that
\begin{align}\label{eq:claim2}
e(S)=\sum_{i=1}^k(d-z_i(\tilde{\mathbf{r}})), 
\end{align}
where $z_i(\cdot)$ is as defined in Proposition~\ref{prop:low_b_gen}.
\end{claim}
Using the above claims, we have 
\begin{align}
\mathsf{a.count}&=e(S)=\sum_{i=1}^k(d-z_i(\tilde{\mathbf{r}}))\label{eq:gfr_1}\\
&\geq \min_{\mathbf{r}\in R} \sum_{i=1}^k(d-z_i(\mathbf{r}))\label{eq:gfr_2}\\
&=\min_{\pi_f} \sum _{i=1}^{k}(d-y_i(\pi_f))\label{eq:gfr_3}\\
&= \sum _{i=1}^k(d-y_i(\pi_f^*))\label{eq:gfr_4}\\
&= \mathcal{M}\label{eq:gfr_5}.
\end{align}
where \eqref{eq:gfr_1} follows from Claim~\ref{clm:gfr}, \eqref{eq:gfr_2} follows from taking the minimum operation, \eqref{eq:gfr_3} follows from the proof of Proposition~\ref{prop:low_b},  \eqref{eq:gfr_4} follows from the optimality of the RFIP, and \eqref{eq:gfr_5} follows from \eqref{eq:file_size_construction}. By Property~2, we have thus proved that the $kd$ packets stored in any set of $k$ nodes can be used to jointly reconstruct the original file of size $\mathcal{M}$. 

The proofs of Claims~\ref{clm:edges_count} and~\ref{clm:gfr} are provided in the following. 




\emph{Proof of Claim~\ref{clm:edges_count}:}
\par We consider {\sc Count} for the set $S'=S\cup \{b\}\backslash a$ and we denote nodes in $S'$ that belong to $N_i$ by $S_i'\stackrel{\Delta}{=}S'\cap N_i$. To avoid confusion when $S'$ is used as input to the subroutine {\sc Count}, we call the new graphs during the counting steps of {\sc Count} by $G'_{i}=(V'_i,E'_i)$, the new vertices by $v'_{i}$, and the new $x_i$ by $x'_i$. Since the subroutine {\sc Count} can be based on any sorting order of nodes in $S$ (and in $S'$) as long as those nodes in $N_{-c}$ come last, we assume that the nodes in $S$ are sorted in a way that node $a$ is the very first node in $S_{-c}$. For convenience, we say that node $a$ is the $i_0$-th node in $S$ and we assume that all the first $(i_0-1)$-th nodes are not in $S_{-c}$ and all the nodes following the $(i_0-1)$-th node are in $S_{-c}$. Namely, $i_0=|S|-|S_{-c}|+1=k+1-|S_{-c}|$.  We now use the same sorting order of $S$ and apply it to $S'$. That is, the $i$-th node of $S$ is the same as the $i$-th node in $S'$ except for the case of $i=i_0$. The $i_0$-th node of $S'$ is set to be node $b$. One can easily check that the sorting orders of $S$ and $S'$ both satisfy the required condition in Step~1 of the subroutine {\sc Count}. 

\par We will run {\sc Count} on both $S$ and $S\cup \{b\}\backslash a$ in parallel and compare the resulting $e(S)$ and $e(S\cup \{b\}\backslash a)$. 

\par It is clear that in rounds 1 to $(i_0-1)$, the subroutine {\sc Count} behaves identically when applied to the two different sets $S$ and $S'=S\cup \{b\}\backslash a$ since their first $(i_0-1)$ vertices are identical. We now consider the $i_0$-th round and argue that the total number of edges in $E'_{i_0}$ incident to $v'_{i_0}$ is equal to the total number of edges incident to $v_{i_0}$ in $E_{i_0}$. Recall that $b$ and $a$ have the same helper sets since they are from the same complete family. Specifically, the edges in $E$ incident to $v_{i_0}=a\in S_{-c}$ that have been counted in the first $(i_0-1)$ rounds are of the form $(u,a)$ for all $u\in\{v_1,v_2,\cdots,v_{i_0-1}\}\cap (S_0\cup S_1\cup \cdots \cup S_{c-1})$. Also note that in the original graph $G$, there are exactly $d$ edges incident to node $a\in S_{-c}$ (some of them are in $\bar{E}$ and some of them in $\tilde{E}$). Therefore, in $E_{i_0}$ (after removing those previously counted edges), there are $(d-|\{v_1,v_2,\cdots,v_{i_0-1}\}\cap (S_0\cup S_1\cup \cdots \cup S_{c-1})|)$ number of edges that are incident to $v_{i_0}$. 

\par Similarly, the edges in $E_{i_0}'$ incident to $v_{i_0}'=b\in S_{c}'$ that have been counted previously are of the form $(u,b)$ for all $u\in \{v_1,v_2,\cdots,v_{i_0-1}\}\cap (S_0\cup S_1\cup \cdots \cup S_{c-1})$ since $v_i'=v_i$ for $1\leq i\leq i_0-1$ and $S_x'=S_x$ for $0\leq x\leq c-1$. Also note that, in the original graph $G'$, there are exactly $d$ edges incident to node $b\in S_{c}'$ (all of them are in $\bar{E'}$). Therefore, in $E_{i_0}'$ (after removing those previously counted edges), there are $(d-|\{v_1,v_2,\cdots,v_{i_0-1}\}\cap (S_0\cup S_1\cup \cdots \cup S_{c-1})|)$ number of edges that are incident to $v_{i_0}'=b$. 

\par We now argue that all the edges in $E_{i_0}$ that are incident to $a$ will contribute to the computation of $x_{i_0}$. The reason is that node $a$ is the first vertex in $S_{-c}$. Therefore, when in the $i_0$-th counting round, no edge of the form $(u,v)$ where $u\in N_0 \backslash S_0$ and $v\in N_{-c}$ has ever been counted in the previous $(i_0-1)$ rounds. Also, since we choose $b\in N_c\backslash S$ to begin with, when running {\sc Count} on $S$, for all $u\in N_0\backslash S_0$ at least one edge, edge $(u,b)$, is not counted during the first $(i_0-1)$ rounds. As a result, for any $u\in N_0\backslash S_0$, in the $i_0$-th round, at least $|\{(u,v): v\in N_{-c}\}|+1=|N_{-c}|+1$ edges incident to $u$ are still in $E_{i_0}$ (not removed in the previous $(i_0-1)$ rounds). This thus implies that the second term of \eqref{eq:inter9} will be non-zero. Therefore, at the $i_0$-th iteration of Step~3 of {\sc Count}, all the edges in $E_{i_0}$ incident to $v_{i_0}=a$ are counted. The $x_{i_0}$ value computed in \eqref{eq:inter9} thus becomes $x_{i_0}= d-|\{v_1,v_2,\cdots,v_{i_0-1}\}\cap (S_0\cup S_1\cup \cdots \cup S_{c-1})|$. 

\par The previous paragraph focuses on the $i_0$-th round when running the subroutine {\sc Count} on $S$. We now consider the $i_0$-th round when running {\sc Count} on $S'$. We argue that all the edges in $E'_{i_0}$ that are incident to $b$ will contribute to the computation of $x'_{i_0}$. The reason is that node $b\in S_{c}'$. Therefore, all edges incident to $b$ belong to $\bar{E'}$. As a result, all the edges in $E'_{i_0}$ that are incident to $b$ will contribute to the computation of $x'_{i_0}$ through the first term in \eqref{eq:inter9}. We thus have $x'_{i_0}= d-|\{v_1,v_2,\cdots,v_{i_0-1}\}\cap (S_0\cup S_1\cup \cdots \cup S_{c-1})|$.

Since $x_{i_0}=x'_{i_0}$, we thus have $e(S)=e(S')$ after the first $i_0$ counting rounds. 

\par We now consider rounds $(i_0+1)$ to $k$. We observe that by our construction $v'_i=v_i\in S'_{-c}\subset S_{-c}$ for $i_0+1 \leq i\leq k$. Moreover, since $v_{i_0}=a\in S_{-c}$ and $v'_{i_0}=b\in S_c'$, both vertices $a$ and $b$ are initially not connected to any vertices in $S_{-c}$ and $S_{-c}'$ respectively (those $v_i$ and $v'_i$ with $i_0+1\leq i\leq k$) since vertices of the same family are not connected. Therefore, replacing the $i_0$-th node $v_{i_0}=a$ by $v_{i_0}'=b$ will not change the value of the first term in \eqref{eq:inter9} when computing $x_i$ for the $i$-th round where $i_0+1\leq i\leq k$. 

\par We now consider the second term of \eqref{eq:inter9}. For any $u\in S_0$, any edge incident to $u$ has been counted in the first $(i_0-1)$ rounds since we assume that when we are running {\sc Count} on the $S$ set, we examine the nodes in $S_{-c}$ in the very last. Therefore, there is no edge of the form $(v_i,u)$ in $E_i$ (resp. $(v_i',u) \in E_i'$) with $u\in S_0$ since those edges have been removed previously. Therefore, the summation over $u\in N_0$ can be replaced by $u\in N_0\backslash S_0$ during the $i_0$-th round to the $k$-th round. On the other hand, for any $u\in N_0\backslash S_0$, if there is an edge connecting $(a,u)\in \tilde{E}$, then by our construction there is an edge $(b,u)\in \bar{E}$. Therefore, in the $i_0$-th round, the same number of edges incident to $u$ is removed regardless whether we are using $S$ as the input to the subroutine {\sc Count} or we are using $S'$ as the input to the subroutine {\sc Count}. As a result, in the beginning of the $(i_0+1)$-th round, for any $u\in N_0$, we have the following equality
\begin{align}
|\{(u,j)\in E_i:j\in N\}|=|\{(u,j)\in E_i':j\in N\}|\label{new:counting-eq}
\end{align}
when $i=i_0+1$. Moreover, for any $u\in N_0\backslash S_0$, we remove one and only one edge $(u,v_i)$ in the $i$-th round. Since $v_i=v_i'$ for all $i=i_0+1$ to $k$, we have \eqref{new:counting-eq} for all $i=i_0+1$ to $k$ as well. The above arguments thus prove that the second term of \eqref{eq:inter9} does not change regardless whether we count over $S$ or $S'$. As a result, $x'_i=x_i$ for $i_0+1\leq i\leq k$. Since $e(S)=e(S')$ for all $k$ rounds of the counting process, we have thus proved \eqref{eq:claim1}.

\emph{Proof of Claim~\ref{clm:gfr}:}
\par For any node set $S$, by iteratively using Claim~\ref{clm:edges_count}, we can construct another node set $S'$ such that $e(S)=e(S')$ while either (Case i) $S'_{-c}=\emptyset$; or (Case ii) $S'_{-c}\neq \emptyset$ and $S'_c=N_c$. As a result, we can assume without loss of generality that we have either (Case i) $S_{-c}=\emptyset$; or (Case ii) $S_{-c}\neq \emptyset$ and $S_c=N_c$ to begin with. 

\par We first consider the former case. Let $\tilde{\mathbf{r}}$ be any vector in $R$ such that its $\tilde{r}_i=v_i$ for $1\leq i\leq k$, i.e., $\tilde{r}_i$ equals the node index of the vertex $v_i$. We will run the subroutine {\sc Count} sequentially for $i=1$ to $k$ and compare the increment of $e(S)$ in each round, denoted by $x_i$ in \eqref{eq:inter9}, to the $i$-th term $(d-z_i(\tilde{\mathbf{r}}))$ in the summation of the RHS of \eqref{eq:claim2}. Consider the $i$-th round of counting for some $1\leq i\leq k$, and assume that the corresponding vertex $v_{i}$ belongs to the $y$-th family, i.e., $v_{i} \in N_y$. Since $S_{-c}=\emptyset$ in this case, we have $v_{i}\notin S_{-c}$ and the second term in \eqref{eq:inter9} is always 0. Therefore, the procedure {\sc Count} is indeed counting the number of edges in $\bar{E}$ that are incident to $S$ without the special conditional counting mechanism in the second term of \eqref{eq:inter9}. Therefore, we have
\begin{align} x_{i}&=|\{(v_{i},j)\in E_{i}\cap\bar{E}:j\in N\}|\nonumber\\
&=d-|\{v_j\notin N_{y}:v_j\in S, 1\leq j\leq i-1\}|, \label{eq:xi_zi}
\end{align}
where $d$ is the number of $\bar{E}$ edges in the original graph $G$ that are incident to $v_{i}$ and $|\{v_j\notin N_{y}:v_j\in S, 1\leq j\leq i-1\}|$ is the number of edges removed during the first $(i-1)$ counting rounds. On the other hand, by the definition of function $z_i(\cdot)$, our construction of $\tilde{\mathbf{r}}$, and the assumption that $S_{-c}=\emptyset$, we always have $|\{v_j\notin N_y:v_j\in S, 1\leq j\leq i-1\}|=z_{i}(\tilde{\mathbf{r}})$. As a result, $x_{i}=(d-z_{i}(\tilde{\mathbf{r}}))$ for $i=1$ to $k$ and our explicitly constructed vector $\tilde{\mathbf{r}}$ satisfies \eqref{eq:claim2}.

\par We now turn our attention to the second case when $S_{-c}\neq \emptyset$ and $S_c=N_c$. Let $\mathbf{r}$ be any vector in $R$ such that its $r_i=v_i$ for $1\leq i\leq k$. Recall that there are $k$ nodes in the set $S$. Define $j^*$ as the value that simultaneously satisfies (i) $k-|S_{-c }|\leq j^* \leq k$ and (ii) there are exactly $d$ entries in the first $j^*$ coordinates of $\mathbf{r}$ that are in $N\backslash N_0$. If no value satisfies the above two conditions simultaneously, set $j^*=k+1$. We now construct another vector $\tilde{\mathbf{r}}$ from $\mathbf{r}$ as follows: Replace the values of the $(j^*+1)$-th coordinate to the $k$-th coordinate of $\mathbf{r}$ by $n$, the node index of the last node in $N_0$ and denote the final vector by $\tilde{\mathbf{r}}$.

\par We will now prove that the above explicit construction of $\tilde{\mathbf{r}}$ satisfies the desired property in \eqref{eq:claim2}. The proof is divided into two cases:

\underline{Case~1:} There exists such a $j^*$ satisfying (i) and (ii). We will run the subroutine {\sc Count} again and compare $x_i$ to the $i$-th term $(d-z_i(\tilde{\mathbf{r}}))$.

We then observe the following facts: 

\begin{enumerate}
\item In {\sc Count}, from $i=1$ to $(k-|S_{-c }|)$. For any $i$ in this range, we must have $FI(v_i)\neq -c$, i.e., the family index of node $v_i$ is not $-c$, since we run the subroutine {\sc Count} using a specific ordering of the nodes in $S$, which examines the nodes in $S_{-c}$ in the very last. As a result, the second term of \eqref{eq:inter9} is always zero. Therefore \eqref{eq:xi_zi} still holds. By the definition of function $z_i(\cdot)$, our construction of $\tilde{\mathbf{r}}$, and the fact that $1\leq i\leq k-|S_{-c}|$ (implying no $v_j\in S_{-c}$ for all $1\leq j\leq i-1$), we get $x_i=d-z_i(\tilde{\mathbf{r}})$ for all $1\leq i\leq k-|S_{-c}|$. 

\item We now consider the case of $i=k-|S_{-c}|+1$ to $j^*$ of Step~3. For any $i$ in this range, we have $v_i\in S_{-c}$. We now argue that $|\{(u,j)\in E_i:j\in N\}|>|N_{-c}|$ for all edges $(u,v_i)\in E_i\cap \tilde{E}$ satisfying $u\in N_0$. The reason is that $(u,v_i)\in E_i$ implies that node $u$ is not counted in the previous $(i-1)$ rounds, i.e., $u\neq v_{i'}$ for all $1\leq i'\leq i-1$. Therefore, an edge of $(u,v)$ is removed if and only if there is a $v=v_j$ for some $v_j$ that is not in $N_0$. Since there are exactly $d$ vertices in $\{v_1,v_2,\dots,v_{j^*}\}$ that are not in $N_0$, it means that the first $(i-1)$ counting rounds where $1\leq i\leq j^*$ can remove at most $(d-1)$ edges incident to such a node $u$. Since node $u$ has $(d+|N_{-c}|)$ number of incident edges in the original graph $G$, we know that the inequality $|\{(u,j)\in E_i:j\in N \}|>|N_{-c}|$ must hold in the $i$-th round. As a result, the second term of \eqref{eq:inter9} is non-zero when $i=k-|S_{-c}|+1$ to $j^*$ and we can thus rewrite 
\begin{align}
x_{i}&=|\{(v_{i},j)\in E_i:j\in N\}|\nonumber\\
&=d-|\{v_j\notin N_{c}\cup N_{-c}:v_j\in S, 1\leq j\leq i-1\}|.\nonumber
\end{align}
By the definition of function $z_i(\cdot)$ and our construction of $\tilde{\mathbf{r}}$, we get $x_i=d-z_i(\tilde{\mathbf{r}})$ for all $k-|S_{-c}|+1\leq i\leq j^*$.  
\item We now consider the $(j^*+1)$-th to the $k$-th round of Step~3. We claim that
\begin{align}
x_i=d-|S_1\cup S_2\cup \cdots \cup S_c|.
\end{align}
The reason behind this is the following. Since $j^*+1\leq i \leq k$, we have $v_i\in S_{-c}$. For any $u\in N_0\backslash S_0$ (those $u\in S_0$ have been considered in the first $(k-|S_{-c}|)$ rounds), there are $(d+|N_{-c}|)$ number edges incident to $u$ in the original graph $G$. On the other hand, since $i\geq j^*+1$ and by our construction, there are $d$ entries in the first $j^*$ coordinates of $\tilde{\mathbf{r}}$ that are are not in $N_0$, we must have removed at least $d$ edges incident to $u$ during the first $(i-1)$ counting rounds as discussed in the previous paragraph. Therefore, the number of incident edges in $E_i$ that are incident to $u\in N_0\backslash S_0$ must be $\leq |N_{-c}|$. The second term of \eqref{eq:inter9} is thus zero. As a result, the $x_i$ computed for $v_i$ will only include those edges in $E_i\cap \bar{E}$ incident to it. Since any $v_i\in S_{-c}$ only has $(d-|N_0|)$ number of edges in $\bar{E}$ to begin with, we have that
\begin{align}
x_i=(d-|N_0|)-|S_1\cup S_2 \cup \cdots \cup S_{c-1}|\nonumber 
\end{align}
where $|S_1\cup S_2 \cup \cdots \cup S_{c-1}|$ is the number of edges in $\bar{E}$ that have been removed during the first $(i-1)$ rounds. Since $S_c=N_c$ in the scenario we are considering and since $|N_c|=|N_0|=n\bmod (n-d)$ in the family repair scheme, we can consequently rewrite $x_i$ as
\begin{align}
x_i=d-|S_1\cup S_2\cup \cdots \cup S_c|\nonumber
\end{align}
for $(j^*+1)\leq i\leq k$. Recall that in the newly constructed $\tilde{\mathbf{r}}$, the values of the $(j^*+1)$-th coordinate to the $k$-th coordinate are $n$, which belongs to $N_0$. Thus, by the definition of function $z_i(\cdot)$, we can see that each of these coordinates only contributes 
\begin{align}
z_i(\tilde{\mathbf{r}})&=|\{\tilde{r}_j\in N\backslash (N_{-c}\cup N_0):1\leq j\leq i-1\}|\nonumber\\
&=|\{\tilde{r}_j\in N\backslash (N_{-c}\cup N_0):1\leq j\leq j^*\}|\label{eq:count-new-CCW1}\\
&=|S_1\cup S_2\cup \cdots \cup S_c|\nonumber
\end{align}
where \eqref{eq:count-new-CCW1} follows from the fact that in the construction of $\tilde{\mathbf{r}}$, the $(j^*+1)$-th to the $k$-th coordinates of $\tilde{\mathbf{r}}$ are always of value $n\in N_0$. Hence, we get $x_i=d-z_i(\tilde{\mathbf{r}})$ for $(j^*+1)\leq i\leq k$.
\end{enumerate}

We have proved for this case that $x_i=d-z_i(\tilde{\mathbf{r}})$ for $i=1$ to $k$. Therefore, we get \eqref{eq:claim2}. 

\underline{Case~2:} No such $j^*$ exists. This means that one of the following two sub-cases is true. Case~2.1: even when choosing the largest $j^*=k$, we have strictly less than $d$ entries that are not in $N_0$. Case~2.2:  Even when choosing the smallest $j^*=k-|S_{-c}|$, we have strictly more than $d$ entries that are not in $N_0$. 

\par Case~2.1 means that we have $<d$ vertices in $S$ that are not in $N_0$, which implies that all vertices in $S$ together do not share more than $d$ edges with any of the vertices in $N_0\backslash S_0$. Therefore, in Step~3 of {\sc Count}, if $v_i\in S_{-c}$, then there will be  $>|N_{-c}|$ edges in $E_i$ that are incident to $u\in N_0\backslash S_0$ since $u$ has $(d+|N_{-c}|)$ number of edges in the original graph $G$ and $<d$ edges are removed in the first $(i-1)$ rounds. As a result, the second term of \eqref{eq:inter9} will be 1 and we count all the edges in $E_i$ incident to $v_i$. By similar arguments as used in a previous proof (when proving the scenario of $k-|S_{-c}|+1\leq i\leq j^*$), we have $x_i=d-z_i(\tilde{\mathbf{r}})$ for all $1\leq i\leq k$ and the proof of this case is complete.
\par Case~2.2 is actually an impossible case. The reason is that for any $1\leq i\leq k-|S_{-c}|$, there are exactly $|S_1|+|S_2|+\cdots + |S_c|$ nodes $v_i$ that are not in $N_0$. And we also have 
\begin{align}
\sum_{m=1}^c |S_m|\leq\sum_{m=1}^c |N_m|=d\nonumber.
\end{align}
This, together with the observation that the first $(k-|S_{-c}|)$ coordinates of $\mathbf{r}$ are transcribed from the distinct nodes in $S_1\cup S_2\cup\cdots\cup S_c$, implies that we cannot have strictly more than $d$ entries that are not in $N_0$ in the first $(k-|S_{-c}|)$ coordinates of $\mathbf{r}$. Case 2.2 is thus an impossible case. 
\par From the above arguments, the proof of Claim~\ref{clm:gfr} is complete. 
\end{IEEEproof}

\section{Conclusion}\label{sec:conc}
In practice, it is natural that the newcomer should access only those ``good'' helpers. This paper has provided a necessary and sufficient condition under which optimally choosing good helpers improves the storage-bandwidth tradeoff. We have also analyzed a new class of low-complexity solutions termed the \emph{family repair scheme}, including its storage-bandwidth tradeoff, the expression of its MBR point, and its (weak) optimality. Moreover, we have constructed an explicit exact-repair code, the \emph{generalized fractional repetition code}, that can achieve the MBR point of that scheme.

\par The main goal of this work is to characterize, for the first time in the literature, when and by how much dynamic helper selection improves RCs. We thus considered the scenario of single failures only in a similar way as in the original RC paper \cite{dimakis2010network}. Since a practical system can easily have multiple failures, as ongoing work, we are studying the helper selection problem under the multiple failures scenario. 


\appendices
\section{Another Example Illustrating The Benefits Of Helper Selection} \label{app:example}

\begin{figure}[h!]
\centering
\subfigure[Arbitrarily choosing the helper nodes is bad.]{
            \label{fig:example_a}
            \includegraphics[width=0.4\textwidth]{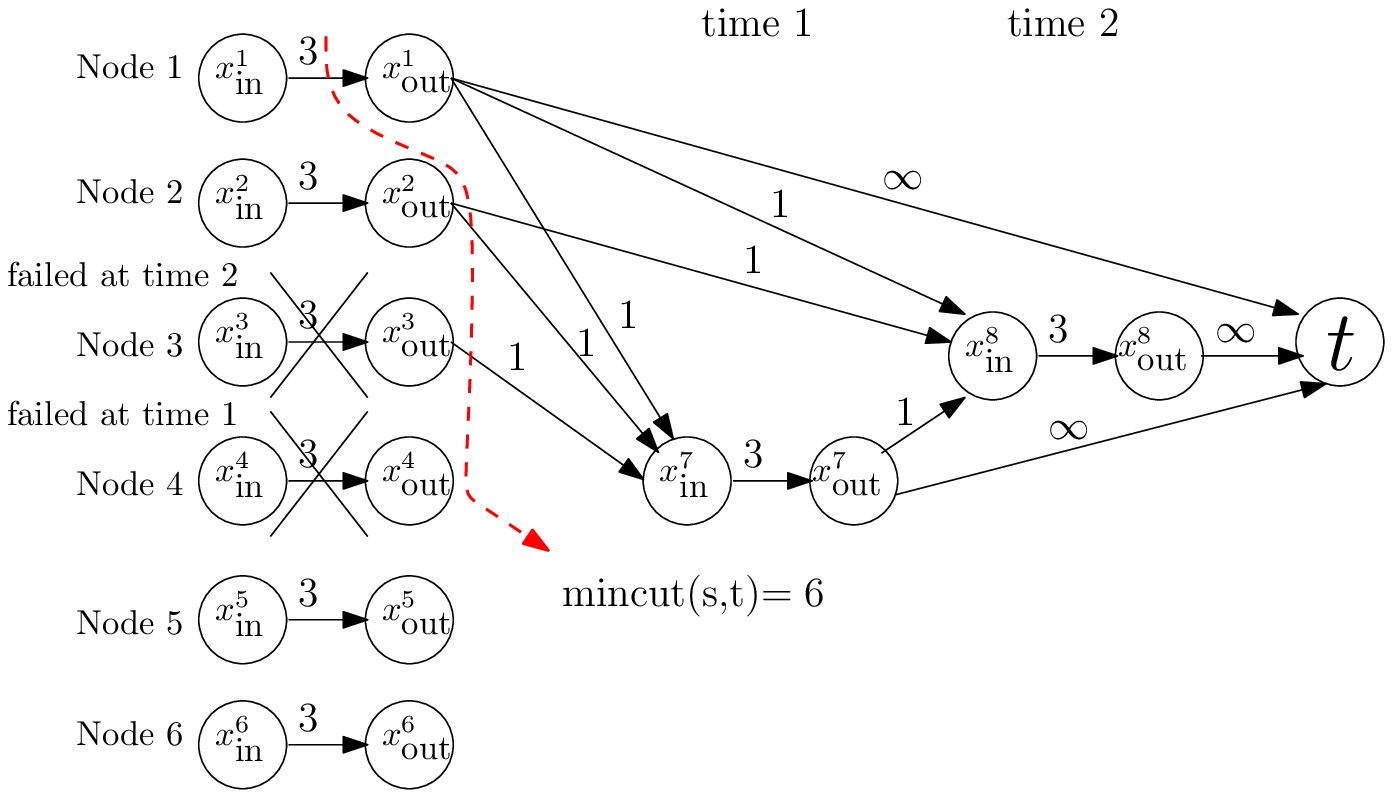} 
        }
\subfigure[Choosing the helper nodes properly is good.]{
            \label{fig:example_b}
            \includegraphics[width=0.4\textwidth]{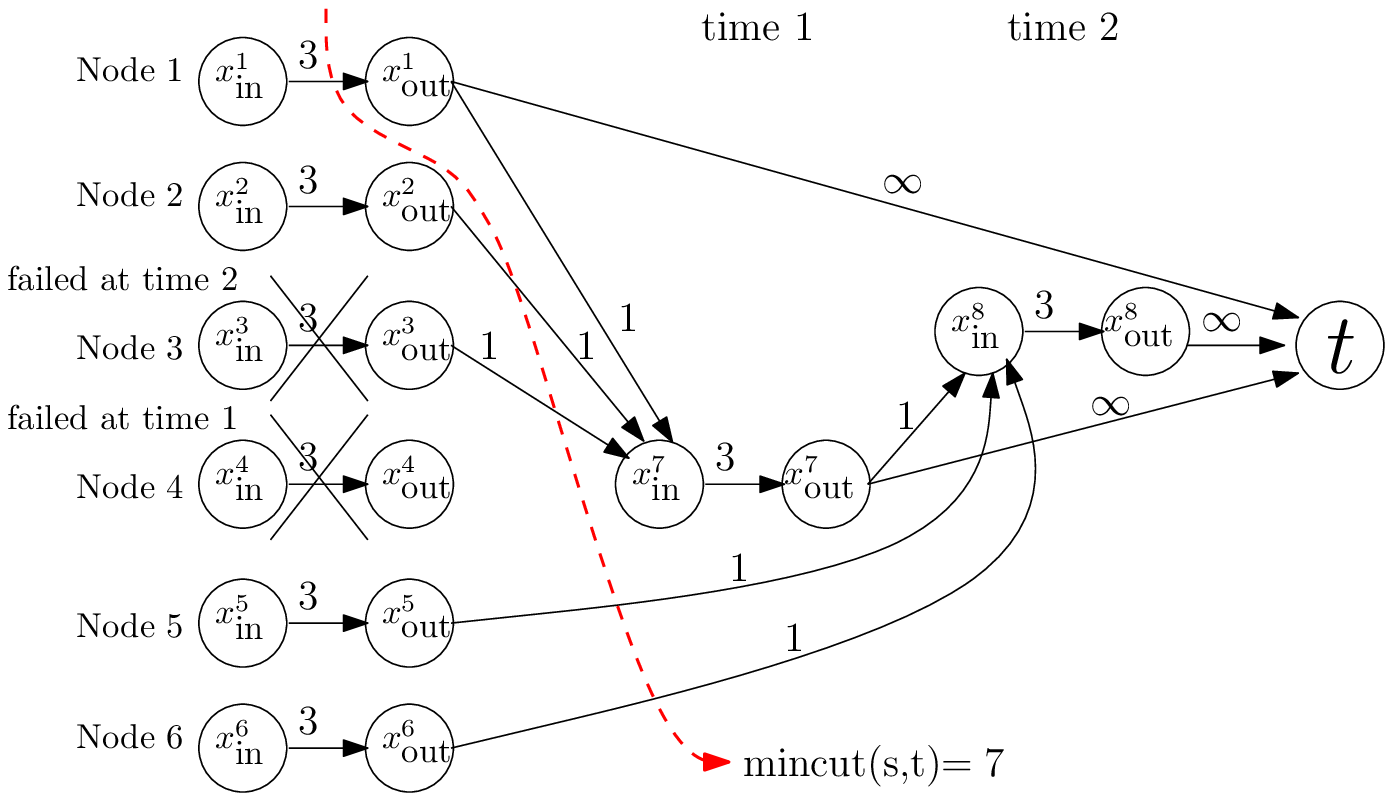}
        }

\caption{An example illustrating the importance of choosing the helper nodes for $(n,k,d,\alpha,\beta)=(6,3,3,3,1)$ and file size $\mathcal{M}=7$.}
\label{fig:example}
\end{figure}

Fig.~\ref{fig:example} shows another example that illustrates how choosing the helpers properly can allow for smaller storage and repair-bandwidth. The parameters of the storage network in this figure are $(n,k,d,\alpha,\beta)=(6,3,3,3,1)$. The goal of this example is to store a data object of size $\mathcal{M}=7$ such that the network can tolerate $n-k =3$ failures. Without loss of generality, we assume that node 4 fails in time 1 and the helpers of the newcomer (replacing node 4) are nodes 1, 2, and 3. Now assume that node 3 fails in time 2. We will demonstrate in the following how the helper choice at time 2 (for replacing node 3) will substantially affect the reliability of the distributed storage network.

\par Choice 1: Suppose the helpers of node 3 in time 2 are nodes 1, 2, and 4. See Fig.~\ref{fig:example_a}. Now we consider the data collector $t$ which would like to reconstruct the original file of size 7 from nodes 1, 3, and 4. By noticing that one of the edge cuts from the virtual source to the data collector has value 6 (see the red dashed curve in Fig.~\ref{fig:example_a}), it is thus impossible for the data collector to reconstruct the original file. In fact, we have from Section~\ref{sec:br_extreme_points} that, when the newcomer chooses its helpers blindly, to protect a file of size $\mathcal{M}=7$, the minimum repair-bandwidth needed is $\beta_{\MBR}=\frac{3.5}{3}$. Therefore, the repair-bandwidth $\beta=1$ (our parameter values are $(n,k,d,\alpha,\beta)=(6,3,3,3,1)$) is not enough to meet the reliability requirement when a BR scheme is used, which agrees with the discussion above.

\par Choice 2: Suppose the helpers of node 3 in time 2 are nodes 4, 5, and 6. See Fig.~\ref{fig:example_b}. Now we consider the same data collector $t$ that accesses nodes 1, 3, and 4. One can verify that the min-cut value from source $s$ to the data collector $t$ is 7, which is equal to the target file size 7. Furthermore, one can check the rest ${6\choose 3}-1=19$ different ways of setting up the data collectors and they all have $\mincut(s,t) \geq 7$. The above observation illustrates that helper selection choice (Choice 2) can strictly improve the min-cut value of the network.

\par The choice of the helpers in this example follows the family repair (FR) scheme described in Section~\ref{subsec:desc_fr}. In Section~\ref{subsec:mbr}, it is proved rigorously that not only we can improve the min-cut value in the end of the first 2 time slots, but the min-cut-value is always $\geq 7$ even after arbitrarily many failure/repair stages with intelligent helper selection for each time slot. We can thus meet the reliability requirement with intelligent helper selection. This example with parameters $(n,k,d,\alpha,\beta)=(6,3,3,3,1)$ is thus another evidence that good helper selection can strictly improve the system performance, i.e., reducing the total repair-bandwidth $\gamma$ from $3.5$ (the smallest possible when BR is used) to $3$ (since our system has $d=3$ and $\beta=1$).

\section{Proof of Proposition~\ref{prop:low_b_gen}} \label{app:low_b_gen}
\par The proof of Proposition~\ref{prop:low_b_gen} below follows the proof of \cite[Lemma 2]{dimakis2010network}.

\par Consider any IFG $G\in \mathcal{G}_A$ where $A$ is a stationary repair scheme. Consider any data collector $t$ of $G$ and call the set of $k$ active output nodes it connects to $V$. Since all the incoming edges of $t$ have infinite capacity, we can assume without loss of generality that the minimum cut $(U,\overline{U})$ satisfies $s\in U$ and $V\subseteq \overline{U}$.

Let $\mathcal{C}$ denote the set of edges in the minimum cut. Let $x^ i_{\out}$ be the chronologically $i$-th output node in $\overline{U}$, i.e., from the oldest to the youngest. Since $V\subseteq\overline{U}$, there are at least $k$ output nodes in $\overline{U}$. We now consider the oldest $k$ output nodes of $\overline{U}$, i.e., $x^1_{\out}$ to $x^k_{\out}$. For $i=1$ to $k$, let $r_i$ denote the node index of $x^i_{\out}$. Obviously, the vector $\mathbf{r}\stackrel{\Delta}{=}(r_1,\cdots, r_k)$ belongs to $R$.

\par Consider $x^1_{\out}$, we have two cases:
\begin{itemize}
\item If $x^1_{\inp}\in U$, then the edge $(x^1_{\inp},x^1_{\out})$ is in $\mathcal{C}$.
\item If $x^1_{\inp}\in \overline{U}$, since $x^1_{\inp}$ has an in-degree of $d$ and $x^1_\text{out}$ is the oldest node in $\overline{U}$, all the incoming edges of $x^1_{\inp}$ must be in $\mathcal{C}$.
\end{itemize}
From the above discussion, these edges related to $x^1_{\out}$ contribute at least a value of $\min((d-z_1(\mathbf{r}))\beta,\alpha)$ to the min-cut value since by definition $z_1(\mathbf{r})=0$.
Now, consider $x^2_{\out}$, we have three cases:
\begin{itemize}
\item If $x^2_{\inp}\in U$, then the edge $(x^2_{\inp},x^2_{\out})$ is in $\mathcal{C}$.
\item If $x^2_{\inp}\in \overline{U}$ and  $r_1\in D_{r_2}$, since one of the incoming edges of $x^2_{\inp}$ can be from $x^1_{\out}$, then at least $(d-1)$ incoming edges of $x^2_{\inp}$ are in $\mathcal{C}$.
\item If $x^2_{\inp}\in \overline{U}$ and  $r_1\notin D_{r_2}$, since no incoming edges of $x^2_{\inp}$ are from $x^1_{\out}$, then all $d$ incoming edges of  $x^2_{\inp}$ are in $\mathcal{C}$.
\end{itemize}
Therefore, these edges related to $x^2_{\out}$ contribute a value of at least $\min ((d-z_2(\mathbf{r}))\beta,\alpha)$ to the min-cut value, where the definition of $z_2(\mathbf{r})$ takes care of the second and the third cases. Consider $x^3_{\out}$, we have five cases:
\begin{itemize}
\item If $x^3_{\inp}\in U$, then the edge $(x^3_{\inp},x^3_{\out})$ is in $\mathcal{C}$.
\item If $x^3_{\inp}\in \overline{U}$ and  $r_1=r_2\in D_{r_3}$, since one of the incoming edges of $x^3_{\inp}$ can be from $x^2_{\out}$, then at least $(d-1)$ incoming edges of $x^3_{\inp}$ are in $\mathcal{C}$. Note that there cannot be an incoming edge of $x^3_{\inp}$ from $x^1_{\out}$ since $x^3_{\inp}$ only connects to active output nodes at the time of repair and $x^1_\text{out}$ is no longer active since $x^2_{\out}$ (of the same node index $r_2=r_1$) has been repaired after $x^1_{\out}$.
\item If $x^3_{\inp}\in \overline{U}$;  $r_1,r_2\in D_{r_3}$; and $r_1\neq r_2$; since one of the incoming edges of $x^3_{\inp}$ can be from $x ^ 1_{\out}$ and another edge can be from $x^2_{\out}$ , then at least $(d-2)$ incoming edges of $x^3_{\inp}$ are in $\mathcal{C}$.
\item If $x^3_{\inp}\in \overline{U}$ and only one of $r_1$ or $r_2$ is in $D_{r_3}$, since one of the incoming edges of $x^3_{\inp}$ is from either $x^1_{\out}$ or $x^2_{\out}$, then at least $(d-1)$ incoming edges of $x^3_{\inp}$ are in $\mathcal{C}$.
\item If $x^3_{\inp}\in \overline{U}$ and $r_1,r_2 \notin D_{r_3}$, then at least $d$ incoming edges of $x^3_{\inp}$ are in $\mathcal{C}$.
\end{itemize}
Therefore, these edges related to $x^3_{\out}$ contribute a value of at least $\min ((d-z_3(\mathbf{r}))\beta,\alpha)$ to the min-cut value, where the definition of $z_3(\mathbf{r})$ takes care of the second to the fifth cases.

In the same manner, we can prove that the chronologically $i$-th output node in $\overline{U}$ contributes at least a value of $\min ((d-z_i(\mathbf{r}))\beta,\alpha)$ to the min-cut value. If we sum all the contributions of the oldest $k$ output nodes of $\overline{U}$ we get \eqref{eq:low_b_gen}, a lower bound on the min-cut value.

\section{Proof of Inequality \eqref{eq:new2}} \label{app:eq:new2}

\par Denote the smallest IFG in $\gf(n,k,d,\alpha,\beta)$ by $G_0$. Specifically, all its nodes are intact, i.e., none of its nodes has failed before. Denote its active nodes arbitrarily by $1,2,\cdots,n$. Consider the family index permutation of the FR scheme $F$ that attains the minimization of the right-hand side of \eqref{eq:new2} and call it $\tilde{\pi}_f$. Fail each active node in $\{1,2,\cdots,n\}$ of $G_0$ exactly once in a way that the sequence of the family indices of the failed nodes is $\tilde{\pi}_f$. Along this failing process, we repair the failed nodes according to the FR scheme $F$. For example, let $(n,d)=(8,5)$ and suppose the minimizing family index permutation is $\tilde{\pi}_f=(1,2,1,-2,0,0,1,2)$. Then, if we fail nodes 1, 4, 2, 6, 7, 8, 3, and 5 in this sequence, the corresponding family index sequence will be $(1,2,1,-2,0,0,1,2)$, which matches the given $\tilde{\pi}_f$. Note that the node failing sequence is not unique in our construction. For example, if we fail nodes 3, 5, 2, 6, 8, 7, 1, and 4 in this sequence, the corresponding family index vector is still $(1,2,1,-2,0,0,1,2)$. Any node failing sequence that matches the given $\tilde{\pi}_f$ will suffice in our construction. We call the resulting new IFG, $G'$.

\par Consider a data collector $t$ in $G'$ that connects to the oldest $k$ newcomers. (Recall that in our construction, $G'$ has exactly $n$ newcomers.)  Now, by the same arguments as in \cite[Lemma 2]{dimakis2010network}, we will prove that $\mincut_{G'}(s,t)=\sum_{i=1}^{k}\min \left(\left(d-y_i(\tilde{\pi}_f)\right)\beta,\alpha\right)$ for the specifically constructed $G'$ and $t$. Number the storage nodes (input-output pair) of the $k$ nodes $t$ is connected to by $1,2,\dots,k$. Define cut $(U,\overline{U})$ between $t$ and $s$ as the following: for each $i\in \{1,\dots,k\}$, if $\alpha \leq (d-y_i(\tilde{\pi}_f))\beta$ then we include $x_{\out}^{i}$ in $\overline{U}$; otherwise, we include both $x_{\out}^{i}$ and $x_{\inp}^{i}$ in $\overline{U}$. It is not hard to see that the cut-value of the cut $(U,\overline{U})$ is equal to $\sum_{i=1}^{k}\min \left(\left(d-y_i(\tilde{\pi}_f)\right)\beta,\alpha\right)$.

Since the left-hand side of \eqref{eq:new2} further takes the minimum over $\mathcal{G}_F$ and all data collectors $t$, we have proved the inequality \eqref{eq:new2}.

\section{Proof of Inequality \eqref{new5}} \label{app:procedure}
\par We prove \eqref{new5} by explicit construction. For any vector $\mathbf{r}\in R$, we will use the following procedure, {\sc Modify}, to gradually modify $\mathbf{r}$ in 4 major steps until the end result is the desired $\mathbf{r'}\in R_2$ that satisfies \eqref{new5}. A detailed example illustrating procedure {\sc Modify} is provided in Appendix~\ref{app:modify_ex} to complement the following algorithmic description of {\sc Modify}.

\par \emph{Step~1:} If there are $i, j\in\{1,\cdots,k\}$ such that $i<j$ and   the $i$-th and the $j$-th coordinates of $\mathbf{r}$ are equal, i.e., $r_i=r_j$, then we can do the following modification. For convenience, we denote the value of $r_i=r_j$ by $h$. Suppose that node $h$ belongs to the $Q$-th family. We now check whether there is any value $\gamma$ satisfying simultaneously (i) $\gamma\in\{1,2,\cdots, n\}\backslash h$; (ii) node $\gamma$ is also in the $Q$-th family; and (iii) $\gamma$ is not equal to any of the coordinates of $\mathbf{r}$. If such $\gamma$ exists, we replace the $j$-th coordinate of $\mathbf{r}$ by $\gamma$. Specifically, after this modification, we will have $r_i=h$ and $r_j=\gamma$.

\par Repeat this step until either there is no repeated $r_i=r_j$, or until no such $\gamma$ can be found.

\par \emph{Step~2:} After finishing Step~1, we perform the following modification. If there still are distinct $i,j\in\{1,\cdots,k\}$ such that $r_i=r_j$ and $i<j$, then we again denote the value of $r_i=r_j$ by $h$. Suppose node $h$ belongs to the $Q$-th family. Consider the following two cases. If the $Q$-th family is the incomplete family, then no further modification will be made.

\par If the $Q$-th family is a complete family, then do the following modification.

Find the largest $j_1\in\{1,\cdots, n\}$ such that node $r_{j_1}=h$ and find the largest $j_2\in \{1,\cdots, n\}$ such that $r_{j_2}$ belongs to the $Q$-th family (the same family of node $h$). If $j_1=j_2$, then we set $\mathbf{r'}=\mathbf{r}$. If $j_1\neq j_2$, then we swap the values of $r_{j_1}$ and $r_{j_2}$ to construct $\mathbf{r'}$. That is, we first set $\mathbf{r'}=\mathbf{r}$ for all coordinates except for the $j_1$-th and the $j_2$-th coordinates, and then set $r'_{j_1}=r_{j_2}$ and $r'_  {j_2}=r_{j_1}$. After we have constructed new $\mathbf{r'}$ depending on whether $j_1=j_2$ or not, we now check whether there is any value $\gamma\in\{1,\cdots, n\}$ satisfying simultaneously (i) node $\gamma$ belongs to a complete family (not necessarily the Q-th family); and (ii) $\gamma$ is not equal to any of the coordinates of $\mathbf{r'}$. If such $\gamma$ exists, we replace the $j_2$-th coordinate of $\mathbf{r'}$ by $\gamma$, i.e., set $r'_{j_2}=\gamma$.

\par Repeat this step until the above process does not change the value of any of the coordinates of $\mathbf{r'}$.

After finishing the above two steps, the current vector $\mathbf{r}$ must be in one of the following cases. Case~1: No two coordinates are equal, i.e., $r_i\neq r_j$ for all pairs $i<j$; Case~2: there exist a pair $i<j$ such that $r_i=r_j$. We have two sub-cases for Case~2. Case~2.1: All such $(i,j)$ pairs must satisfy that node $r_i$ belongs to a complete family. Case~2.2: All such $(i,j)$ pairs must satisfy that node $r_i$ belongs to the incomplete family. Specifically, the above construction (Steps~1 and 2) has eliminated the sub-case that some $(i,j)$ pair has $r_i=r_j$ belonging to a complete family and some other $(i,j)$ pair has $r_i=r_j$ belonging to the incomplete family. The reason is as follows. Suppose some $(i,j)$ pair has $r_i$ belonging to a complete family. Since we have finished Step~2, it means that any node $\gamma$ that belongs to a complete family must appear in one of the coordinates of $\mathbf{r}$. Since there are $(n-d)\left\lfloor\frac{n}{n-d}\right\rfloor$ number of nodes belonging to complete families, at least $(n-d)\left\lfloor\frac{n}{n-d}\right\rfloor+1$ number of coordinates of $\mathbf{r}$ must refer to a node in a complete family (since $r_i$ and $r_j$ have the same value). Therefore, there are at most $n-\left((n-d)\left\lfloor\frac{n}{n-d}\right\rfloor+1\right) =(n\bmod (n-d))-1$ number of coordinates of $\mathbf{r}$ referring to a node in the incomplete family. However, if we have another $(i',j')$  pair has $r_{i'}=r_{j'}$ belonging to the incomplete family, then it means that the coordinates of $\mathbf{r}$ can refer to at most $(n\bmod(n-d))-2$ distinct nodes of the incomplete family (since $r_{i'}$ and $r_{j'}$  are equal). Since there are $n\bmod(n-d)$ distinct nodes in the incomplete family, there must exist a $\gamma$ value such that node $\gamma$ belongs to the incomplete family and $\gamma$ does not appear in any one of the coordinates of $\mathbf{r}$. This contradicts the fact that we have exhausted Step~1 before moving on to Step~2.

\par We now consider Cases~1, 2.1, and 2.2, separately. If the $\mathbf{r}$ vector is in Case~1, then such $\mathbf{r}$ belongs to $R_2$ and our construction is complete. If $\mathbf{r}$ belongs to Case~2.2, then do Step~3. If $\mathbf{r}$ belongs to Case~2.1, do Step~4.

\emph{Step~3:}  We use $(i,j)$ to denote the pair of values such that $r_i=r_j$ and $i<j$. Denote the value of $r_i=r_j$ by $h$. Since we are in Case~2.2, node $h$ belongs to the incomplete family. Find the largest $j_1\in\{1,\cdots, n\}$ such that node $r_{j_1}=h$ and find the largest $j_2\in \{1,\cdots, n\}$ such that $r_{j_2}$ belongs to the incomplete family. If $j_1=j_2$, then we keep $\mathbf{r}$ as is. If $j_1\neq j_2$, then we swap the values of $r_{j_1}$ and $r_{j_2}$. Recall that we use $c\stackrel{\Delta}{=}\left\lfloor \frac{n}{n-d}\right\rfloor$ to denote the family index of the last complete family. We now choose arbitrarily a $\gamma$ value from $\{(n-d)\left(c-1\right)+ 1,\dots,(n-d)c\}$. Namely, $\gamma$ is the index of a node of the last complete family. Fix the $\gamma$ value. We then replace $r_{j_2}$ by the arbitrarily chosen $\gamma$.

\par If the value of one of the coordinates of $\mathbf{r}$ (before setting $r_{j_2}=\gamma$) is $\gamma$, then after setting $r_{j_2}=\gamma$ we will have some $i\neq j_2$ satisfying $r_i=r_{j_2}=\gamma$. In this case, we start over from Step~1. If none of the coordinates of $\mathbf{r}$ (before setting $r_{j_2}=\gamma$) has value $\gamma$, then one can easily see that after setting $r_{j_2}=\gamma$ there exists no $i<j$ satisfying ``$r_i=r_j$ belong to a complete family'' since we are in Case~2.2 to begin with. In this case, we are thus either in Case~1 or Case~2.2. If the new $\mathbf{r}$ is now in Case~1, then we stop the modification process. If the new $\mathbf{r}$ is still in Case~2.2, we will then repeat this step (Step~3).

\emph{Step~4:} We use $(i,j)$ to denote the pair of values such that $r_i=r_j$ and $i<j$. Denote the value of $r_i=r_j$ by $h$. Since we are in Case~2.1, node $h$ belongs to a complete family. Suppose $h$ is in the $Q$-th complete family. Find the largest $j_1\in\{1,\cdots, n\}$ such that node $r_{j_1}=h$ and find the largest $j_2\in \{1,\cdots, n\}$ such that $r_{j_2}$ belongs to the $Q$-th complete family. If $j_1=j_2$, then we keep $\mathbf{r}$ as is. If $j_1\neq j_2$, then we swap the values of $r_{j_1}$ and $r_{j_2}$. We now find a $\gamma$ value such that (i) node $\gamma$ belongs to the incomplete family; and (ii) $\gamma$ is not equal to any of the coordinates of $\mathbf{r}$. Note that such $\gamma$ value always exists. The reason is that since we are now in Case~2.1 and we have finished Step~2, it means that any node $\gamma$ that belongs to a complete family must appear in one of the coordinates of $\mathbf{r}$. Therefore, there are at least $(n-d)\left\lfloor \frac{n}{n-d}\right\rfloor+1$ number of coordinates of $\mathbf{r}$ referring to a node in one of the complete families. This in turn implies that there are at most $n-\left((n-d)\left\lfloor\frac{n}{n-d}\right\rfloor+1\right)=(n\bmod (n-d))-1$ number of coordinates of $\mathbf{r}$ referring to a node in the incomplete family. Since there are $n\bmod (n-d)$ distinct nodes in the incomplete family, there must exist a $\gamma$ value such that node $\gamma$ belongs to the incomplete family and $\gamma$ does not appear in any one of the coordinates of $\mathbf{r}$.

\par Once the $\gamma$ value is found, we replace the $j_2$-th coordinate of $\mathbf{r}$ by $\gamma$, i.e., $r_{j_2}=\gamma$. If the new $\mathbf{r}$ is now in Case~1, then we stop the modification process. Otherwise, $\mathbf{r}$ must still be in Case~2.1 since we replace $r_{j_2}$ by a $\gamma$ that does not appear in $\mathbf{r}$ before. In this scenario, we will then repeat this step (Step~4).

\par An example demonstrating the above iterative process is provided in Appendix~\ref{app:modify_ex}.

\par To prove that this construction is legitimate, we need to prove that the iterative process ends in a finite number of time. To that end, for any vector $\mathbf{r}$, define a non-negative function $T(\mathbf{r})$ by
\begin{align}
T(\mathbf{r})&=|\{(i,j):i<j,r_i=r_j\text{ is a complete family node}\}|+\nonumber\\
&2|\{(i,j):i<j,r_i=r_j\text{ is an incomplete family node}\}|.\nonumber
\end{align}

\par One can then notice that in this iterative construction, every time we create a new $\mathbf{r}'$ vector that is different from the input vector $\mathbf{r}$, the value of $T(\mathbf{r})$ decreases by at least 1. As a result, we cannot repeat this iterative process indefinitely. When the process stops, the final vector $\mathbf{r}'$ must be in Case~1. Therefore, the procedure {\sc Modify} converts any vector $\mathbf{r}\in R$ to a new vector $\mathbf{r'}\in R_2$ such that all coordinate values of $\mathbf{r'}$ are distinct. What remains to be proved is that along the above 4-step procedure, the inequality \eqref{new5} always holds. That is, the value of $\sum_{i=1}^{k}\min ((d-z_i(\mathbf{r}))\beta,\alpha)$ is non-increasing along the process. The detailed proof of the non-increasing $\sum_{i=1}^{k}\min ((d-z_i(\mathbf{r}))\beta,\alpha)$ will be provided shortly. From the above discussion, we have proved \eqref{new5}.

In the rest of this appendix, we prove the correctness of {\sc Modify}. For each step of {\sc Modify}, we use $\mathbf{r}$ to denote the input (original) vector and $\mathbf{w}$ to denote the output (modified) vector. In what follows, we will prove that the $\mathbf{r}$ and $\mathbf{w}$ vectors always satisfy
\begin{align} \label{eq:modified}
\sum_{i=1}^{k}\min ((d-z_i(\mathbf{w}))\beta,\alpha) \leq \sum_{i=1}^{k}\min ((d-z_i(\mathbf{r}))\beta,\alpha).
\end{align}
\par In Step~1 of the procedure, suppose that we found such $\gamma$. Denote the vector after we replaced the $j$-th coordinate with $\gamma$ by $\mathbf{w}$. We observe that for $1\leq m\leq j$, we will have $z_m(\mathbf{r})=z_m(\mathbf{w})$ since $r_m=w_m$ over $1\leq m\leq j-1$ and the new $w_j=\gamma$ belongs to the $Q$-th family, the same family as node $r_j$. For $j+1\leq m\leq k$, we will have $z_m(\mathbf{w})\geq z_m(\mathbf{r})$. The reason is that by our construction, we have $w_j=\gamma \neq r_j= r_i=w_i$. For any $m>j$, $z_m(\mathbf{r})$ only counts the repeated $r_i=r_j$ once. Therefore, $z_m(\mathbf{w})$ will count the same $w_i$ as well. On the other hand, $z_m(\mathbf{w})$ may sometimes be larger than $z_m(\mathbf{r})$, depending on whether the new $w_j \in D_{w_m}$ or not. The fact that $z_m(\mathbf{w})\geq z_m(\mathbf{r})$ for all $m=1$ to $k$ implies  \eqref{eq:modified}.

\par In Step~2, if $j_1=j_2$, then we will not swap the values of $r_{j_1}$ and $r_{j_2}$. On the other hand, $j_1=j_2$ also means that $r_{j_1}=r_{j_2}=h$. In this case, $\mathbf{w}$ is modified from $\mathbf{r}$ such that $w_{j_2}=\gamma$ if such a $\gamma$ is found. For $1\leq m\leq j_2-1$, $z_m(\mathbf{w})=z_m(\mathbf{r})$ since $r_m=w_m$ over this range of $m$. We now consider the case of  $m=j_ 2$. Suppose node $\gamma$ belongs to the $Q_{\gamma}$-th family. We first notice that by the definition of $z_m(\cdot)$ and the definition of the family repair scheme, $(z_m(\mathbf{w})-z_m(\mathbf{r}))$ is equal to the number of distinct nodes in the $Q$-th family that appear in the first $(j_2-1)$ coordinates of $\mathbf{r}$ minus the number of distinct nodes in the $Q_{\gamma}$-th family that appear in the first $(j_2-1)$ coordinates of $\mathbf{w}$.  For easier reference, we call the former ${\mathsf{term1}}$ and the latter ${\mathsf{term2}}$ and we will quantify these two terms separately.

\par Since we start Step~2 only after Step~1 cannot proceed any further, it implies that all distinct $(n-d)$ nodes of family $Q$ must appear in $\mathbf{r}$ otherwise we should continue Step~1 rather than go to Step 2. Then by our specific construction of $j_2$, all distinct $(n-d)$ nodes of family $Q$ must appear in the  first $(j_2-1)$-th coordinates of $\mathbf{r}$. Therefore ${\mathsf{term1}}=(n-d)$. Since there are exactly $(n-d)$ distinct nodes in the $Q_{\gamma}$-th family, by the definition of ${\mathsf{term2}}$, we must have ${\mathsf{term2}}\leq (n-d)$. The above arguments show that ${\mathsf{term2}}\leq {\mathsf{term1}}=(n-d)$, which implies the desired inequality $z_m(\mathbf{w})-z_m(\mathbf{r})\geq 0$ when $m=j_2$.

\par
We now consider the case when $m>j_2$. In this case, we still have $z_m(\mathbf{w})\geq z_m(\mathbf{r})$. The reason is that by our construction, we have $w_{j_2}=\gamma \neq r_{j_2}= r_i=w_i$. For any $m>j_2$, $z_m(\mathbf{r})$ only counts the repeated $r_i=r_{j_2}$ once. Therefore, $z_m(\mathbf{w})$ will count the same $w_i$ as well. On the other hand, $z_m(\mathbf{w})$ may sometimes be larger than $z_m(\mathbf{r})$, depending on whether the new $w_{j_2} \in D_{w_m}$ or not. The fact that $z_m(\mathbf{w})\geq z_m(\mathbf{r})$ for all $1\leq m\leq k$ implies \eqref{eq:modified}.

\par Now, we consider the case when $j_1\neq j_2$, which implies that $r_{j_1}=h\neq r_{j_2}$ and Step~2 swaps the $j_1$-th and the $j_2$-th coordinates of $\mathbf{r}$. Note that after swapping, we can see that if we apply the same $j_1$ and $j_2$ construction to the {\em new} swapped vector, then we will have $j_1=j_2$. By the discussion in the case of $j_1=j_2$, we know that replacing the value of $r_{j_2}$ by $\gamma$ will not decrease the value $z_m(\mathbf{w})$ for any $m=1$ to $k$ and \eqref{eq:modified} still holds. As a result, we only need to prove that swapping the $j_1$-th and the $j_2$-th coordinates of $\mathbf{r}$ does not decrease the value of $z_m(\mathbf{r})$.

\par To that end, we slightly abuse the notation and use $\mathbf{w}$ to denote the resulting vector after swapping the $j_1$-th and the $j_2$-th coordinates of $\mathbf{r}$ (but before replacing $r_{j_2}$ by $\gamma$). For the case of $1\leq m\leq j_1$, we have $z_m(\mathbf{w})=z_m(\mathbf{r})$ since for $1\leq m\leq j_1-1$, $r_m=w_m$, and both $r_{j_1}$ and $w_{j_1}=r_{j_2}$ are from the same family $Q$. For $j_1+1\leq m \leq j_2-1$, we have $z_m(\mathbf{w})\geq z_m(\mathbf{r})$. The reason is as follows. We first observe that  $w_{j_1}=r_{j_2} \neq r_{j_1}= r_i=w_i$. For any $j_1+1\leq m \leq j_2-1$, $z_m(\mathbf{r})$ only counts the repeated $r_i=r_{j_1}$ once (since by our construction of $j_1$ we naturally have $j_1> i$). Therefore, $z_m(\mathbf{w})$ will count the same $w_i$ as well. On the other hand, $z_m(\mathbf{w})$ may sometimes be larger than $z_m(\mathbf{r})$, depending on whether the new $w_{j_1} \in D_{w_m}$ or not. We thus have $z_m(\mathbf{w})\geq z_m(\mathbf{r})$ for $j_1+1\leq m\leq j_2-1$.

\par For the case of $m=j_2$, we notice that $w_{j_2}=r_{j_1}$ and $r_{j_2}$ are from the same $Q$-th family. Therefore, we have $z_m(\mathbf{w})= z_m(\mathbf{r})$. For the case of $j_2+1\leq m\leq k$, we argue that $z_m(\mathbf{w})=z_m(\mathbf{r})$. This is true because of the definition of $z_m(\cdot)$ and the fact that both $j_1<m$ and $j_2<m$. In summary, we have proved $z_m(\mathbf{w})\geq z_m(\mathbf{r})$ for $m=1$ to $k$, which implies \eqref{eq:modified}.

\par In Step~3, we first consider the case of $j_1=j_2$, which means that $r_{j_1}=r_{j_2}$ is replaced with $\gamma$, a node from the last complete family. For $1\leq m\leq j_1-1$, since we have $r_m=w_m$ for all $1\leq m\leq j_1-1$, we must have $z_m(\mathbf{r})=z_m(\mathbf{w})$. We now consider the case of $m=j_1$. By the definition of $z_m(\cdot)$ and the definition of the family repair scheme, $(z_m(\mathbf{w})-z_m(\mathbf{r}))$ is equal to the number of distinct nodes in the incomplete family that appear in the first $(j_1-1)$ coordinates of $\mathbf{r}$ minus the number of distinct nodes in the last complete family that simultaneously (i) belong to the helper set of the incomplete family and (ii) appear in the first $(j_1-1)$ coordinates of $\mathbf{w}$. For easier reference, we call the former $\mathsf{term1}$ and the latter $\mathsf{term2}$ and we will quantify these two terms separately. 

\par Since we have finished executing Step~1, it means that all $n\bmod(n-d)$ nodes in the incomplete family appear in the vector $\mathbf{r}$. By our construction of $j_1$,  all  $n\bmod(n-d)$ nodes in the incomplete family must appear in the first $(j_1-1)$ coordinates of $\mathbf{r}$. Therefore, $\mathsf{term1}=n\bmod(n-d)$. Since there are exactly $n\bmod(n-d)$ distinct nodes in the last complete family that belong to the helper set of the incomplete family, by the definition of ${\mathsf{term2}}$, we must have ${\mathsf{term2}}\leq n\bmod(n-d)$. The above arguments show that ${\mathsf{term2}}\leq {\mathsf{term1}}=n\bmod(n-d)$, which implies the desired inequality $z_m(\mathbf{w})-z_m(\mathbf{r})\geq 0$.

\par For the case of  $j_1+1=j_2+1\leq m$, we also have  $z_m(\mathbf{w})\geq z_m(\mathbf{r})$.  The reason is that by our construction, we have $w_{j_2}=\gamma \neq r_{j_2}= r_i=w_i$. For any $m>j_2$, $z_m(\mathbf{r})$ only counts the repeated $r_i=r_{j_2}$ once. Therefore, $z_m(\mathbf{w})$ will count the same $w_i$ as well. On the other hand, $z_m(\mathbf{w})$ may sometimes be larger than $z_m(\mathbf{r})$, depending on whether the new $w_{j_2} \in D_{w_m}$ or not. We have thus proved that $z_m(\mathbf{w})\geq z_m(\mathbf{r})$ for all $m=1$ to $k$, which implies \eqref{eq:modified}.

\par We now consider the case of $j_1\neq j_2$. Namely, we swap the $j_1$-th and the $j_2$-th coordinates of $\mathbf{r}$ before executing the rest of Step~3. We can use the same arguments as used in proving the swapping step of Step~2 to show that after swapping, we still have $z_m(\mathbf{w})\geq z_m(\mathbf{r})$ for all $m=1$ to $k$, which implies \eqref{eq:modified}. The proof of Step~3 is complete.

\par In Step~4, we again consider the case of $j_1=j_2$ first. In this case, $r_{j_1}=h$ is replaced with $\gamma$, a node of the incomplete family. For $1\leq m \leq j_1-1$, $z_m(\mathbf{w})=z_m(\mathbf{r})$ since $w_m=r_m$ over this range of $m$. For $m=j_1$, we have to consider two cases. If the $Q$-th family is the last complete family, then $(z_m(\mathbf{w})-z_m(\mathbf{r}))$ is equal to the number of distinct nodes in the $Q$-th family that simultaneously (i) belong to the helper set of the incomplete family and (ii) appear in the first $(j_1-1)$ coordinates of $\mathbf{r}$, minus the number of distinct nodes in the incomplete family that appear in the first $(j_1-1)$ coordinates of $\mathbf{w}$. For easier reference, we call the former $\mathsf{term1}$ and the latter $\mathsf{term 2}$. If, however, the $Q$-th family is not the last complete family, then $(z_m(\mathbf{w})-z_m(\mathbf{r}))$ is equal to the difference of another two terms. We slightly abuse the notation and refer again to the two terms as $\mathsf{term1}$ and $\mathsf{term2}$ where $\mathsf{term1}$ is the number of distinct nodes in the $Q$-th family that appear in the first $(j_1-1)$ coordinates of $\mathbf{r}$ and $\mathsf{term2}$ is the number of distinct nodes in the last complete family that simultaneously (i) does not belong to the helper set of the incomplete family and (ii) appear in the first $(j_1-1)$ coordinates of $\mathbf{w}$ plus the number of distinct nodes in the incomplete family that appear in the first $(j_1-1)$ coordinates of $\mathbf{w}$. 

\par We will now quantify these two terms separately. Since we have finished executing Step~1 and by the construction of $j_1$, all  $(n-d)$ nodes in the $Q$-th family must appear in the first $(j_1-1)$ coordinates of $\mathbf{r}$, which are the same as the first $(j_1-1)$ coordinates of $\mathbf{w}$. Therefore, the value of $\mathsf{term1}$ is $n\bmod(n-d)$ if the $Q$-th family is the last complete family or $(n-d)$ if it is one of the first $c-1$ complete families. We now quantify $\mathsf{term2}$. For when the $Q$-th family is the last complete family, since there are exactly $n\bmod(n-d)$ distinct nodes in the incomplete family, by the definition of ${\mathsf{term2}}$, we must have ${\mathsf{term2}}\leq n\bmod(n-d)$. When the $Q$-th family is not the last complete family, ${\mathsf{term2}}\leq (n-d)$ since the number of distinct nodes in the incomplete family is $n\bmod(n-d)$ and the number of distinct nodes in the last complete family that do not belong to the helper set of the incomplete family is $(n-d-n\bmod(n-d))$ and their summation is $\leq n-d$. The above arguments show that ${\mathsf{term2}}\leq {\mathsf{term1}}$ for both cases, which implies the desired inequality $z_m(\mathbf{w})-z_m(\mathbf{r})\geq 0$ for $m=j_1$.

\par For $j_1+1\leq m\leq k$, since $r_{j_1}=h=r_i$ was a repeated node, then it was already not contributing to $z_m(\mathbf{r})$ for all $m>j_1$. Thus, $z_{m}(\mathbf{w})\geq z_m(\mathbf{r})$ for all $m=j_1+1$ to $k$. (Please refer to the $j_1+1\leq m$ case in Step~3 for detailed elaboration.) In summary, after Step~4, assuming $j_1=j_2$, we have $z_{m}(\mathbf{w})\geq z_m(\mathbf{r})$ for all $m=1$ to $k$, which implies \eqref{eq:modified}.

\par Finally, we consider the case of $j_1\neq j_2$. Namely, we swap the $j_1$-th and the $j_2$-th coordinates of $\mathbf{r}$ before executing the rest of Step~4. We can use the same arguments as used in proving the swapping step of Step~2 to show that the inequality \eqref{eq:modified} holds after swapping. The proof of Step~4 is thus complete.

\section{An Illustrative Example for the {\sc Modify} Procedure} \label{app:modify_ex}

\par For illustration, we apply the procedure {\sc Modify} to the following example with $(n,d)=(8,5)$ and some arbitrary $k$. Recall that family 1 contains nodes $\{1,2,3\}$, family 2 (last complete family) contains  nodes $\{4,5,6\}$, and the incomplete family, family 0, contains nodes $\{7,8\}$. Suppose the initial $\mathbf{r}$ vector is $\mathbf{r}=(1,2,2,2,4,7,7,7)$. We will use {\sc Modify} to convert $\mathbf{r}$ to a vector $\mathbf{r}'\in R_2$

\par We first enter Step~1 of the procedure. We observe\footnote{We also observe that $r_2=r_3=2$ and we can choose $i=2$ and $j=3$ instead. Namely, the choice of $(i,j)$ is not unique. In {\sc Modify}, any choice satisfying our algorithmic description will work. } that $r_3=r_4=2$ ($i=3$ and $j=4$) and node 2 belongs to the first family. Since node 3 is also in family 1 and it is not present in $\mathbf{r}$,  we can choose $\gamma=3$. After replacing $r_4$ by 3, the resulting vector is $\mathbf{r}=(1,2,2,3,4,7,7,7)$. Next, we enter Step~1 for the second time. We observe that $r_7=r_8=7$. Since node 8 is in family 0 and it is not present in $\mathbf{r}$, we can choose $\gamma=8$. The resulting vector is $\mathbf{r}=(1,2,2,3,4,7,7,8)$. Next, we enter Step~1 for the third time. For the new $\mathbf{r}$, we have $r_2=r_3=2$ and $r_6=r_7=7$, but for both cases we cannot find the desired $\gamma$ value. As a result, we cannot proceed any further by Step~1. For that reason, we enter Step 2. 

\par We observe that for $r_2=r_3=2$, we find $j_1=3$, the last coordinate of $\mathbf{r}$ equal to $2$, and $j_2=4$, the last coordinate of $\mathbf{r}$ that belongs to family~1.  By Step~2, we swap $r_{3}$ and $r_{4}$, and the resultant vector is $\mathbf{r}=(1,2,3,2,4,7,7,8)$. Now, since node 5 belongs to family 2, a complete family, and it is not present in $\mathbf{r}$, we can choose $\gamma=5$. After replacing $r_{j_2}$ by $\gamma$, the resultant vector is $\mathbf{r}=(1,2,3,5,4,7,7,8)$. Next, we enter Step~2 for the second time. Although $r_6=r_7=7$, we notice that node 7 is in family 0. Therefore, we do nothing in Step~2.

\par After Step~2, the latest $\mathbf{r}$ vector is $\mathbf{r}=(1,2,3,5,4,7,7,8)$, which belongs to Case~2.2. Consequently, we enter Step~3. In Step~3, we observe that $j_1=7$, the last coordinate of $\mathbf{r}$ being 7, and $j_2=8$, the last coordinate of $\mathbf{r}$  that belongs to the incomplete family, family 0. Thus, we swap $r_{7}$ and $r_{8}$, and the resultant vector is $\mathbf{r}=(1,2,3,5,4,7,8,7)$. Now, we choose arbitrarily a $\gamma$ value from $\{4,5,6\}$, the last complete family. Suppose we choose\footnote{We can also choose $\gamma=4$ or $5$. For those choices, the iterative process will continue a bit longer but will terminate eventually.} $\gamma =6$. The resultant vector is $\mathbf{r}=(1,2,3,5,4,7,8,6)$. Since we have no other repeated nodes of family 0, the procedure finishes at this point. Indeed, we can see that the final vector $\mathbf{r'}=(1,2,3,5,4,7,8,6)\in R_2$, which has no repeated nodes and is the result expected.

\section{Proof of Proposition~\ref{prop:mbr}}\label{app:mbr_proof}
For fixed $(n,k,d)$ values, define function $g$ as
\begin{align}
g(\alpha,\beta)=\min_{G\in\mathcal{G}_F} \min_{t\in\DC(G)}\mincut_G(s,t).
\end{align}
We first note that by \eqref{eq:low_b}, we must have $g(d\beta,\beta)=m\beta$ for some integer $m$. The value of $m$ depends on the $(n,k,d)$ values and the minimizing family index permutation $\pi_f$, but does not depend on $\beta$. We then define $\beta^*$ as the $\beta$ value such that $g(d\beta,\beta)= \mathcal{M}$. We will first prove that $\beta_{\MBR}=\beta^*$ by contradiction. Suppose $\beta_{\MBR}\neq \beta^*$. Since $(\alpha,\beta)=(d\beta^*,\beta^*)$ is one way that can satisfy $g(\alpha,\beta)=\mathcal{M}$, the minimum-bandwidth consumption $\beta_{\MBR}$ must satisfy $\beta_{\MBR}\leq \beta^*$. Therefore, we must have $\beta_{\MBR}<\beta^*$. However, we then have the following contradiction.
\begin{align}
\mathcal{M}\leq g(\alpha_{\MBR},\beta_{\MBR})\leq g(\infty, \beta_{\MBR})&=\nonumber \\
g(d \beta_{\MBR}, \beta_{\MBR})&<g(d\beta^*,\beta^*)= \mathcal{M}, \label{new:MBR}
\end{align}
where the first inequality is by knowing that $(\alpha_{\MBR},\beta_{\MBR})$ satisfies the reliability requirement; the second inequality is by the definition of $g(\alpha,\beta)$; the first equality is by \eqref{eq:low_b}; and the third inequality (the only strict inequality) is by the fact that $g(d\beta,\beta)=m\beta$ for all $\beta$ and by the assumption of $\beta_{\MBR}<\beta^*$; and the last equality is by the construction of $\beta^*$.

\par The above arguments show that $\beta_{\MBR}=\beta^*$. To prove that $\alpha_{\MBR}=d\beta^*$, we first prove
\begin{align}
g(\alpha,\beta)<g(d\beta,\beta), \mbox{ if } \alpha<d\beta. \label{eq:g1}
\end{align}
The reason behind \eqref{eq:g1} is that (i) $k\geq 1$ and we thus have at least one summand in the RHS of \eqref{eq:low_b}; and (ii) the first summand is always $\min(d\beta,\alpha)$ since $y_1(\pi_f)=0$ for any family index permutation $\pi_f$. Suppose $\alpha_{\MBR}\neq d\beta^*$. Obviously, we have $\alpha_{\MBR}\leq d\beta^*$ by the construction of $\beta^*$. Therefore, we must have $\alpha_{\MBR}<d\beta^*$. However, we then have the following contradiction
\begin{align}
\mathcal{M}\leq g(\alpha_{\MBR},\beta_{\MBR}) < g(d\beta^*,\beta^*)= \mathcal{M}, \label{new:MBR2}
\end{align}
where the first inequality is by knowing that $(\alpha_{\MBR},\beta_{\MBR})$ satisfies the reliability requirement, the second inequality is by \eqref{eq:g1}, and the equality is by the construction of $\beta^*$.

The above arguments prove that $\alpha_{\MBR}=d\beta_{\MBR}$.  This also implies that when considering the MBR point, instead of finding a $\pi_f$ that minimizes \eqref{eq:low_b}, we can focus on finding a $\pi_f$ that minimizes 
\begin{align}
\sum_{i=1}^k (d-y_i(\pi_f)) \label{eq:new15}
\end{align}
instead, i.e., we remove the minimum operation of \eqref{eq:low_b} and ignore the constant $\beta$, which does not depend on $\pi_f$. We are now set to show that $\pi_f^*$ is the minimizing family index permutation at the MBR point. 
\par First, define
\begin{align} \label{eq:y_def}
y_{\text{offset}}(\pi_f)=\sum_{i=1}^k (i-1-y_i(\pi_f)).
\end{align}
Notice that a family index permutation that minimizes $y_{\text{offset}}(\cdot)$ also minimizes \eqref{eq:new15}. Therefore, any minimizing family index permutation for \eqref{eq:new15}, call it $\pi_f^{\min}$, must satisfy
\begin{align} \label{eqn:y}
y_{\text{offset}}(\pi_f^{\min})=\min_{\forall \pi_f} y_{\text{offset}}(\pi_f).
\end{align}

\par Consider the following two cases:

\underline{Case 1:} $n\bmod(n-d)=0$, i.e., we do not have an incomplete family.

Consider any family index permutation $\pi_f$ and let $l_j$ be the number of the first $k$  coordinates of $\pi_f$ that have value $j$. Recall that there is no incomplete family in this case. Suppose the $i$-th coordinate of $\pi_f$ is $m$. Then, we notice that the expression ``$(i-1)-y_i(\pi_f)$'' counts the number of appearances of the value $m$ in the first $i-1$ coordinates of $\pi_f$ (recall that there is no incomplete family in this case). Therefore, we can rewrite \eqref{eq:y_def} by
\begin{align} \label{eq:y_families}
y_{\text{offset}}(\pi_f)=\sum_{i=1}^{l_1}(i-1) + \sum_{i=1}^{l_2}(i-1) + \dots + \sum_{i=1}^{l_{\frac{n}{n-d}}}(i-1).
\end{align}
We now prove the following claim.

\begin{claim} \label{clm:diverse} The above equation implies that a family index permutation is a minimizing permutation $\pi_f^{\min}$ if and only if
\begin{align} \label{eq:diverse}
|l_i-l_j|\leq1 \text{ for all $i,j$ satisfying }1\leq i,j\leq \frac{n}{n-d}.
\end{align}
\end{claim}

\begin{IEEEproof}
We first prove the only if direction by contradiction. The reason is as follows. If $l_i>l_j+1$ for some $1\leq i,j\leq \frac{n}{n-d}$, then we consider another family permutation $\pi_f'$ and denote its corresponding $l$ values by $l'$, such that $l'_i=l_i-1$, $l'_j=l_j+1$, and all other $l$s remain the same. Clearly from \eqref{eq:y_families}, such $\pi_f'$ will result in strictly smaller $y_{\text{offset}}(\pi_f')<y_\text{offset}(\pi_f)$. Note that such $\pi_f'$ with the new $l'_i=l_i-1$, $l'_j=l_j+1$ always exists. The reason is the following. By the definition of $l_j$ and the fact that $\pi_f$ is a family index permutation, we have $0\leq l_j\leq (n-d)$ for all $j=1,\cdots, \frac{n}{n-d}$. The inequality $l_i>l_j+1$ then implies $l_i\geq 1$ and $l_j\leq (n-d)-1$. Therefore, out of the first $k$ coordinates of $\pi_f$, at least one of them will have value $i$; and out of the last $(n-k)$ coordinates of $\pi_f$, at least one of them will have value $j$. We can thus swap arbitrarily one of the family indices $i$ from the first $k$ coordinates with another family index $j$ from the last $(n-k)$ coordinates and the resulting $\pi_f'$ will have the desired $l_i'$ and $l_j'$.

\par We now prove the if direction. To that end, we first observe that the equality
$\sum_{i=1}^{\frac{n}{n-d}}l_i=k$ always holds because of our construction of $l_i$. Then  \eqref{eq:diverse} implies that we can uniquely decide the {\em distribution} of $\{l_i:i=1,\cdots,\frac{n}{n-d}\}$ even though we do not know what is the minimizing permutation $\pi_f^{\min}$ yet. For example, if $\frac{n}{n-d}=3$, $k=5$, $l_1$ to $l_3$ satisfy \eqref{eq:diverse}, and the summation $l_1+l_2+l_3$ is $k=5$, then among $l_1$, $l_2$, and $l_3$, two of them must be 2 and the other one must be 1. On the other hand, we observe that the value of $y_{\text{offset}}(\cdot)$ depends only on the distribution of $\{l_i\}$, see \eqref{eq:y_families}. As a result, the above arguments prove that any $\pi_f$ satisfying \eqref{eq:diverse} is a minimizing $\pi_f^{\min}$.

\end{IEEEproof}

\par Finally, by the construction of the RFIP $\pi_f^*$, it is easy to verify that the RFIP $\pi_f^*$ satisfies \eqref{eq:diverse}. Therefore, the RFIP $\pi_f^*$ is a minimizing permutation for this case.

\underline{Case 2:} $n\bmod(n-d)\neq 0$, i.e., when we do have an incomplete family.
In this case, we are again interested in minimizing \eqref{eq:new15}, and equivalently minimizing \eqref{eq:y_def}. To that end, we first prove the following claim.

\begin{claim}\label{clm:rfip_incomplete} Find the largest $1\leq j_1\leq k$ such that the $j_1$-th coordinate of $\pi_f$ is 0. If no such $j_1$ can be found, we set $j_1=0$.  Find the smallest $1\leq j_2\leq k$ such that the $j_2$-th coordinate of $\pi_f$ is a negative number if no such $j_2$ can be found, we set $j_2=k+1$. We claim that if we construct $j_1$ and $j_2$ based on a $\pi_f$ that minimizes $\sum_{i=1}^k (d-y_i(\pi_f))$, we must have $j_1<j_2$.
\end{claim}

\begin{IEEEproof}
We prove this claim by contradiction. Consider a minimizing family index permutation $\pi_f$ and assume $j_2<j_1$. This means, by our construction, that $1\leq j_2<j_1\leq k$.  Since the $j_2$-th coordinate of $\pi_f$ is a negative number by construction, $y_{j_2}(\pi_f)$ counts all coordinates before the $j_2$-th coordinate of $\pi_f$ with values in $\{1,2,\cdots,c-1,0\}$, i.e., it counts all the values before the $j_2$-th coordinate except for the values $c$ and $-c$, where $c$ is the family index of the last complete family. Thus, knowing that there are no $-c$ values before the $j_2$-th coordinate of $\pi_f$, we have that
\begin{align} \label{eq:y_j2_pi}
y_{j_2}(\pi_f)=j_2-1-\lambda^{[1,j_2)}_{\{c\}},
\end{align}
where $\lambda^{[1,j_2)}_{\{c\}}$ is the number of $c$ values before the $j_2$-th coordinate. Similarly, since the $j_1$-th coordinate is 0, we have that $y_{j_1}(\pi_f)$ counts all coordinates before the $j_1$-th coordinate of $\pi_f$ with values in $\{1,2,\cdots,c\}$, i.e., it counts all the values before the $j_1$-th coordinate except for the values $-c$ and $0$. Thus, we have that 
\begin{align}
y_{j_1}(\pi_f)&=j_1-1-\lambda^{[1,j_1)}_{\{0\}}-\lambda^{[1,j_1)}_{\{-c\}}
\end{align}
where $\lambda^{[1,j_1)}_{\{0\}}$ is the number of 0 values preceding the $j_1$-th coordinate in $\pi_f$ and $\lambda^{[1,j_1)}_{\{-c\}}$ is the number of $-c$ values preceding the $j_1$-th coordinate in $\pi_f$. Now, swap the $j_2$-th coordinate and the $j_1$-th coordinate of $\pi_f$, and call the new family index permutation $\pi_f'$. Specifically, $\pi_f'$ has the same values as $\pi_f$ on all its coordinates except at the $j_2$-th coordinate it has the value 0 and at the $j_1$-th coordinate it has the value $-c$. For $1\leq m\leq j_2-1$, we have that $y_m(\pi_f')=y_m(\pi_f)$ since the first $j_2-1$ coordinates of the two family index permutations are equal. Moreover, since there are no negative values before the $j_2$-th coordinate of $\pi_f'$, we have that
\begin{align}\label{eq:y_j2}
y_{j_2}(\pi_f')=j_2-1-\phi^{[1,j_2)}_{\{0\}},
\end{align}
where $\phi^{[1,j_2)}_{\{0\}}$ is the number of 0 values in $\pi_f'$ preceding the $j_2$-th coordinate. 

\par For $j_2+1\leq m\leq j_1-1$, if the $m$-th coordinate of $\pi_f'$ is either $c$ or $-c$, then $y_m(\pi_f')=y_m(\pi_f)+1$; otherwise, $y_m(\pi_f')=y_m(\pi_f)$. The reason behind this is that the function $y_m(\pi_f')$ now has to take into account the new 0 at the $j_2$-th coordinate when the $m$-th coordinate is either $c$ or $-c$. When the value of the $m$-th coordinate is in $\{1,\cdots, c-1\}$, then by the definition of $y_m(\cdot)$, we have $y_m(\pi_f')=y_m(\pi_f)$. The last situation to consider is when the value of the $m$-th coordinate is $0$. In this case, we still have $y_m(\pi_f')=y_m(\pi_f)$  since $y_m(\pi_f)$ already does not count the value on the $j_2$-th coordinate of $\pi_f$ since it is a negative value. 

\par Denote the number of $c$ and $-c$ values from the $(j_2+1)$-th coordinate to the $(j_1-1)$-th coordinate of $\pi_f'$ by $\phi^{(j_2,j_1)}_{\{c,-c\}}$. We have that
\begin{align}\label{eq:y_j1_pip}
y_{j_1}(\pi_f')=j_1-1-\lambda^{[1,j_2)}_{\{c\}} -\phi^{(j_2,j_1)}_{\{c,-c\}},
\end{align}
since the $j_1$-th coordinate of $\pi_f'$ has a $-c$ value. Finally, for $j_1+1\leq m\leq n$, we have that $y_m(\pi_f')=y_m(\pi_f)$ since the order of the values preceding the $m$-th coordinate in a permutation does not matter for $y_m(\cdot)$. By the above, we can now compute the following difference
\begin{align}
\sum_{i=1}^k&(d-y_i(\pi_f))-\sum_{i=1}^k(d-y_i(\pi_f')) \nonumber\\
&= \sum_{i=1}^k (y_i(\pi_f')-y_i(\pi_f))\nonumber\\
&= \sum_{i=j_2}^{j_1} (y_i(\pi_f')-y_i(\pi_f))\label{eq:CCWnew3}\\
&=(y_{j_2}(\pi_f')-y_{j_2}(\pi_f))+ \phi^{(j_2,j_1)}_{\{c,-c\}}+(y_{j_1}(\pi_f')-y_{j_1}(\pi_f)) \label{eq:diff_1}\\
&=\left(\lambda^{[1,j_2)}_{\{c\}}-\phi^{[1,j_2)}_{\{0\}}\right)+\phi^{(j_2,j_1)}_{\{c,-c\}}+\nonumber\\
&\qquad\qquad\qquad\left(\lambda^{[1,j_1)}_{\{0\}}+\lambda^{[1,j_1)}_{\{-c\}}-\lambda^{[1,j_2)}_{\{c\}} -\phi^{(j_2,j_1)}_{\{c,-c\}}\right) \label{eq:CCWnew4}\\
&=\lambda^{[1,j_1)}_{\{0\}}+\lambda^{[1,j_1)}_{\{-c\}}-\phi^{[1,j_2)}_{\{0\}}\nonumber\\
&>0\label{eq:diff_2},
\end{align}
where \eqref{eq:CCWnew3} follows from $y_i(\pi_f')=y_i(\pi_f)$ for all $i<j_2$ and for all $i>j_1$; \eqref{eq:diff_1} follows from our analysis about $y_i(\pi_f')=y_i(\pi_f)+1$ when the $i$-th coordinate of $\pi_f$ belongs to $\{-c,c\}$ and $y_i(\pi_f')=y_i(\pi_f)$ otherwise, and there are thus $\phi^{(j_2,j_1)}_{\{c,-c\}}$ coordinates between the $(j_2+1)$-th coordinate and the $(j_1-1)$-th coordinate of $\pi_f'$ that satisfy $y_i(\pi_f')=y_i(\pi_f)+1$; \eqref{eq:CCWnew4} follows from \eqref{eq:y_j2_pi} to \eqref{eq:y_j1_pip}; and \eqref{eq:diff_2} follows from the facts that $\lambda^{[1,j_1)}_{\{0\}} \geq \lambda^{[1,j_2)}_{\{0\}}=\phi^{[1,j_2)}_{\{0\}}$ and that $\lambda^{[1,j_1)}_{\{-c\}} \geq 1$ since we have a $-c$ value at the $j_2$-th coordinate of $\pi_f$. By \eqref{eq:diff_2}, we have that $\pi_f'$ has a strictly smaller ``$\sum_{i=1}^k(d-y_i(\cdot))$''. As a result, the case of $j_1>j_2$ is impossible. 

\par By the construction of $j_1$ and $j_2$, it is obvious that $j_1\neq j_2$. Hence, we must have $j_1<j_2$. The proof of this claim is complete.
\end{IEEEproof}

\par Claim~\ref{clm:rfip_incomplete} provides a necessary condition on a minimizing permutation vector. We thus only need to consider permutations for which $j_1<j_2$. That is, instead of taking the minimum over all $\pi_f$, we now take the minimum over only those $\pi_f$ satisfying $j_1<j_2$.

This observation is critical to our following derivation. The reason is that if we consider a permutation $\pi_f$ that has $1\leq j_2<j_1\leq k$, then the expression ``$(j_1 - 1)- y_{j_1}(\pi_f )$'' is not equal to the number of appearances of the value $0$ in the first $j_1-1$ coordinates of $\pi_f$ (recall that by our construction the $j_1$-th coordinate of $\pi_f$ is 0). Instead, by the definition of $y_{i}(\cdot)$, $(j_1-1)-y_{j_1}(\pi_f)$ is the number of appearances of the values 0 {\em and} $-c$ in the first $(j_1-1)$ coordinates of $\pi_f$. Therefore, we cannot rewrite \eqref{eq:y_def} as \eqref{eq:y_families} if $1\leq j_2<j_1\leq k$. 

\par On the other hand, Claim~\ref{clm:rfip_incomplete} implies that we only need to consider those $\pi_f$ satisfying $j_1<j_2$. We now argue that given any $\pi_f$ satisfying $j_1<j_2$, for all $i=1$ to $k$, the expression $(i-1)-y_{i}(\pi_f)$ is now representing the number of appearances of $m$ and $-m$ in the first $(i-1)$ coordinates of $\pi_f$, where $m$ is the {\em absolute value} of the $i$-th coordinate of $\pi_f$. The reason is as follows. Let $m$ denote the absolute value of the $i$-th coordinate of $\pi_f$. If $m\neq 0$, then by the definition of $y_i(\pi_f)$, we have that $(i-1)-y_{i}(\pi_f)$ represents the number of appearances of $m$ in the first $(i-1)$ coordinates of $\pi_f$. If $m=0$, then by the definition of $y_i(\pi_f)$, we have that $(i-1)-y_{i}(\pi_f)$ represents the number of appearances of 0 and $-c$ in the first $(i-1)$ coordinates of $\pi_f$. However, by the construction of $j_1$, we have $i\leq j_1$. Since $j_1<j_2$, we have $i<j_2$. This implies that in the first $(i-1)$ coordinates of $\pi_f$, none of them is of value $-c$. As a result, we have that $(i-1)-y_{i}(\pi_f)$ again represents the number of appearances of 0 in the first $(i-1)$ coordinates of $\pi_f$. 

\par
We now proceed with our analysis while only considering those $\pi_f$ satisfying $j_1<j_2$ as constructed in Claim~\ref{clm:rfip_incomplete}. Let $l_j$ be the number of the first $k$  coordinates of $\pi_f$ that have values $j$ or $-j$. We can then rewrite \eqref{eq:y_def} by
\begin{align} \label{eq:new-y_families}
y_{\text{offset}}(\pi_f)= \sum_{i=1}^{l_0}&(i-1)+\sum_{i=1}^{l_1}(i-1)+ \nonumber\\
& \sum_{i=1}^{l_2}(i-1) + \dots + \sum_{i=1}^{l_{\left\lfloor\frac{n}{n-d}\right\rfloor}}(i-1).
\end{align}
The above equation implies that a family index permutation is a minimizing permutation $\pi_f^{\min}$ if and only if either
\begin{align}\label{eq:diverse2_1}
\begin{cases}l_0=n\bmod(n-d),\\
|l_i-l_j|\leq1 \text{ for all $i,j$ satisfying}~1\leq i,j\leq c,\\
l_i\geq l_0 \text{ for all $i$ satisfying}~1\leq i\leq c.
\end{cases}
\end{align}
or
\begin{align} \label{eq:diverse2_2}
|l_i-l_j|\leq1, \text{for all $i,j$ satisfying}~0\leq i,j\leq c. 
\end{align}
If we compare \eqref{eq:diverse2_1} and \eqref{eq:diverse2_2} with \eqref{eq:diverse} in Claim~\ref{clm:diverse}, we can see that \eqref{eq:diverse2_2} is similar to \eqref{eq:diverse}. The reason we need to consider the situation described in \eqref{eq:diverse2_1} is that the range of $l_0$ is from 0 to $n\bmod(n-d)$ while the range of all other $l_i$s is from 0 to $(n-d)$. Therefore, we may not be able to make $l_0$ as close to other $l_i$s (within a distance of 1) as we would have hoped for due to this range discrepancy. For some cases, the largest $l_0$ we can choose is $n\bmod(n-d)$, which gives us the first scenario when all the remaining $l_i$s are no less than this largest possible $l_0$ value. If $l_0$ can also be made as close to the rest of $l_i$s, then we have the second scenario.

The proof that \eqref{eq:diverse2_1} and \eqref{eq:diverse2_2} are the if-and-only-if condition on $\pi_f^{\min}$ can be completed using the same arguments as in the proof of Claim~\ref{clm:diverse}. Finally, notice that the RFIP $\pi_f^*$ satisfies \eqref{eq:diverse2_1} or \eqref{eq:diverse2_2} and has $j_1<j_2$. As a result, $\pi_f^*$ must be one of the minimizing permutations $\pi_f^\text{min}$. The proof of this proposition is hence complete.

\section{Proof of Proposition~\ref{prop:msr}} \label{app:msr_proof}
We first consider the case when $d\geq k$. We have $\alpha_{\MSR}\geq\frac{\mathcal{M}}{k}$ since otherwise the MSR point cannot satisfy \eqref{eq:condition} even when plugging in $\beta=\infty$ in \eqref{eq:low_b}. Define
\begin{align}\label{eq:max-y-msr}
y_{\max}\stackrel{\Delta}{=}\max_{\forall \pi_f} \max_{1\leq i\leq k} y_i(\pi_f).
\end{align}
By \eqref{eq:low_b}, we have that the $(\alpha,\beta)$ pair
\begin{align}
(\alpha,\beta)=\left(\frac{\mathcal{M}}{k}, \frac{\mathcal{M}}{k(d-y_{\max})}\right)
\end{align}
satisfies \eqref{eq:condition} since $(d-y_i(\pi_f))\beta\geq (d-y_{\max})\beta= \frac{\mathcal{M}}{k}=\alpha$. Therefore, $\frac{\mathcal{M}}{k}$ is not only a lower bound of $\alpha_{\MSR}$ but is also achievable, i.e.,  $\alpha_{\MSR}=\frac{\mathcal{M}}{k}$. Now, for any $(\alpha,\beta)$ pair satisfying
\begin{align}
(\alpha,\beta)=\left(\frac{\mathcal{M}}{k},\beta\right)
\end{align}
for some $\beta<\frac{\mathcal{M}}{k(d-y_{\max})}$, we argue that \eqref{eq:condition} does not hold anymore. The reason is the following. When $\alpha=\frac{\mathcal{M}}{k}$ and $\beta<\frac{\mathcal{M}}{k(d-y_{\max})}$, we plug in the $\pi_f^{\circ}$  vector that maximizes \eqref{eq:max-y-msr} into \eqref{eq:low_b}. Therefore, for at least one $i^{\circ}\leq k$, we will have $(d-y_{i^{\circ}}(\pi_f^{\circ}))\beta<\alpha=\frac{\mathcal{M}}{k}$. This implies ``$\eqref{eq:low_b}< \mathcal{M}$'' when evaluated using $\pi_f^{\circ}$.  By taking the minimum over all $\pi_f$, we still have ``$\eqref{eq:low_b}< \mathcal{M}$''.  Therefore, the above choice of $(\alpha,\beta)$ cannot meet the reliability requirement at the MSR point. As a result, we have $\beta_{\MSR}=\frac{\mathcal{M}}{k(d-y_{\max})}$.

\par We now argue that $y_{\max}=k-1$. According to the definition of function $y_i(\cdot)$, $y_i\leq k-1$. Recall that the size of a helper set is $d$, which is strictly larger than $k-1$. We can thus simply set the values of the first $(k-1)$ coordinates of $\pi_f$ to be the family indices of the $(k-1)$ distinct helpers (out of $d$ distinct helpers) of a node and place the family index of this node on the $k$-th coordinate. Such a permutation $\pi_f$ will have $y_k(\pi_f)=k-1$. Therefore, we have proved that $\beta_{\MSR}=\frac{\mathcal{M}}{k(d-k+1)}$.

We now consider the remaining case in which $d<k$. To that end, we first notice that for any $(n,k,d)$ values we have $\left\lfloor \frac{n}{n-d}\right\rfloor \geq 1$ number of complete families. Also recall that family 1 is a complete family and all families $\neq 1$ are the helpers of family 1, and there are thus $d$ number of nodes in total of family index $\neq 1$. We now consider a permutation $\pi_f^{\circ}$ in which all its first $d$ coordinates are family indices not equal to 1 and its last $(n-d)$ coordinates are of family index 1. Observe that if we evaluate the objective function of the right-hand side of \eqref{eq:low_b} using $\pi_f^{\circ}$, out of the $k$ summands, of $i=1$ to $k$, we will have exactly $d$ non-zero terms since (i) by the definition of $y_i(\cdot)$, we always have $y_i(\pi_f^{\circ})\leq (i-1)$ and therefore, when $i\leq d$, we always have $(d-y_i(\pi_f^{\circ}))\geq 1$; (ii) whenever $i>d$, the corresponding term $y_i(\pi_f^{\circ})=d$ due to the special construction of the $\pi_f^{\circ}$. As a result, when a sufficiently large $\beta$ is used, we have 
\begin{align}
\sum_{i=1}^k \min((d-y_i(\pi_f^{\circ}))\beta,\alpha) = d \alpha.
\end{align}

The above equality implies $\alpha_{\MSR}\geq \frac{\mathcal{M}}{d}$. Otherwise if $\alpha_{\MSR}<\frac{\mathcal{M}}{d}$, then we will have ``$\eqref{eq:low_b} <\mathcal{M}$'' when using the aforementioned $\pi_f^{\circ}$, which implies that  ``$\eqref{eq:low_b} <\mathcal{M}$''  holds still when minimizing over all $\pi_f$. This contradicts the definition that $\alpha_{\MSR}$ and $\beta_{\MSR}$ satisfy the reliability requirement.

\par On the other hand, we know that $\alpha_{\MSR}=\frac{\mathcal{M}}{d}$ and $\beta_{\MSR}=\frac{\mathcal{M}}{d}$ for the BR scheme when $d<k$, see \eqref{eq:msr_br_alpha}. Since the performance of the FR scheme is not worse than that of the BR scheme, we have $\alpha_{\MSR}=\frac{\mathcal{M}}{d}$ and $\beta_{\MSR}\leq \frac{\mathcal{M}}{d}$ for the FR scheme. Hence, the proof is complete. 

\section{Proof of Corollary~\ref{cor:low_b}} \label{app:cor_proof}
First consider the case when $d\geq k-1=\left\lceil \frac{n}{n-d}\right\rceil$. Since there are $\left\lceil \frac{n}{n-d}\right\rceil$ number of families (complete plus incomplete families) and $k=\left\lceil \frac{n}{n-d}\right\rceil+1$, any family index permutation has at least one pair of indices of the same family in its first $k$ coordinates. Using \eqref{eq:low_b}, this observation implies that 
\begin{align}
\min_{G\in\gf}&\min_{t\in\DC(G)}\mincut(s,t)\nonumber\\
&=\min_{\forall \pi_f} \sum_{i=1}^{k}\min \left(\left(d-y_i(\pi_f)\right)\beta,\alpha\right) \geq \min_{2\leq m\leq k} C_m.
\end{align}
Now define $\pi_f^{[m]}$ as a family index permutation such that its first $k$ coordinates, in this order, are $1,2,\cdots,m-1,1,m+1,\cdots,c,0$ if $n\bmod(n-d)\neq 0$ and define $\pi_f^{[m]}$ as $1,2,\cdots,m-1,1,m+1,\cdots,c$ if $n\bmod(n-d)=0$. Since all the $k$ coordinates have different values except the first coordinate and the $m$-th coordinate have equal value $1$, and since they have no $-c$ value, we have
\begin{align}
\sum_{i=1}^k \min\left(\left(d-y_i\left(\pi_f^{[m]}\right)\right)\beta,\alpha\right)=C_m.
\end{align}
Thus, we get the equality in \eqref{eq:low_b_spec}.

\par We now consider the case when $d<k-1=\left\lceil \frac{n}{n-d}\right\rceil$. Before proceeding, we first argue that among all $(n,k,d)$ values satisfying \eqref{eq:ccw1}, the only possible cases of having $d\leq \left\lceil \frac{n}{n-d}\right\rceil-1$ are either $d=1$ or $d=n-1$. The reason behind this is the following. Suppose $d\leq \left\lceil \frac{n}{n-d}\right\rceil-1$. For any $2\leq d\leq n-2$, we have 
\begin{align}
0\leq \left\lceil\frac{n}{n-d}\right\rceil-1-d&=\left\lceil1+\frac{d}{n-d}\right\rceil-1-d\nonumber\\
&=\left\lceil\frac{d}{n-d}\right\rceil-d\nonumber\\
&\leq \left\lceil\frac{d}{2}\right\rceil-d \label{eq:corr1_1}\\
&=\begin{cases} -\frac{d}{2}, & \mbox{if } d\mbox{ is even} \\ \frac{1-d}{2}, & \mbox{if } d\mbox{ is odd} \end{cases}\nonumber\\
&<0\label{eq:corr1_2},
\end{align}
where we get \eqref{eq:corr1_1} by our assumption that $d\leq n-2$ and \eqref{eq:corr1_2} follows from the assumption that $d\geq 2$. The above contradiction implies either $d=1$ or $d=n-1$.
Since Corollary~\ref{cor:low_b} requires $d\geq 2$, the only remaining possibility is $d=n-1$. However, $k$ will not have a valid value since in this case we have $d=n-1<k-1$, which implies $k>n$, an impossible paramemter value violating \eqref{eq:ccw1}. Hence, the proof is complete. 


\section{Proof of Corollary~\ref{cor:mbr_plus}}\label{app:mbr_plus_proof}
Consider first the case when $n\bmod(2d)\neq 0$. Without loss of generality, assume that $n_B=n_{\text{remain}}$ and $n_b=2d$ for $b=1$ to $B-1$, i.e., the indices $b=1$ to $B-1$ correspond to the regular groups and the index $b=B$ corresponds to the remaining group. Now, applying the same reasoning as in the proof of Proposition~\ref{prop:mbr} to \eqref{eq:low_b_plus}, we have that $\alpha_{\MBR}=\gamma_{\MBR}=d\beta_{\MBR}$ for the family-plus repair scheme as well. In the following, we will prove that (i) if $k\leq 2d$, then one minimizing $\mathbf{k}$ vector can be constructed by setting $k_b=0$ for $b=1$ to $B-1$ and $k_B=k$; (ii) if $k>2d$, then we can construct a minimizing $\mathbf{k}$ vector by setting $k_B=\min(n_{\text{remain}},k)$ and among all $b=1$ to $B-1$, at most one $k_b$ satisfies $0<k_b<2d$.


\par To prove this claim, we first notice that since we are focusing on the MBR point, we can assume $\alpha$ is sufficiently large. Therefore, we can replace the minimizing permutation for each summand of \eqref{eq:low_b_plus} by the RFIP (of $(n,d)=(2d,d)$ for the summand $b=1$ to $B-1$ and of $(n,d)=(n_{\text{remain}},d)$ for summand $b=B$) using the arguments in the proof of Proposition~\ref{prop:mbr}. Therefore, we can rewrite \eqref{eq:low_b_plus} by
\begin{align}\label{eq:inter5}
\eqref{eq:low_b_plus}=
 \min_{\mathbf{k}\in K} \sum_{b=1}^{B} \sum_{i=1}^{k_b}(d-y_i(\pi_b))\beta
\end{align}
where $\pi_b$ is the RFIP of $(n,d)=(2d,d)$ for $b=1$ to $B-1$ and the RFIP of $(n,d)=(n_{\text{remain}},d)$ for $b=B$. Note that for $(n,d)=(2d,d)$, in the FR scheme we have 2 complete families and no incomplete family and the RFIP in this case is $\pi_1^*=(1,2,1,2,\cdots,1,2)$. As a result, $\pi_b=\pi_1^*$ for all $b=1$ to $B-1$. For $(n,d)=(n_{\text{remain}},d)$, we have one complete family and one incomplete family and the RFIP in this case is 
\begin{align}
\pi_2^*=(\overbrace{1,0,1,0,\cdots,1,0}^{2d \text{ coordinates}},\overbrace{-1,-1,\cdots,-1}^{(n_{\text{remain}}-2d) \text{ coordinates}}).
\end{align} 
We thus have $\pi_B=\pi_2^*$. We now argue that a vector $\mathbf{k^*}$ satisfying conditions (i) and (ii) stated above minimizes \eqref{eq:inter5}. Note first that both $y_i(\pi_1^*)$ and $y_i(\pi_2^*)$ are non-decreasing with respect to $i$ according to our construction of the RFIP. Also, we always have $y_i(\pi_1^*)=y_i(\pi_2^*)$ for all $1\leq i\leq 2d$. 

\par We are now ready to discuss the structure of the optimal $\mathbf{k}$ vector. Since for each $b=1$ to $B$, we are summing up the first $(d-y_i(\pi_b))$ from $i=1$ to $k_b$ and in total there are $\sum_b k_b=k$ such terms, \eqref{eq:inter5} implies that to minimize \eqref{eq:low_b_plus} we would like to have as many terms corresponding to ``large $i$'' as possible in the summation $\sum_b k_b=k$ terms. If $k\leq2d$, this can be done if and only if we set all $k_b$ to 0 except for one $k_b$ value to be $k$, which is our construction (i). If $k>2d$, this can be done if and only if we set $k_B=\min(n_{\text{remain}},k)$ and, for $b=1$ to $B-1$, we set all $k_b$ to either $2d$ or $0$ except for one $k_b$. 

\par Knowing that $\mathbf{k^*}$ is of this special form, we can compute the RHS of \eqref{eq:low_b_plus} by
\begin{align}
\text{RHS of \eqref{eq:low_b_plus}}&=\left\lfloor \frac{k-\min(n_{\text{remain}},k)}{2d} \right\rfloor \text{sum}^{(1)} \nonumber\\
&+ \text{sum} ^{(2)}+\text{sum} ^{(3)},
\end{align}
where $\left\lfloor \frac{k-\min(n_{\text{remain}},k)}{2d} \right\rfloor$ is the number of $b$ from 1 to $B-1$ with $k_b=2d$ in the minimizing vector $\mathbf{k^*}$; $\text{sum}^{(1)}$ is the contribution to the min-cut value from those groups with $k_b=2d$, which is equal to $\sum_{i=1}^{2d}(d-y_i(\pi_1^*))\beta$; $\text{sum}^{(2)}$ is the contribution to the min-cut value from the single regular group with $k_b=(k-\min(n_{\text{remain}},k))\bmod (2d)$, which is equal to $\sum_{i=1}^{k_b}(d-y_i(\pi_1^*))\beta$; and $\text{sum}^{(3)}$ is the contribution to the min-cut value from the remaining group (group $B$), which is equal to 
\begin{align}
\text{sum}^{(3)}=\sum_{i=1}^{\min(n_{\text{remain}},k)}(d-y_i(\pi_2^*))\beta.
\end{align}
By plugging in the expressions of the RFIPs $\pi_1^*$ and $\pi_2^*$, we have
\begin{align}
\text{sum}^{(1)}&= \sum_{i=0}^{2d-2}\left(d-i+\left\lfloor \frac{i}{2}\right\rfloor\right)\beta=d^2\beta, \nonumber\\
\text{sum}^{(2)}&=\sum_{i=0}^{q}\left(d-i+\left\lfloor\frac{i}{2}\right\rfloor\right) \beta, \text{ and}\nonumber\\
\text{sum}^{(3)}&=\sum_{i=0}^{\min(k,2d-1)-1}\left(d-i+\left\lfloor\frac{i}{2}\right\rfloor\right)\beta,\label{eq:sum3}
\end{align}
where $q=((k-\min(n_{\text{remain}},k))\bmod(2d))-1=((k-n_{\text{remain}})^+\bmod(2d))-1$ and \eqref{eq:sum3} follows from the fact that $y_j(\pi_2^*)=d$ when $j\geq 2d$ and $n_{\text{remain}}\geq 2d+1$. The minimum repair-bandwidth $\beta_{\MBR}$ thus satisfies \eqref{eq:gamma_plus}.

\par Now, for the case when $n\bmod(2d)=0$, in a similar fashion, we can prove that a $\mathbf{k}$ vector minimizes the right-hand side of \eqref{eq:low_b_plus} at the MBR point if and only if there is at most one $b\in\{1,\cdots,B\}$ such that $0<k_b<2d$. By setting $\pi_b=\pi_1^*$ for all $b$ in \eqref{eq:inter5}, recall that $\pi_1^*$ is the RFIP for $(n,d)=(2d,d)$, we get  
\begin{align}
\text{RHS of \eqref{eq:low_b_plus}}&=d^2\left\lfloor \frac{k}{2d}\right\rfloor\beta+\sum_{i=0}^{(k\bmod(2d))-1}\left(d-i+\left\lfloor\frac{i}{2}\right\rfloor\right)\beta,
\end{align}
and thus $\beta_{\MBR}$ satisfies \eqref{eq:gamma_plus} for this case too.
The proof is hence complete.

\section{Proof of Proposition~\ref{prop:weak}} \label{app:weak_proof}
\par We first show that whenever $\alpha=d\beta$, we have
\begin{align} \label{eq:fpr_vs_fr}
\min_{G\in\mathcal{G}_{F^+}}\min_{t\in \DC (G)} \mincut_G(s,t)&\geq\nonumber\\ 
\min_{G\in\mathcal{G}_{F}}&\min_{t\in \DC (G) } \mincut_G(s,t),
\end{align}
where $\mathcal{G}_F$ is the collection of IFGs of an FR scheme $F$. That is, when $\alpha=d\beta$, the additional step of partitioning nodes into sub-groups in the family-plus scheme will monotonically improve the performance when compared to the original FR scheme without partitioning.

\par When $n<4d$, the family-plus repair scheme collapses to the FR scheme since each group of the family-plus scheme needs to have at least $2d$ nodes and when $n<4d$ we can have at most 1 group. Thus, trivially, we have \eqref{eq:fpr_vs_fr} when $n<4d$. Now, we consider the case when $n\geq 4d$. 

\par We first consider the original FR scheme (the RHS of \eqref{eq:fpr_vs_fr}). In this case, the FR scheme has $\left\lfloor \frac{n}{n-d}\right\rfloor=1$ complete family and one incomplete family. The corresponding RFIP $\pi_f^*$ is thus
\begin{align}
\pi_f^*=(\overbrace{1,0,1,0,\cdots, 1,0}^{2d \text{ coordinates}}, \overbrace{-1,-1,\cdots, -1}^{(n-2d) \text{ coordinates}})\nonumber.
\end{align}
By Proposition~\ref{prop:mbr}, we have

\begin{align}
\min_{G\in\mathcal{G}_{F}}\min_{t\in \DC (G)}\mincut_G(s,t)&=\nonumber\\ \sum_{i=0}^{\min(k,2d-1)-1}&\left(d-i+\left\lfloor\frac{i}{2}\right\rfloor\right)\beta,\label{eq:optimality_fr}
\end{align}
where \eqref{eq:optimality_fr} from the fact that $y_j(\pi_f^*)=d$ when $j\geq 2d$. 

\par We now turn our focus to the family-plus repair scheme. Consider first the case when $n\bmod(2d)=0$. If $k<2d$, we have by \eqref{eq:gamma_plus} and \eqref{eq:optimality_fr} that \eqref{eq:fpr_vs_fr} is true since the third term on the LHS of \eqref{eq:gamma_plus} is the RHS of \eqref{eq:optimality_fr}. If $k\geq 2d$, we again have by \eqref{eq:gamma_plus} and \eqref{eq:optimality_fr} that \eqref{eq:fpr_vs_fr} is true since the second term on the LHS of \eqref{eq:gamma_plus} is no less than the RHS of \eqref{eq:optimality_fr}. Now, consider the case when $n\bmod(2d)\neq 0$. Similarly, we have by \eqref{eq:gamma_plus} and \eqref{eq:optimality_fr} that \eqref{eq:fpr_vs_fr} is true since the first term on the LHS of \eqref{eq:gamma_plus} is the RHS of \eqref{eq:optimality_fr}.

\par We are now ready to prove \eqref{eq:weak}. If neither (i) nor (ii) of Proposition~\ref{prop:comparison} is true, we must have one of the three cases: (a) $d\geq 2$ and $k> \left\lceil \frac{n}{n-d}\right\rceil$; (b) $d=1$, $k> 2$, and even $n$; and (c) $d=1$, $k>3$, and odd $n$. For case (a), since $d> \left\lceil\frac{n}{n-d}\right\rceil-1$ whenever $2\leq d\leq n-2$ (see the proof of Corollary~\ref{cor:low_b} in Appendix~\ref{app:cor_proof}), we have that $\min(d+1,k)> \left\lceil\frac{n}{n-d}\right\rceil$. Considering the FR scheme, we thus have that among the first $\min(d+1,k)$ indices of a family index permutation $\pi_f$ there is at least one family index that is repeated. Jointly, this observation, Proposition~\ref{prop:low_b}, the MBR point formula in \eqref{eq:gamma}, and \eqref{eq:fpr_vs_fr} imply \eqref{eq:weak} when $\alpha=d\beta$. Note that $d=n-1$ is not possible in case (a) since we will have $k>\left\lceil\frac{n}{n-d}\right\rceil=n$, which violates \eqref{eq:ccw1}. For both cases (b) and (c), since $n\geq k$ by \eqref{eq:ccw1}, we have $n\geq 4$. The construction of the family-plus scheme thus will generate at least 2 groups. That is, the value of $B$ in Proposition~\ref{prop:low_b_plus} must satisfy $B\geq 2$. Moreover, in case (b), we have no remaining group since $n$ is even. Therefore, since $k>2$, for any $\mathbf{k}\in K$ defined in Proposition~\ref{prop:low_b_plus}, there are at least two distinct $b$ values with $k_b\geq 1$. In case (c), we have $k>3=n_{\text{remain}}$ (note that $n_{\text{remain}}=3$ since we have that $2d+1\leq n_{\text{remain}}\leq 4d-1$ by construction). Therefore, similarly, for any $\mathbf{k}\in K$ defined in Proposition~\ref{prop:low_b_plus}, there are at least two distinct $b$ values with $k_b\geq 1$.

\par Using the above observation (at least two distinct $b$ values having $k_b\geq 1$) and Proposition~\ref{prop:low_b_plus},  we have that in both cases (b) and (c)
\begin{align}
\min_{G\in\gfp} \min_{t\in\DC(G)} \mincut_G(s,t)\geq 2\min(d\beta,\alpha)> \min(\beta,\alpha),
\end{align}
where the first inequality follows from (i) considering only those $b$ values with $k_b\geq 1$; (ii) plugging in the min-cut formula in Proposition~\ref{prop:low_b}; and (iii) only counting the first term ``$i=1$'' when summing up for all $i=1$ to $k_b$. The second inequality follows from the assumption that $d=1$ in both cases (b) and (c) and the fact that both $\beta$ and $\alpha$ must be strictly positive. By noticing that for cases (b) and (c) the RHS of \eqref{eq:weak} is indeed $\min(\beta,\alpha)$, the proof is complete.

\section{Proof of Lemma~\ref{lem:gfr_existence}} \label{app:gfr_existence}

\begin{figure}[h!]
\centering
\includegraphics[width=0.45\textwidth]{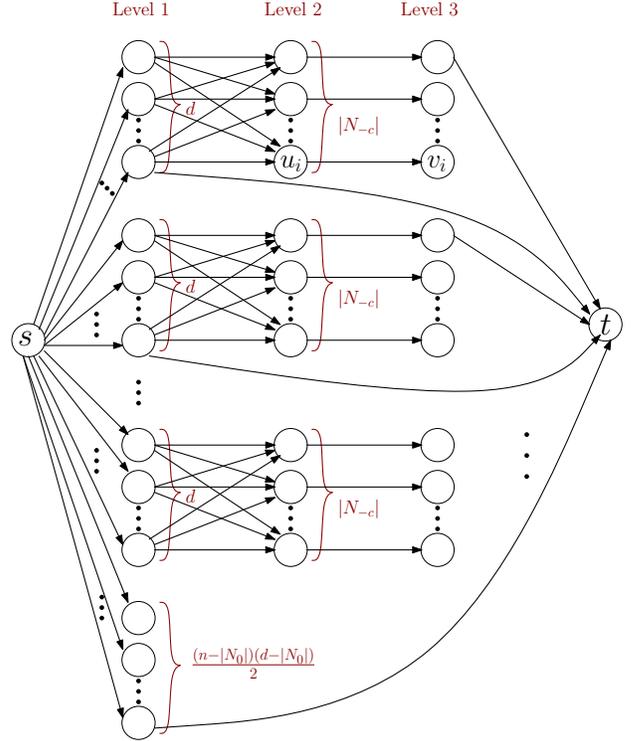} 
\caption{The graph of the proof of Lemma~2.}
\label{fig:gfr_existence}
\end{figure}

To prove this lemma, we model the problem using a finite directed acyclic graph and then we invoke the results from random linear network coding \cite{ho2006random}. The graph has a single source vertex $s$ that is incident to $|\bar{E}|=|{\sf IJ}^{[1]}|+|{\sf IJ}^{[2]}|=\frac{(n-|N_0|)(d-|N_0|)}{2}+d|N_0|$ other vertices with edges of capacity 1. We call these vertices \emph{level 1} vertices. Among these level 1 vertices, we form $|N_0|$ disjoint groups and each group consists of $d$ arbitrarily chosen distinct vertices. The idea is that each group of them is associated with a vertex in $N_0$. Note that there are $d|N_0|$ vertices forming $|N_0|$ groups while there are still $\frac{(n-|N_0|)(d-|N_0|)}{2}$ vertices that do not form any group at all. See Fig.~\ref{fig:gfr_existence} for illustration.  

\par Now, in addition to the source $s$ and the level 1 vertices, we add $|N_0|\cdot |N_{-c}|$ new node pairs $(u_i,v_i)$ for all $1\leq i\leq |N_0|\cdot |N_{-c}|$. Each $(u_i,v_i)$ is connected by an edge of capacity 1. We call the $u_i$ nodes, level 2 vertices and the $v_i$ nodes level 3 vertices. We partition the new node pairs (edges) into $|N_0|$ groups and each group consists of $|N_{-c}|$ edges. We then associate each group of $|N_{-c}|$ edges to one group of $d$ level 1 vertices created previously. See Fig.~\ref{fig:gfr_existence} for illustration. Finally, for the level 1, level 2, and level 3 vertices belonging to the same group (there are $|N_0|$ groups in total), we connect all the level 1 vertices in this group and all the level 2 vertices in this group by an edge with infinite capacity.  

\par We now describe the relationship of the newly constructed graph in Fig.~\ref{fig:gfr_existence} to the graph representation of the generalized fractional repetition code. For easier reference, we use “the graph in Fig.~\ref{fig:gfr_existence}” to refer to the newly constructed graph; and use “the graph in Fig.~\ref{fig:gfr_graph}” to refer to the graph representation of the generalized fractional repetition codes. There are $|N_0|$ groups in the graph of Fig.~\ref{fig:gfr_existence} and each group corresponds to one node in $N_0$ of the graph of Fig.~\ref{fig:gfr_graph}. We notice that there are $|\bar{E}|=\frac{(n-|N_0|)(d-|N_0|)}{2}+d|N_0|$ number of level 1 vertices in the graph of Fig.~\ref{fig:gfr_existence} and $|\bar{E}|=\frac{(n-|N_0|)(d-|N_0|)}{2}+d|N_0|$ number of edges in $\bar{E}$ of the graph of Fig.~\ref{fig:gfr_graph}. As a result, we map each level 1 vertex bijectively to an edge in $\bar{E}$. There are $|N_0|\cdot |N_{-c}|$ number of level 3 vertices in the graph of Fig.~\ref{fig:gfr_existence} and there are $|N_0|\cdot |N_{-c}|$ number of $\tilde{E}$ edges in the graph of Fig.~\ref{fig:gfr_graph}. As a result, we map each level 3 vertex bijectively to an edge in $\tilde{E}$.

\par We now focus on the graph of Fig.~\ref{fig:gfr_existence}. Assume that source $s$ has a file of $\mathcal{M}$ packets. We perform random linear network coding (RLNC) \cite{ho2006random} on the graph of Fig.~\ref{fig:gfr_existence} assuming a sufficiently large finite field $\text{GF}(q)$ is used. After we have finished the RLNC-based code construction on the graph of Fig.~\ref{fig:gfr_existence}, we now describe how to map the construction back to the edges in the graph of Fig.~\ref{fig:gfr_graph}. Specifically, the coded packet corresponding to $(s,u)$ where $u$ is a level 1 vertex in the graph of Fig.~\ref{fig:gfr_existence} is assigned to the edge $e\in \bar{E}$ (in the graph of Fig.~\ref{fig:gfr_graph}) corresponding to node $u$. We now consider the coded packets corresponding to $(u,v)$ where $u$ is a level 2 vertex and $v$ is a level 3 vertex in the graph of Fig.~\ref{fig:gfr_existence}. Without loss of generality, we assume that $(u,v)$ belongs to the $i_0$-th group in Fig.~\ref{fig:gfr_existence} and $v$ is the $j_0$-th level 3 vertex in this group. Then, we assign the coded packets on the edge $(u,v)$ to the edge $e\in \tilde{E}$ (in the graph of Fig.~\ref{fig:gfr_graph}) that connects the $i_0$-th node in $N_0$ and the $j_0$-th node in $N_{-c}$.  

\par In the following, we will prove that the above code construction (from the RLNC-based code in the graph of Fig.~\ref{fig:gfr_existence} to the generalized fractional repetition codes in the graph of Fig.~\ref{fig:gfr_graph}) satisfies Lemma~\ref{lem:gfr_existence}. 

\par To prove that the above construction satisfies Property~1, we notice that any coded packet $\tilde{P}_{(i_0,j_0)}$ corresponding to some $(i_0,j_0)\in {\sf IJ}^{[3]}$ in the graph of Fig.~\ref{fig:gfr_graph} is now mapped from a $(u,v)$ edge in Fig.~\ref{fig:gfr_existence} where $u$ is a level 2 vertex; $v$ is a level 3 vertex; $(u,v)$ belongs to the $i_0$-th group in Fig.~\ref{fig:gfr_existence}; and $v$ is the $j_0$-th level 3 vertex in this group. By the graph construction in Fig.~\ref{fig:gfr_existence}, such a coded packet is a linear combination of the coded packets in Fig.~\ref{fig:gfr_existence} from source $s$ to vertex $\tilde{u}$ where the $\tilde{u}$ vertices are the level 1 vertices corresponding to the $i_0$-th group. Since those packets along $(s,\tilde{u})$ are the $P_{(j_1,i_0)}$ packets for all $j_1$ satisfying $(j_1,i_0)\in {\sf IJ}^{[2]}$ in the graph of Fig.~\ref{fig:gfr_graph}, we have thus proved Property~1: Namely, any coded packet $\tilde{P}_{(i_0,j_0)}$ corresponding to some $(i_0,j_0)\in {\sf IJ}^{[3]}$ is a linear combination of the packets $P_{(j_1,i_0)}$ for all $j_1$ satisfying $(j_1,i_0)\in {\sf IJ}^{[2]}$.

\par To prove that the above construction satisfies Property~2, for any subset of edges in the graph of Fig.~\ref{fig:gfr_graph}, we place a sink node $t$ in the graph of Fig.~\ref{fig:gfr_existence} that connects to the corresponding set of level 1/level 3 vertices in Fig.~\ref{fig:gfr_existence} using edges of infinite capacity. See Fig.~\ref{fig:gfr_existence} for illustration of one such $t$. One can quickly verify that the min-cut-value from the source $s$ to the sink $t$ in the graph of Fig.~\ref{fig:gfr_existence} is the $\mathsf{a.count}$ value computed from the given subset of edges in the graph of Fig.~\ref{fig:gfr_graph}. As a result, with a sufficiently large finite field $\text{GF}(q)$, any sink $t$ satisfying $\mincut(s,t)=\mathsf{a.count}(t)\geq \mathcal{M}$ can successfully reconstruct the original file with close-to-one probability. Since the sink $t$ accesses only level 1 and level 3 vertices, the $P_{(i,j)}$ packets in the graph of Fig.~\ref{fig:gfr_graph} that correspond to the level 1 vertices in the graph of Fig.~\ref{fig:gfr_existence} and the $\tilde{P}_{(i,j)}$ packets in the graph of Fig.~\ref{fig:gfr_graph} that correspond to the level 3 vertices in the graph of Fig.~\ref{fig:gfr_existence} jointly can reconstruct the original file of size $\mathcal{M}$. Property~2 is thus also satisfied. 

\par By the above arguments, the proof of Lemma~\ref{lem:gfr_existence} is complete. 

\bibliography{paper}
\bibliographystyle{IEEEtranS}

\end{document}